\documentclass[a4paper]{iopart}
\usepackage{latexsym}
\usepackage{amsmath}           
\usepackage{amsfonts}          
\usepackage{amssymb}           

\usepackage[english]{babel}

\newif\ifpdf
\ifx\pdfoutput\undefined
\pdffalse 
\else
\pdfoutput=1 
\pdftrue
\fi

\ifpdf
\usepackage[pdftex]{graphicx}
\usepackage{epstopdf}
\else
\usepackage[dvips]{graphicx}
\fi

\newcommand{\ef}[1]{\, #1}     

\newcommand{\highlight}[1]{\makebox[0pt][c]
            {\rule{0pt}{11pt}\framebox{$\boldsymbol{#1}$}}}

\newcommand{\ldotsfitti}{\makebox[3pt][l]{.}\makebox[3pt][l]{.}\makebox[3pt][l]{.}}

\newcommand{\negg}[1]{\overline{#1}}
\newcommand{\bor}{\vee}
\newcommand{\band}{\wedge}
\newcommand{\xor}{\dot{\vee}}

\newcommand{\ket}[1]{\left| #1 \right\rangle}

\newcommand{\myappline}{\raisebox{3pt}{\rule{5.2cm}{0.5pt}} \\}
\newcommand{\mydictitem}[6]{\scriptsize \makebox[4mm][c]{#1}%
\makebox[3.cm][r]{#2}\makebox[1.1cm][r]{#3}%
\makebox[4.5cm][r]{$#4$}\makebox[0.6cm][r]{$#5$}
\quad \makebox[3.cm][l]{#6}\\}

\newcommand{\be}{\begin{equation}}
\newcommand{\ee}{\end{equation}}

\newcommand{\references}{
\section*{References}}

\newcommand{\eval}[1]{\left\langle {#1} \right\rangle}

\newcommand{\reval}[1]{\overline{#1}}

\begin{document}

\jl{1}

\def\PACS{{\it ACM-class: F.2.2}}

\article{}{One-in-Two-Matching Problem is NP-complete}


\author
{
  Sergio Caracciolo$^{[1]}$,
  Davide Fichera$^{[1]}$,
  Andrea Sportiello$^{[2]}$
}
\address{
\makebox[4mm][l]
{$[1]$}
Universit\`a degli Studi di Milano - Dip. di Fisica and INFN, \\
\makebox[4mm][l]{}
via Celoria 16, I-20133 Milano, Italy \\
\makebox[4mm][l]{$[2]$}
R.~Peierls Centre for Theoretical Physics, University of Oxford, \\
\makebox[4mm][l]{}
1 Keble Road, Oxford OX1 3NP, England \\
\makebox[2cm][l]{Mail address:}
\tt{Serg{}io.Carac{}ciolo@{}mi.infn.it} \\
\makebox[1.95cm][l]{}
\tt{David{}e.Fiche{}ra@mi.in{}fn.it} \\
\makebox[1.95cm][l]{}
\tt{And{}rea.Spo{}rtiell{}o@s{}ns.it}
}


\date{28${}^{\textrm{th}}$ April 2006}

\begin{abstract}
2-dimensional Matching Problem, which requires to find a matching of
left- to right-vertices in a balanced $2n$-vertex bipartite graph, is
a well-known polynomial problem, while various variants, like the
3-dimensional analogoue (3DM, with triangles on a tripartite graph),
or the Hamiltonian Circuit Problem (HC, a restriction to ``unicyclic''
matchings) are among the main examples of NP-hard problems, since the
first Karp reduction series of 1972. The same holds for the
weighted variants of these problems, the Linear Assignment Problem
being polynomial, and the Numerical 3-Dimensional Matching and
Travelling Salesman Problem being NP-complete.

In this paper we show that a small modification of the 2-dimensional
Matching and Assignment Problems in which for each $i \leq n/2$ it is
required that either $\pi(2i-1)=2i-1$ or $\pi(2i)=2i$, is a
NP-complete problem. The proof is by linear reduction from SAT (or
NAE-SAT), with the size $n$ of the Matching Problem being four times
the number of edges in the factor graph representation of the boolean
problem.  As a corollary, in combination with the simple linear
reduction of One-in-Two Matching to 3-Dimensional Matching, we show
that SAT can be linearly reduced to 3DM, while the original Karp
reduction was only cubic.
\end{abstract}

\noindent
\PACS



\section{Introduction}
\label{sec.intro}

Consider the problems encodable in the framework of the theory of
Turing machines (say, the problems which could be translated into a
computer program). Say that an instance of a problem has size $N$ if
it can be encoded into a sequence of $N$ bits, within a given
dictionary depending only on the problem.  In a few words, Complexity
Theory deals with the classification of these problems according to
the asymptotic behaviour in $N$ of the time of solution in the most
difficult instance of size at most $N$. This kind of analysis is
called \emph{worst-case analysis}, and is sided by an
\emph{average-case analysis}, where the time of the
solution is the average in a given measure over the ensemble of
instances of size $N$.

We can restrict our attention, from this large class of problems, to
problems of a special kind, such that, for any size $N$, a known
finite set $\mathcal{S}_N$ of ``feasible solutions'' (or
``configurations'') exist, and, given an instance, the problem is to
find the solution minimizing a certain integer-valued \emph{cost
function} (or \emph{objective function}); or, if the cost function is
valued in $\{ \textrm{True}, \textrm{False} \}$, to find a solution
evaluated to $\textrm{True}$ or a certificate that all solutions are
evaluated to $\textrm{False}$. Problems in the first and second class
are called respectively \emph{optimization} and \emph{decision
problems}. If, for each feasible solution, the time required for the
evaluation of its cost is finite, any of these problems allows at
least for one finite-time algorithm, the one which just tries
sequentially all the feasible solutions.

The class P is the class of problems such that the worst-case
complexity is bounded by a given polynomial in $N$.  Up to exotic
cases, you can think to the class of problems such that, for each
problem, a given algorithm has been explicitly provided, and has been
proven to have a worst-case complexity bounded by the given
polynomial. This class has a pleasant property: suppose that two
problems $A$ and $B$ are given such that:
\begin{itemize}
\item problem B is known to be of polynomial complexity, with a
  certain polynomial $P_{B}(N)$;
\item an \emph{encoding} map exists,
  which encodes a size-$N$
  instance of problem $A$ into an instance of problem $B$ of size at
  most $Q(N)$, in a time $o(P_B(Q(N)))$;
\item a \emph{decoding} map exists,
  which decodes a whatever
  optimal solution of an instance of problem $B$, obtained from the
  encoding above, into an optimal solution of the original problem A,
  in a time $o(P_B(Q(N)))$;
\end{itemize}
Then the algorithm consisting of the encoding, followed by problem-$B$
algorithm, and finally the decoding, results into a polynomial-time
algorithm for problem $A$, with polynomial $P_A(N)=P_B(Q(N))$. This
idea leads to the concept of \emph{polynomial reduction} of problems.

In the theory of Turing machines, the traditional paradigm is sided by
the definition of 
\emph{non-deterministic Turing machines}~\cite{GJ, Sipser}.
The reader used to programming can think to these machines as
ordinary computer programs, with the formal rule that the time
of a forked process is the maximum of the two forked times (as if
the forked processes were running on different processors), instead of
the sum of the two times (as if
the forked processes were running sequentially on a unique processor).
The class NP of problems is the class such that the complexity,
calculated within the rule above, is polynomial (indeed, NP stands
for \emph{non-deterministic--polynomial-time}).
The idea of polynomial reduction holds also
for this class of problems, as it can be applied separately to each
branch of the process.

Then, the crucial theorem by Cook~\cite{cook71} states that any NP
problem can be polynomially reduced to the \emph{Boolean
Satisfiability Problem}, i.e.~to the decision problem given by
$n$ boolean variables $u_i$ (i.e.~\emph{literals} on the alphabet 
$\{ \textrm{True}, \textrm{False} \}$), and a statement in disjunctive
normal form with $m$ clauses (i.e.~a string of the kind $(u_1 \bor u_5
\bor \negg{u}_7) \band (u_2 \bor \negg{u}_3 \bor u_9 \bor u_{13})
\band \ldots$ with $m$ parentheses), a solution being a boolean
assignment to the $n$ variables which satisfies the statement.
Despite many efforts, it has not been determined whether the class NP
coincides with the class P (P$\,=\,$NP), or is definitely wider
(P$\,\subsetneq\,$NP), although the present intuition is maybe in the
direction of stating that P$\,\subsetneq\,$NP. Thus, understanding the
roots of complexity commons to problems in the NP class, in order to
compare them to the characteristics of the P class, has become an
important direction of research of last decades.

In particular, since every NP problem is polynomially equivalent to
Boolean Satisfiability, it is also polynomially equivalent to any
other problem which encodes Boolean Satisfiability through a (chain
of) polynomial reductions. This induces the concept of
\emph{NP-completeness} for problems with this characteristic: via Cook
theorem and the corresponding chain of reductions, a hypothetic
polynomial-time algorithm for any NP-complete problem would
automatically provide a polynomial-time algorithm for \emph{all}
problems in the NP class, thus proving that P$\,=\,$NP.  Since the
first papers by Karp~\cite{Karp72, Karp75}, the family of NP-complete
problems has grown, up to lead to the famous large collection in Garey
and Johnson book~\cite{GJ}, and, as can be easily argued, has still
grown in the 25 following years~\cite{tutti}.

\subsection{Motivations and a digression}

In this paper we describe a new NP-complete problem, the One-in-Two
Matching, and give a polynomial reduction from Boolean Satisfiability.
A detailed description, section by section, of the contents of the
paper, is postponed to the conclusions, in section~\ref{sec.conc}, while
here we mostly concentrate on the motivations.

What is the reason, in present days, for such a work? Of course, the
fact is interesting by itself, in the idea of making the list of
NP-complete problems still wider (a larger list of NP-complete
problems gives a larger number of possible starting points for a
reduction proof, and thus makes easier the task of determing whether a
new problem is NP-complete).  As we will see, as a side result we show
that One-in-Two Matching Problem induces a chain of reductions from
Boolean Satisfiability to the important 3-Dimensional Matching Problem
(3DM in Garey-Johnson~\cite{GJ}) which avoids the complicated and
size-demanding original reduction by Karp~\cite{Karp72}.

Another motivation, which was indeed the original one, is that
One-in-Two Matching and Assignment give a hint on the structural
reason why Hamiltonian Circuit (HC) and Travelling Salesman (TSP) are
NP-complete, although their analogue Matching and Assignment are
polynomial, and intrinsecally simple for what concerns the energetic
landscape of configurations~\cite{KarpSipser, shah, MezLenka}. 
Indeed, configurations of HC and TSP are permutations
$\pi$ composed of a single cycle. It is trivial to impose that $\pi$
has no fixed points (just taking infinite weights on the diagonal),
while the remaining restriction to have no cycles of length $2 \leq
\ell < n$ must be at the root of the complexity discrepancy among the
two problems.

In the definition of One-in-Two problems, put at infinity the
non-diagonal block elements, $w_{2i-1,2i}=w_{2i,2i-1}=+\infty$, make
the diagonal weight very favourite, $w_{ii} \to w_{ii} - \Delta$ with
$\Delta \to +\infty$, and then transpose all the row pairs
$(2i-1,2i)$. Then, the One-in-Two constraints is equivalent to forbid
all the length-2 cycles among $2i-1$ and $2i$ (which are a small
subset of all the cycles forbidden in TSP). On the other side, the
matching $\pi(2i)=2i-1$, $\pi(2i-1)=2i$ constitutes the obvious ground
state of pure Assignment.

So, although \emph{random} TSP instances have good heuristics and
efficient approximants, based on the connection with
Assignment~\cite{KarpPatchTSP, KarpTSP}, this new ensemble of random
TSP instances would be hard in the average case, as, instead of just
condensing a few ($\mathcal{O}(\ln n)$) relevant long cycles, it must
first choice how to disentangle $\mathcal{O}(n)$ robust short
cycles. Roughly speaking, as a cycle of length $\ell$ can be broken in
$\ell$ points, and the set of cycle lengths is a partition of $n$, the
average complexity of this procedure scales with 
$\exp(n/\reval{\ell}\, \ln \reval{\ell})$, and has a finite maximum
for $\reval{\ell} = \mathcal{O}(1)$.  Similar arguments are depicted at
the end of section \ref{ssec.oneintwodef}~\cite{ourTSPinprep}.

Indeed, the existence of a NP-completeness proof for One-in-Two
problems suggests that this narrow subset of the set of extra
constraints of HC and TSP already contains the core of extra
complexity of these problems. Also remark that, although the chain of
reductions from SAT to Hamiltonian Circuit (3-SAT $\to$
Vertex-Covering $\to$ HC) is beautiful and elegant~\cite{GJ}, our
direct reduction to One-in-Two Matching is much cheaper.

\emph{A posteriori,} maybe the main motivation for this work lies in
how similar the problem is to situations arising in real-world
experiments, to be compared to graph-, number- or set-theoretical
problems, which require a level of mathematical abstraction. This
makes One-in-Two Matching particularly suitable for a Quantum
Adiabatic Algorithm (QAA) implementation
\cite{farhi00, farhi01, vazirani, vandam, Aaronson}. 

Quantum Adiabatic techniques are an alternative to more traditional
Quantum Computing ideas.
The former would solve a given NP decision
problem, measuring observables on a quantum system which \emph{directly}
encodes the desired objective function (more precisely, a
final-time Hamiltonian would reproduce the objective function, while
an initial-time Hamiltonian is such that the ground state can be
prepared, and the intermediate evolution is sufficiently slow that, by
the quantum adiabatic theorem, with finite probability
the final state encodes a solution of the problem).

The latter, instead, would implement a quantum algorithm in
a sequence of logical gates, in a kind of ``quantum logic circuit'',
employing quantum bits (\emph{qubits}, i.e.~two-state quantum systems)
interacting only locally, in the single gate.  On one side, the
existence of Quantum Error Correction techniques is what makes the
mainstream Quantum Computing idea promising, as it would protect us
from the unavoidable decoherence of single qubits (anyhow,
cfr.~\cite{farhichilds} for Error Correction on QAA). 
The drawback is the
requirement of a large number of identical quantum gates, and quantum
analog of electric wires, to be arranged in a complex circuit, all
within a support that does not add too much decoherence: 
while modern ``classic'' electronics allows for extremely complicated
circuit patterns, we lack for a quantum device with a similar
flexibility.

Although the possibility that a QAA paradigm, as well as any other
classical or quantum \emph{gedankenexperiment}, could solve NP
complete problems in polynomial time, is quite remote~\cite{Aaronson},
it has shown promising features on less ambitious tasks (cfr.~mainly
\cite{vandam}). So, the Quantum Adiabatic idea remains a promising
field, provided that the mapping to the physical device is done
starting from an ``experimentally simple'' NP decision problem.  We
can outline some points which could describe this vague idea:

\begin{description}
\item[modularity:] the easiest realizations are the ones in which it
  is required to design a
  very small number of elementary objects, arranged in a regular
  pattern. This \emph{excludes} problems encoded on general graphs,
  unless we allow to deal with the (generally much larger) array which
  encodes the graph structure, i.e.~for example the adjacency matrix
  of the graph, or the incidence matrix between vertices and edges.
  (So, the adjacency structure of a graph, required in a
  graph-theoretical NP-complete problem, would be not cheaper than our
  array structure).

\item[low dimensionality:] two-dimensional arrays of electronical
  devices are experimentally much more accessible than
  three-dimensional ones, and infinitely more accessible than
  higher-dimensional ones.  Conversely, one-dimensional arrays, with
  elements having a short-range interaction, are probably
  insufficient, as the corresponding ground state is at all times a
  tree state \cite{Aaronson_thesis}.  Our array would be
  two-dimensional. Cfr.~\cite{vandam} for a related discussion.

\item[reusability:] the ``hard and expensive'' experimental work for
  producing the device should allow to build a machine which could
  encode \emph{any} instance of the chosen NP-complete problem, up to
  a given size. Then, there should exist a relatively ``cheap and
  fast'' machine, allowing to set up the specific instance, such that
  a given device can be used many times for different instances, and
  also for the many variants of the instance coming from the small
  changes of parameters (for example, variations in time) in the
  real-life problem one is interested into. In our case, the set of
  all constraints should be implemented on the device, and the set of
  weights should be encoded in tunable local ``external fields'', of
  some physical nature, on the entries of the grid.
\item[small encoding:] the theoretical idea of polynomial reduction
  could be uneffective in real-life applications 
  for large polynomials (large coefficients
  and/or large order). One should search for a NP-complete problem
  whose reduction from the original SAT universal problem is as small
  as possible, and it is strongly adviced to be \emph{linear} in the
  encoding size of the SAT instance,
  i.e.~approximatively the number of logic operators (AND and OR) in
  the logic statement. We will see how we have a linear encoding from
  generic SAT, and, in section~\ref{sec.more}, how the encoding could
  be still improved in real-life applications.
\item[no fine-tuning:] the Hamiltonian should not reproduce the cost
  function through an exact cancellation of dominant terms, or a
  fine-tuning of contributions coming from different physical terms (say,
  like reproducing a 2-SAT clause by mean of a two-body interaction
  among two spins, and the contribution of an external magnetic field
  \emph{with a specific fine-tuned value} which makes $(T,T)$, $(T,F)$
  and $(F,T)$ of degenerate cost), because this is unlikely to be
  faithfully reproduced by the experimental device. The unavoidable
  degeneracies of the cost function in a complex decision problem must
  derive uniquely from symmetry properties of the device
  components. This point will be shortly discussed in
  section~\ref{sec.qaa}.
\item[locality:] it is severely hard to reproduce non-local
  constraints, like the one of Hamiltonian Path or of Number
  Partitioning Problems, in a physical
  system, just because most of the interactions under experimental
  control have a short range. Our new One-in-Two constraint is highly
  local, as it involves occupation numbers of two $(i,i)-(i+1,i+1)$ 
  neighbours on the grid; the traditional Matching constraint can
  be reproduced by many mechanisms, of which the most simple is just
  creating at the beginning one ``row''-particle for each row
  (and resp.~for columns), and implement a dynamic with conserved
  number of particles (cfr.~section~\ref{sec.qaa}).
\end{description}

\noindent
So, as it will be clear from the paper, contrarily to many other
NP-complete problems, our One-in-Two Matching seems to meet a large
number of these requirements, and it is maybe not hazardous to
auspicate that it could be a viable candidate for a large-size
experimental implementation of the Quantum Adiabatic Algorithm, if any
will ever be pursued.\footnote{It is worth remarking that QAA
  procedure on small instances has already been tried
  in~\cite{vandamExp}.}

\subsection{Definition of 2-Dimensional Matching and Linear Assignment
Problems}
\label{ssec.defMatch}

Given an unoriented bipartite graph $G(V_1, V_2; E)$, with $V_1$ and
$V_2$ being the sets of the two kinds of vertices, both of size $n$,
and $E \subseteq V_1 \times V_2$ the set of edges,
the \emph{Matching Problem}~\cite{LP} asks for a spanning subgraph
$G(V_1,V_2;M)$, $M \subseteq E$ with cardinality $n$, such that
each vertex has degree exactly 1, or for a certificate that
such a subgraph does not exist. A set $M$ satisfying this requirement is
called a \emph{perfect matching} over $G$. Here and in the following,
we describe a problem in the scheme
\[
\begin{array}{l|cl}
\multicolumn{3}{l}{\textrm{\bf Problem name:}} \\
\hline
\rule{0pt}{14pt}
\textrm{instance} && \textrm{feasible solution;} \\
\textrm{description.} && \textrm{condition to satisfy.}
\end{array}
\]
then the scheme corresponding to Matching is
\[
\begin{array}{l|cl}
\multicolumn{3}{l}{\textrm{\bf Matching Problem:}} \\
\hline
\rule{0pt}{14pt}
G(V_1, V_2; E)                 && M \subseteq E ; \\
\textrm{with } |V_1|=|V_2|=n.  && |M|=n, \quad 
                                 \forall v \in V_1 \cup V_2 \quad \deg_M(v)=1.
\end{array}
\]
This problem is a specific case of the more general Linear Assignment
Problem, in which
integer weights $w(e)$ are associated to the edges, a threshold
value $k$
is given, and the search is restricted to
perfect matchings $M$ such that the sum of weights on the edges
of $M$ is smaller than $k$~\footnote{Equivalently, one can restrict to 
consider the complete balanced bipartite graph with $2n$ vertices,
$\mathcal{K}_{n,n}$, and set $w(e)=+\infty$ for edges in 
$E(\mathcal{K}_{n,n}) \smallsetminus E(G)$.}.
\[
\begin{array}{l|cl}
\multicolumn{3}{l}{\textrm{\bf Linear Assignment Problem:}} \\
\hline
\rule{0pt}{14pt}
G(V_1, V_2; E)                 && M \subseteq E ; \\
\textrm{with } |V_1|=|V_2|=n;  && |M|=n, \quad 
                                 \forall v \in V_1 \cup V_2 \quad
                                 \deg_M(v)=1, \\
w : E \to \mathbb{Z};          && \sum_{e \in M} w(e) \leq k. \\
k \in \mathbb{Z}.              &&
\end{array}
\]
More precisely, the weights could be also infinite, 
i.e.~$w : E \to \mathbb{Z} \cup \{+\infty\}$, with the natural formal
rules $n+\infty=+\infty+n=+\infty+\infty=+\infty$ and $+\infty>k$.
Then, one can assume without loss of generality that $G$ is the
complete balanced bipartite graph.

In traditional notations, Matching Problem is resumed in the case 
$w(e)=1$ for edges in $E(G)$ and $w(e)=0$ otherwise, $\leq$ being
replaced by $\geq$, and $k=n$.

A convenient representation of these problems is via the 
$n \times n$ matrix $W=\{ w_{ij} \}$ of the weights. A feasible
matching $M$ is then described by a matrix $X=\{ x_{ij} \}$,
with $x_{ij} = 0, 1$ and exactly one element equal to 1 per row and
per column, such that $x_{ij}=1$ if edge $e=(i,j)$ is in $M$. 
The cost function for $M$ is restated into
\be
C_W(X)= \sum_{i,j} w_{ij} x_{ij} = \tr W X^{\rm T}
\ef.
\ee
Another convenient representation of feasible matchings is via
permutations in the symmetric group over $n$ elements, 
$\pi \in \mathcal{S}_n$, where $\pi(i)=j$ if edge $(i,j)$ is in
$M$. In this notation the cost function reads
\be
C_W(\pi)= \sum_{i} w_{i \pi(i)}
\ef.
\ee
An example of problem instance and solution could be (on the left,
items $w_{i \pi(i)}$ are written in bold)
\begin{equation*}
W=
\begin{pmatrix}
3 & 7 & \highlight{2} & 4 & 1 & 1 \\
\highlight{1} & 6 & 1 & 7 & 8 & 2 \\
3 & \highlight{3} & 2 & 5 & 6 & 3 \\
4 & 2 & 8 & 6 & \highlight{2} & 5 \\
5 & 5 & 1 & 6 & 3 & \highlight{4} \\
4 & 9 & 8 & \highlight{1} & 4 & 3 
\end{pmatrix}
\textrm{ with $k=15$;}
\qquad
\begin{array}{ll}
\{ \pi(i) \}_{i=1, \ldots, 6}=
\{ 3,1,2,5,6,4 \} \ef; \\
\rule{0pt}{15pt}C(\pi)=13.
\end{array}
\end{equation*}
Given a whatever Assignment instance $W$,
many algorithms allow to find in polynomial time the optimal
assignment $\pi^*$, and its cost $C^*=C(\pi^*)$, for example the
\emph{Hungarian Algorithm}~\cite{kuhn, LP}.

\subsection{Definition of One-in-Two Matching and One-in-Two
Assignment Problems}
\label{ssec.oneintwodef}

Now we can define the 
\emph{One-in-Two Matching} and \emph{Assignment Problems} 
as the variant of Matching (resp.~Assignment) Problem in which,
assumed that the dimension $2n$ of the matrix is even,
the set of allowed partitions $\pi$ is restricted to include only the
ones such that, for each $i=1, \ldots, n$, either $\pi(2i-1)=2i-1$ or
$\pi(2i)=2i$. 

Thus the description of One-in-Two Matching and Assignment 
could be resumed in the tables
\[
\begin{array}{l|cl}
\multicolumn{3}{l}{\textrm{\bf One-in-Two Matching Problem:}} \\
\hline
\rule{0pt}{14pt}
G(V_1, V_2; E)                 && M \subseteq E ; \\
\textrm{with } |V_1|=|V_2|=n;  && |M|=n, \quad 
                                 \forall v \in V_1 \cup V_2 \quad
                                 \deg_M(v)=1, \\
\textrm{partition of $\{V_1; V_2\}$ into} && \\
\textrm{quadruplets $q=(v,v';u,u')$.} && 
\forall q \quad 
\big( (v,u) \in M \big) \ \xor\ \big( (v',u') \in M \big) .
\\
\end{array}
\]
\[
\begin{array}{l|cl}
\multicolumn{3}{l}{\textrm{\bf One-in-Two Assignment Problem:}} \\
\hline
\rule{0pt}{14pt}
G(V_1, V_2; E)                 && M \subseteq E ; \\
\textrm{with } |V_1|=|V_2|=n;  && |M|=n, \quad 
                                 \forall v \in V_1 \cup V_2 \quad
                                 \deg_M(v)=1, \\
\textrm{partition of $\{V_1; V_2\}$ into} && \\
\textrm{quadruplets $q=(v,v';u,u')$;} && 
\forall q \quad 
\big( (v,u) \in M \big) \ \xor\ \big( (v',u') \in M \big) ,
\\
w : E \to \mathbb{Z};          && \sum_{e \in M} w(e) \leq k. \\
k \in \mathbb{Z}.              &&
\end{array}
\]
Remark that, when proven that One-in-Two Assignment is NP-complete, we
will also have a proof that the variant with 
$\big( (v,u) \in M \big) \ \bor\ \big( (v',u') \in M \big)$
instead of 
$\big( (v,u) \in M \big) \ \xor\ \big( (v',u') \in M \big)$
is NP-complete. Indeed, given an instance of the OR problem, shifting
the diagonal weights to $w_{ii} \to w_{ii} + \Delta$, 
and $k \to k + n \Delta$, we have that $C_W(M)-k = n' \Delta$, with
$n'$ the number of blocks with two matched elements, and in the limit
$\Delta \to +\infty$ we recover the analogous XOR problem. On the
contrary, in the limit $\Delta \to -\infty$ the problem becomes
trivial.

The costs of the diagonal elements do not play any role, and
can be fixed to zero. Indeed,
if for some $i \leq n$ we have $w_{2i, 2i}=w_{2i-1, 2i-1}= +\infty$,
no finite-cost
assignment exists, while if only one of the two is infinite (say,
$w_{2i-1, 2i-1}$), we are forced
to fix the permutation on the other one ($w_{2i, 2i}$), and thus
the elements with $i$ or $j$ equal to $2i$ never play a role: we
would have had an identical cost function if $w_{2i-1, 2i-1}$ were zero, and
$w_{2i,j}=w_{j,2i}=+\infty$ for each $j \neq 2i$:
\be
W=
\left(
\begin{array}{cc|cc}
+\infty & w_{12} & w_{13} & \hdots \\
 w_{21} & \highlight{w_{22}} & w_{23} & \hdots \\
\hline
 w_{31} & w_{32} & w_{33} & \hdots \\
 \vdots & \vdots & \vdots & \ddots
\end{array}
\right)
\equiv
\left(
\begin{array}{cc|cc}
      0 & +\infty & w_{13} & \hdots \\
+\infty & \highlight{w_{22}} & +\infty & \hdots \\
\hline
w_{31} & +\infty & w_{33} & \hdots \\
 \vdots & \vdots & \vdots & \ddots
\end{array}
\right)
\ef.
\ee
Thus, without loss of generality, we can assume that $w_{ii}$ is
finite for each $i$.  Then, from the invariance of Linear Assignment,
one easily convince himselfs that an equivalent instance (i.e.~an
instance with identical cost function, up to an overall constant) can
be produced, with $w_{ii}=0$ for each $i \leq 2n$, and that the values
$w_{2i, 2i-1}$ and $w_{2i-1, 2i}$ never appear in allowed
matchings. For this reason, we will assume in the following that
$w_{ii}=0$, and denote the four elements in the $n$ diagonal blocks of
size 2
with special symbols $*$ and $\cdot$, instead that with a weight
value.
For example,
an instance with $2n=6$ could be
\begin{equation*}
W=
\left(
\,
\begin{array}{cccccc}
\cline{1-2}
\multicolumn{1}{|c}{*} & \multicolumn{1}{c|}{\cdot} & {2} & 4 & 1 & 1 \\
\multicolumn{1}{|c}{\cdot} & \multicolumn{1}{c|}{*}  & 1 & 7 & 8 & 2 \\
\cline{1-4}
3 & {3} & \multicolumn{1}{|c}{*} & \multicolumn{1}{c|}{\cdot}
& 6 & 3 \\
4 & 2 & \multicolumn{1}{|c}{\cdot} & \multicolumn{1}{c|}{*} &
{2} & 5 \\
\cline{3-6}
5 & 5 & 1 & 6 & \multicolumn{1}{|c}{*} & \multicolumn{1}{c|}{\cdot} \\
4 & 9 & 8 & {1} & \multicolumn{1}{|c}{\cdot} &
\multicolumn{1}{c|}{*} \\
\cline{5-6}
\end{array}
\,
\right)
\textrm{ with $k=10$;}
\end{equation*}
The choice of representation with $*$s is done for mnemonic reasons: at
sight, one knows that a valid matching should use exactly one $*$ per block.
For example, a valid matching with weight 9 could be the following (on the
right side, elements
$i$ such that $\pi(i)=i$ are underlined in order to highlight the
one-in-two constraint satisfaction)
\begin{equation*}
\left(
\,
\begin{array}{cccccc}
\cline{1-2}
\multicolumn{1}{|c}{*} & \multicolumn{1}{c|}{\cdot} &
\highlight{2} & 
4 & 1 & 1 \\
\multicolumn{1}{|c}{
\cdot} & \multicolumn{1}{c|}{{\highlight{*}}}  & 1 & 7 & 8 & 2 \\
\cline{1-4}
3 & 3 & \multicolumn{1}{|c}{*} & \multicolumn{1}{c|}{\cdot}
& 6 & \highlight{3} \\
4 & 2 & \multicolumn{1}{|c}{\cdot} & 
\multicolumn{1}{c|}{{\highlight{*}}} & 2 & 5 \\
\cline{3-6}
5 & 5 & 1 & 6 & \multicolumn{1}{|c}{{\highlight{*}}} & 
\multicolumn{1}{c|}{\cdot} \\
\highlight{4} & 9 & 8 & 1 & \multicolumn{1}{|c}{\cdot} &
\multicolumn{1}{c|}{*} \\
\cline{5-6}
\end{array}
\,
\right)
\qquad
\begin{array}{ll}
\{ \pi(i) \}_{i=1, \ldots, 6}=
\{ 3,\underline{2};6,\underline{4};\underline{5},1 \} \ef; \\
\rule{0pt}{15pt}C(\pi)=9.
\end{array}
\end{equation*}
Clearly, a one-in-two matching can be described by a choice of the
elements kept fixed by the permutation (i.e.~the $*$ chosen in each block),
times an allowed choice of assignment in the
$n$-dimensional minor matrix resulting from the removal of the
fixed rows and columns. Thus, allowed matchings are in bijection with
pairs $(\vec{\sigma}, \pi)$, where 
$\vec{\sigma} \in \{ 0,1\}^n$ 
and
$\pi \in \mathcal{S}_{n}$, with all $\pi(i) \neq i$.
For example, the previous configuration could
be described as 
$( \vec{\sigma}, \{\pi(i) \}_{i=1,\ldots, 3} )
=((0,0,1), \{2,3,1 \} )$.
This fact suggests a na\"ive interpretation for
the potential hardness of this variant of Assignment: for
any choice of fixed elements (the $*$s), the problem of finding the optimal
assignment is polynomial (just put $+\infty$ on the remaining $*$s,
and use Hungarian Algorithm on the resulting Linear Assignment instance),
nonetheless one should perform a search among these
$2^n$ possible choices, which, in absence of a sufficiently strong
correlation or a skill mathematical structure, could make the search
exponential in size. We will come back on this point in
section~\ref{sec.roots}.

\section{Proof of linear reduction from Boolean Satisfiability
  problems}
\label{sec.redproof}

Here we prove that also One-in-Two Matching, less general w.r.t.~the
analogue One-in-Two Assignment, allows for linear reduction from
arbitrary instances of SAT, 3-SAT or NAE-3-SAT Problems.

A SAT (or NAE-SAT) instance with $n$ literals and $m$ clauses can be
encoded into a bipartite graph $G(V_{\ell}, V_{c}; E)$, with 
$V_{\ell}$ being the set of literals $\{u_i \}_{i=1, \ldots, n}$ and
$V_c$ the set of clauses $\{ C_a \}_{a=1, \ldots, m}$, and a map 
$s: E \to \{ \pm 1 \}$ which states whether the literal enters negated or
not, i.e.~if $u_i \in C_a$ then $s(i,a)=+1$, while if 
$\negg{u}_i \in C_a$ then $s(i,a)=-1$.

A satisfiability instance can be presented in the form of
a factor graph, i.e.~the bipartite graph above, where
vertices in $V_{\ell}$ are denoted by small circles, and
vertices in $V_{c}$ by small squares, and an edge is drawn
in solid line if $s(e)=+1$, and dashed if $s(e)=-1$.
The instance is satisfied by a given boolean assignment if
a local constraint is satisfied on each clause vertex.
For the SAT problem, given a clause with $k$ incident variables,
the number of solid-edge true neighbours plus dashed-edge false
neighbours must be at least 1. For NAE-SAT it must be at least 1, and
at most $k-1$.

In order to perform our reduction, it is easier to first perform a
decoration on the graph: for each variable, introduce an auxiliary
variable per incident edge (in a sense, the literal ``as seen from the
clause''), then substitute the original variable node by a
``consistency check'' clause (drawn as a small triangle), which
ensures that the boolean values on the copies of the variable
coincide. A small example of SAT factor graph, and the corresponding
decorated graph, could be the following:
\[
\setlength{\unitlength}{20pt}
\begin{picture}(14.5,3.6)(0,0)
\put(0.,0.){\includegraphics[bb=23 4 457 107, clip=true, scale=.666]
{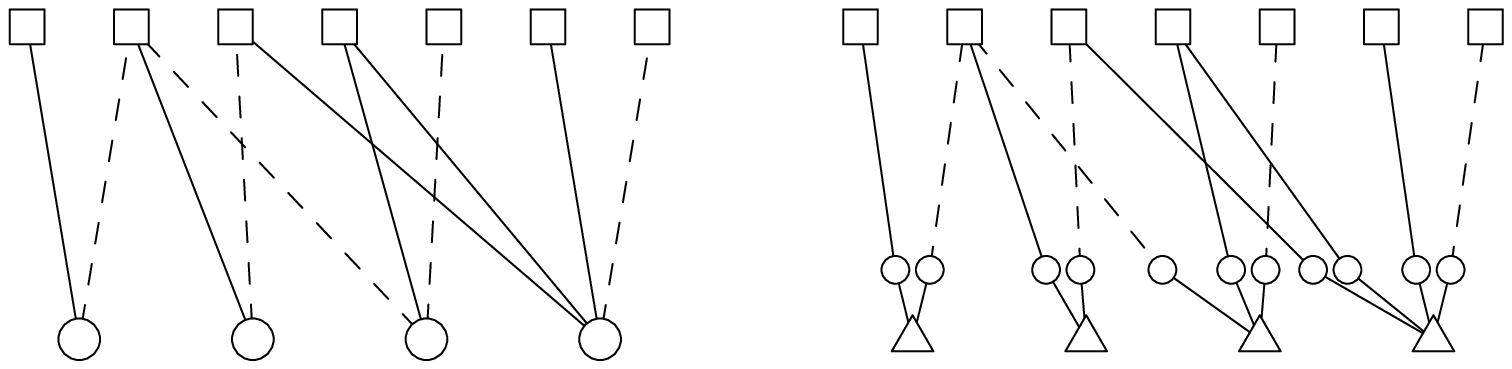}}
\end{picture}
\]
In our reduction, we have two $2 \times 2$ blocks
$\big( \begin{smallmatrix} * & \cdot \\ \cdot & * \end{smallmatrix}
\big)$ for each edge $(i,a) \in E(G)$, and thus the entries of the matrix
$W$ are labeled by an index $(i,a)^{\pm}_{1,2}$, where $(1,2)$ stands
for the first and second block, and $\pm$ stands for the
first and second index inside the block. 
Suppose to order arbitrarily the edges 
$e_{\alpha}=(i_{\alpha}, a_{\alpha})$, then
for helping visualization, we will
assume that the entries of $W$ are ordered as 
\be
((i_1, a_1)^+_1, (i_1, a_1)^-_1, \ldots, 
  (i_k, a_k)^+_1, (i_k, a_k)^-_1,
\,
  (i_1, a_1)^+_2, (i_1, a_1)^-_2, \ldots, 
  (i_k, a_k)^+_2, (i_k, a_k)^-_2)
\ef,
\ee
that is, first all the 1-blocks,
then, in the same order, all the 2-blocks.

The choice of $*$s in the matching corresponds to the sequence of
boolean assignments for the literals, ``as they are seen from the
clause'', i.e.~for the literals in the decorated instance.  So we need
a ``truth-setting'' structure,
which ensures that all these values coincide (in other words,
implements the ``consistency check'' clause), and a
``satisfaction-testing'' structure,
which checks that each boolean clause in the original formula is
satisfied.  The truth-setting structure is encoded in the set of 1s in
the entry pairs with $i=j$, while the satisfaction-testing structure
is encoded in the set of 1s in the entry pairs with $a=b$.  More
precisely, entries $w=1$ can appear in the off-block matrix elements
at index pairs $((i,a)^{\pm}_{\alpha}, (j,b)^{\pm}_{\beta})$ only in
one of the two cases:

\begin{tabular}{ll}
$i=j$, $\alpha=2$ and $\beta=1$
&
(truth-setting structure) \\
$a=b$, $\alpha=1$ and $\beta=2$
&
(satisfaction-testing structure)
\end{tabular}

\noindent
We start describing the truth-setting structures.  For each variable
$i$, call $A(i)$ the set of adjacent clauses, and choose an arbitrary
cyclic ordering on this set. For $i \in V_{\ell}$ and $a,b \in A(i)$
we state
\be
w_{(i,a)^{\sigma}_{\alpha}, (i,b)^{\tau}_{\beta}}
=
\left\{
\begin{array}{ll}
1 & 
  \begin{array}{llll}
  a=b,   & \sigma=\tau=-, & \alpha=2, & \beta=1; \\
  a=b-1, & \sigma=\tau=+, & \alpha=2, & \beta=1;
  \end{array} \\
0 & \rule{0pt}{5mm}\textrm{ otherwise.}
\end{array}
\right.
\ee
that is, the minor of $W$ restricted to indices with fixed $i$ looks like
\begin{align}
W|_{\textrm{fixed $i$}} &=
\left(
\begin{array}{c|c}
I_* & 0 \\
\hline
\rule{0pt}{12pt}
W' & I_*
\end{array}
\right)
\ef;
&&
\\
I_* &=
\begin{pmatrix}
\makebox[10pt][c]{$\begin{smallmatrix} * & \cdot  \\ \cdot & *
  \end{smallmatrix}$} 
& 0 & 0 & \makebox[0pt][c]{$\hdots$} & 
0 \\
0 & \makebox[10pt][c]{\rule{0pt}{12pt}$\begin{smallmatrix} * & \cdot
    \\ \cdot & * \end{smallmatrix}$} & 0 & & 0 \\
0 & 0
& \makebox[10pt][c]{\rule{0pt}{12pt}$\begin{smallmatrix} * & \cdot  \\
    \cdot & * \end{smallmatrix}$} & & 0 \\
\vdots &&& \makebox[0pt][c]{$\ddots$} & \vdots \\
0 & 0 & 0 & 
\cdots &
\makebox[10pt][c]{\rule{0pt}{12pt}$\begin{smallmatrix} * & \cdot  \\
    \cdot & * \end{smallmatrix}$}
\end{pmatrix}
&
W' &=
\begin{pmatrix}
\makebox[10pt][c]{$\begin{smallmatrix} 0 & 0     \\ 0 & 1
  \end{smallmatrix}$} 
& 0 & 0 & \makebox[0pt][c]{$\hdots$} & 
\makebox[10pt][c]{$\begin{smallmatrix} 1 & 0     \\ 0 & 0
  \end{smallmatrix}$} \\
\makebox[10pt][c]{\rule{0pt}{12pt}$\begin{smallmatrix} 1 & 0     \\ 0 & 0
  \end{smallmatrix}$} 
& \makebox[10pt][c]{$\begin{smallmatrix} 0 & 0     \\ 0 & 1
  \end{smallmatrix}$} & 0 & & 0 \\
0 & \makebox[10pt][c]{\rule{0pt}{12pt}$\begin{smallmatrix} 1 & 0     \\ 0 & 0
  \end{smallmatrix}$} 
& \makebox[10pt][c]{$\begin{smallmatrix} 0 & 0     \\ 0 & 1
  \end{smallmatrix}$} & & 0 \\
\vdots && \makebox[0pt][c]{$\ddots$} & \makebox[0pt][c]{$\ddots$} & \vdots \\
0 & \hdotsfor{2} & 
\makebox[10pt][c]{$\begin{smallmatrix} 1 & 0     \\ 0 & 0
  \end{smallmatrix}$} & 
\makebox[10pt][c]{$\begin{smallmatrix} 0 & 0     \\ 0 & 1
  \end{smallmatrix}$}
\end{pmatrix}
\ef.
\end{align}
Remark that in all rows with sub-index 2 and columns with sub-index 1
(i.e.~all rows ands columns in $W'$) we have exactly one allowed entry
beyond the $*$ element on the diagonal. 
In order to see how the truth-setting procedure works,
consider what happens if we choose the top-left $*$ in the first
block (i.e.~we choose ``$\sigma_{(i,a_1)} = 1$'' in the string of
$\vec{\sigma}_{(i,a)}$ for the ``variables seen from the clauses''). At
the beginning the matrix is (a circle means ``element chosen in the
matching'', a bar means ``element not choosen in the matching'')
\[
W=
\left(
\raisebox{-40pt}{\setlength{\unitlength}{8pt}
\begin{picture}(12,11.5)(0,-1)
\put(0,9){$*$}
\put(1,8){$*$}
\put(2,7){$*$}
\put(3,6){$*$}
\put(4,4.4){$\ddots$}
\put(6,3){$*$}
\put(7,2){$*$}
\put(8,1){$*$}
\put(9,0){$*$}
\put(10,-1.6){$\ddots$}
\put(1,2){$1$}
\put(3,0){$1$}
\put(0,0.8){$1$}
\put(4,-1.6){$\ddots$}
\put(2,-1.6){$\ddots$}
\thinlines
\put(-0.2,9.8){\line(1,0){2}}
\put(-0.2,9.8){\line(0,-1){2}}
\put(1.8,7.8){\line(-1,0){2}}
\put(1.8,7.8){\line(0,1){2}}
\put(1.8,7.8){\line(1,0){2}}
\put(1.8,7.8){\line(0,-1){2}}
\put(3.8,5.8){\line(-1,0){2}}
\put(3.8,5.8){\line(0,1){2}}
\put(7.8,1.8){\line(-1,0){2}}
\put(7.8,1.8){\line(0,1){2}}
\put(7.8,1.8){\line(1,0){2}}
\put(7.8,1.8){\line(0,-1){2}}
\put(9.8,-0.2){\line(-1,0){2}}
\put(9.8,-0.2){\line(0,1){2}}
\put(1.8,1.8){\line(-1,0){2}}
\put(1.8,1.8){\line(0,1){2}}
\put(1.8,1.8){\line(1,0){2}}
\put(1.8,1.8){\line(0,-1){2}}
\put(3.8,-0.2){\line(-1,0){4}}
\put(3.8,-0.2){\line(0,1){2}}
\put(-0.2,-0.2){\line(0,1){4}}
\put(-0.2,3.8){\line(1,0){12}}
\put(5.8,-1.8){\line(0,1){12}}
\put(0.3,9.3){\circle{1}}
\end{picture}}
\right)
\]
while, after six logical implication we have
\[
W=
\left(
\raisebox{-40pt}{\setlength{\unitlength}{8pt}
\begin{picture}(12,11.5)(0,-1)
\put(0,9){$*$}
\put(1,8){$*$}
\put(2,7){$*$}
\put(3,6){$*$}
\put(4,4.4){$\ddots$}
\put(6,3){$*$}
\put(7,2){$*$}
\put(8,1){$*$}
\put(9,0){$*$}
\put(10,-1.6){$\ddots$}
\put(1,2){$1$}
\put(3,0){$1$}
\put(0,0.8){$1$}
\put(4,-1.6){$\ddots$}
\put(2,-1.6){$\ddots$}
\thinlines
\put(-0.2,9.8){\line(1,0){2}}
\put(-0.2,9.8){\line(0,-1){2}}
\put(1.8,7.8){\line(-1,0){2}}
\put(1.8,7.8){\line(0,1){2}}
\put(1.8,7.8){\line(1,0){2}}
\put(1.8,7.8){\line(0,-1){2}}
\put(3.8,5.8){\line(-1,0){2}}
\put(3.8,5.8){\line(0,1){2}}
\put(7.8,1.8){\line(-1,0){2}}
\put(7.8,1.8){\line(0,1){2}}
\put(7.8,1.8){\line(1,0){2}}
\put(7.8,1.8){\line(0,-1){2}}
\put(9.8,-0.2){\line(-1,0){2}}
\put(9.8,-0.2){\line(0,1){2}}
\put(1.8,1.8){\line(-1,0){2}}
\put(1.8,1.8){\line(0,1){2}}
\put(1.8,1.8){\line(1,0){2}}
\put(1.8,1.8){\line(0,-1){2}}
\put(3.8,-0.2){\line(-1,0){4}}
\put(3.8,-0.2){\line(0,1){2}}
\put(-0.2,-0.2){\line(0,1){4}}
\put(-0.2,3.8){\line(1,0){12}}
\put(5.8,-1.8){\line(0,1){12}}
\put(0.3,8.7){\vector(0,-1){6.5}}
\put(1,1.2){\vector(1,0){6.5}}
\put(9.3,1.5){\vector(0,-1){0.9}}
\put(9.3,1.5){\line(-1,0){0.5}}
\put(8.8,0.3){\vector(-1,0){4.8}}
\put(3.3,1){\vector(0,1){4.6}}
\put(3.5,7.4){\line(0,-1){0.6}}
\put(3.5,7.4){\vector(-1,0){0.8}}
\put(0.3,9.3){\circle{1}}
\put(2.3,7.3){\circle{1}}
\put(8.3,1.3){\circle{1}}
\put(3.3,0.35){\circle{1}}
\qbezier(3,6)(3,6)(3.6,6.7)
\qbezier(9,0)(9,0)(9.6,0.7)
\qbezier(0,0.9)(0,0.9)(0.6,1.6)
\end{picture}}
\right)
\]
Remarking that the choice of 1s in the matrix has a cyclic ordering
w.r.t.~clause indices, with a simple induction one deduces that the
boolean assignments are ``$\sigma_{i,a_k} = 1$'' for all clauses $a_k$
incident with variable $i$. A similar statement can be done in the
case ``$\sigma_{i,a_1} = 0$'' (just exchange circles with bars in the
diagrams above).  So, we have determined that only consistent boolean
assignment of literals are allowed.

Now we can build the structures for the satisfaction testing.
We are interested in the restriction of W to a given fixed index $a$,
which is of the form
\begin{align}
W|_{\textrm{fixed $a$}} &=
\left(
\begin{array}{c|c}
I_* & W'' \\
\hline
\rule{0pt}{12pt}
0 & I_*
\end{array}
\right)
\ef.
\end{align}
We describe the matrix $W''$ for a clause of length $k$ with all
unnegated literals, $C_a= u_{i_1} \vee \ldots \vee u_{i_{k}}$.  All
the other cases can be trivially inferred~\footnote{Define the
transposition matrix $(T^{(k)})_{ij}=1$ if $i=j \not\in \{ 2k, 2k-1
\}$, if $i=2k$ and $j=2k-1$ or if $i=2k-1$ and $j=2k$, and zero
otherwise. I.e., matrix $T^{(k)}$, acting on the left, transpose rows
$2k-1$ and $2k$, while acting on the right transpose the corresponding
columns.  Then, if matrix $W''$ encodes a clause involving $(u_1,
\ldots, u_k, \ldots, u_{\ell})$, the matrix $T^{(k)} W'' T^{(k)}$
encodes the same clause on literals $(u_1, \ldots, \negg{u}_k, \ldots,
u_{\ell})$.}.
Choose a whatever literal index (say, $i_1$) among the
neighbours of the clause $a$, and set 
\be
\label{eq.prescr0}
w_{(i,a)^{\sigma}_1, (j,a)^{\tau}_2}=
\left\{
\begin{array}{ll}
1 & 
   \begin{array}{l}
   \textrm{if $i=j$ and $\sigma=\tau=-1$;} \\
   \textrm{if $i=j\neq i_1$ and $\sigma=\tau=+1$;} \\
   \textrm{if $i=i_1$, $j\neq i_1$, $\sigma=+1$ and $\tau=-1$;} \\
   \textrm{if $i\neq i_1$, $j=i_1$, $\sigma=-1$ and $\tau=+1$;}
   \end{array} \\
0 & \rule{0pt}{5mm}\textrm{ otherwise}
\end{array}
\right.
\ee
that is, in an extensive representation of the matrix,
\be
\label{eq.Wsatchk}
W''
= \left(
\begin{array}{ccccccc}
0 & \multicolumn{1}{c|}{0} & 0 & 1 & 0 & 1 &
\ldots \\
0 & \multicolumn{1}{c|}{1} & 0 & 0 & 0 & 0 & \ldots \\
\cline{1-4}
0 & \multicolumn{1}{c|}{0} & 1 & \multicolumn{1}{c|}{0} & 0 &0 &
\makebox[0pt][c]{$\ldots$} \\
1 & \multicolumn{1}{c|}{0} & 0 & \multicolumn{1}{c|}{1} & 0 &0 &
\makebox[0pt][c]{$\ldots$} \\
\cline{3-6}
0 & 0 & 0 & \multicolumn{1}{c|}{0} & 1 & \multicolumn{1}{c|}{0} \\
1 & 0 & 0 & \multicolumn{1}{c|}{0} & 0 & \multicolumn{1}{c|}{1} \\
\cline{5-6}
\vdots & \vdots &\vdots &\vdots &&&\ddots
\end{array}
\right)
\ee
which indeed makes the game, as one can easily check.
Indeed, if all literals are negated, we are left with matrix minor
\[
W''_{\textrm{(F,F,\ldots,F)}}
=
\left(
\begin{array}{c|c}
0 & {\bf 0} \\
\hline
{\bf 0} & I_{k-1}
\end{array}
\right)
\ef,
\]
which clearly does not allow for any valid matching, while, if the
first literal is true we have
\[
W''_{\textrm{(T,\ldots)}}
=
\left(
\begin{array}{c|c}
1 & {\bf \cdot} \\
\hline
{\bf \cdot} & I_{k-1}
\end{array}
\right)
\ef,
\]
which allows at least for the diagonal matching, $\pi(j)=j$ for all
$j=1, \ldots, k$,
and if the first literal is false, but one of the others (say, the
$h$-th) is true, we have
\[
W''_{\textrm{(F,\ldots,T,\ldots)}}
=
\left(
\begin{array}{c|ccccc}
0 & {\bf \cdot}
& 1 & {\bf \cdot} \\
\hline
{\bf \cdot}
& I_{h-2} & {\bf 0} & 0 \\
1 & {\bf 0} & 1 & {\bf 0} \\
{\bf \cdot} & 0 & {\bf 0} & I_{k-h}
\end{array}
\right)
\ef,
\]
which allows at least for the matching $\pi(1)=h$, $\pi(h)=1$ and
$\pi(j)=j$ otherwise.
Similarly, for a NAE-$k$-SAT clause 
$C_a= (u_{i_1} \vee \ldots \vee u_{i_{k}}) 
\wedge (\negg{u}_{i_1} \vee \ldots \vee \negg{u}_{i_{k}})$ 
we can choose
\be
\label{eq.prescr0NAE}
w_{(i,a)^{\sigma}_1, (j,a)^{\tau}_2}=
\left\{
\begin{array}{ll}
1 & 
   \begin{array}{l}
   \textrm{if $i=j\neq i_k$ and $\sigma=\tau=-1$;} \\
   \textrm{if $i=j\neq i_1$ and $\sigma=\tau=+1$;} \\
   \textrm{if $i=i_1$, $j\neq i_1$, $\sigma=+1$ and $\tau=-1$;} \\
   \textrm{if $i\neq i_1$, $j=i_1$, $\sigma=-1$ and $\tau=+1$;} \\
   \textrm{if $i=i_k$, $j\neq i_k$, $\sigma=-1$ and $\tau=+1$;} \\
   \textrm{if $i\neq i_k$, $j=i_k$, $\sigma=+1$ and $\tau=-1$;}
   \end{array} \\
0 & \rule{0pt}{5mm}\textrm{ otherwise}
\end{array}
\right.
\ee
corresponding to the matrix minor
\be
\label{eq.Wnaesatchk}
W''
= \left(
\begin{array}{ccccccccc}
0 & \multicolumn{1}{c|}{0} & 0 & 1 & 0 & 1 &
\ldots & 0 & 1 \\
0 & \multicolumn{1}{c|}{1} & 0 & 0 & 0 & 0 & \ldots & 0 & 0 \\
\cline{1-4}
0 & \multicolumn{1}{c|}{0} & 1 & \multicolumn{1}{c|}{0} & 0 &0 &
\makebox[0pt][c]{$\ldots$} & 0 & 1 \\
1 & \multicolumn{1}{c|}{0} & 0 & \multicolumn{1}{c|}{1} & 0 &0 &
\makebox[0pt][c]{$\ldots$} & 0 & 0 \\
\cline{3-6}
0 & 0 & 0 & \multicolumn{1}{c|}{0} & 1 & \multicolumn{1}{c|}{0} &&0&1\\
1 & 0 & 0 & \multicolumn{1}{c|}{0} & 0 & \multicolumn{1}{c|}{1} &&0&0\\
\cline{5-6}
\vdots & \vdots &\vdots &\vdots &&&\ddots&\vdots &\vdots \\
\cline{8-9}
0&0&0&0&0&0& \cdots & \multicolumn{1}{|c}{1} & 0 \\
1&0&1&0&1&0& \cdots & \multicolumn{1}{|c}{0} & 0
\end{array}
\right)
\ee
This completes the reduction proof for SAT and NAE-SAT (and then, in
particular, for 3-SAT and NAE-3-SAT). 
Indeed, the proposed encodings of a SAT and NAE-SAT clause are a
special case of the most general $k$-literal clause, in which the
number of true literals must be in the range 
$\{h_{\textrm{min}}, \ldots, h_{\textrm{max}} \}$, 
(also having as a special case the 1-in-3-SAT problem, which
is NP-complete~\cite{GJ}), for which a general
encoding is shown in section~\ref{sec.more}.


It is common in reduction proofs that multiple appearence of a literal
in a clause requires some care
(truth-setting and satisfaction-testing
structures could damagely interfere), and one should make some
standard comment on
the fact that this case can be excluded with small effort from any SAT
instance. 
This does not happen in our case. One can understand this from
the fact that the 1s in the two structures appear in different
rows and columns, and logical implications which allow to test the
performance of the structures involve only these rows/columns. 

Also remark that the reduction is linear not only in
matrix size w.r.t.~the original factor-graph size (the dimension $n$ of
the Matching matrix is four times the number of edges in the factor
graph), but also in the number $\mathcal{N}$ of non-zero entries in the matrix,
which is indeed very sparse. Each literal of coordination $k$ requires
a truth-setting structure with $2k$ entries, while each clause of
length $k$ requires a satisfaction-testing structure with $4k-3$
entries ($6k-8$ for a NAE clause), thus we have that, for a factor
graph $G(V_{\ell}, V_{c}; E)$

\begin{tabular}{lcl}
\rule{0pt}{14pt}
$\mathcal{N} = 6|E(G)| - 3 |V_{c}(G)|$
&& SAT problem; \\
\rule{0pt}{12pt}
$\mathcal{N} = 8|E(G)| - 8 |V_{c}(G)|$
&& NAE-SAT problem.
\end{tabular}

\noindent
Finally, we also remark that the instances of One-in-Two Matching
obtained as reduction from Satisfiability Problems (or, as we will see
in section~\ref{sec.more}, from problems with clauses having more
generic truth tables) are of a particular kind: the $1$s are contained
only in the top-right and bottom-left $2n \times 2n$ quadrants 
(the matrices $W'$ and $W''$). We call \emph{Bipartite} One-in-Two
Matching this specialized problem.

\section{Proof of linear reduction from One-in-Two Matching to 3
Dimensional Matching}
\label{sec.3DM}

We recall the definition of the 3-Dimensional Matching
Problem~\cite{GJ}:
\[
\begin{array}{l|cl}
\multicolumn{3}{l}{\textrm{\bf 3-Dimensional Matching Problem:}} \\
\hline
\rule{0pt}{14pt}
G(V_1, V_2, V_3; T)                 && M \subseteq T ; \\
\textrm{with } T \subseteq V_1 \times V_2 \times V_3 
                                    && |M|=n, \\
\textrm{and } |V_1|=|V_2|=|V_3|=n.  &&  \forall v \in V_1 \cup V_2 \cup V_3 \quad
                                 \deg_T(v)=1. \\

\end{array}
\]
The object $G$ could be called a ``tripartite hypergraph'': indeed,
instead of edges, it contains hyper-edges with three endpoints, one
for each set of vertices. Equivalently to what has been done in
section~\ref{ssec.defMatch}, another representation turns out to be
useful, in which the allowed elements for a matching are encoded in an
array of zeroes and ones. Now the array is three-dimensional. Thus,
define
$W=\{ w_{ijk} \}_{i,j,k=1, \ldots, n}$ such that $w_{ijk}=1$ if
$(i,j,k) \in T$, with $i \in V_1$, $j \in V_2$ and $k \in V_3$, and
$w_{ijk}=0$ otherwise.
A feasible
matching $M$ is then described by an array $X=\{ x_{ijk} \}$,
with $x_{ijk} = 0, 1$, such that $x_{ijk}=1$ if the triangle
$t=(i,j,k)$ is in $M$, and then
exactly one element of the array is equal to 1 per $i$, $j$
or $k$ fixed, i.e.
\begin{subequations}
\begin{align}
\forall \ i=1, \ldots, n &\qquad
\sum_{j,k} x_{ijk} = 1 \ef;
\\
\forall \ j=1, \ldots, n &\qquad
\sum_{i,k} x_{ijk} = 1 \ef;
\\
\forall \ k=1, \ldots, n &\qquad
\sum_{i,j} x_{ijk} = 1 \ef.
\end{align}
\end{subequations}
The cost function for $M$ is then restated into
\be
\label{eq.67354}
C_W(X)= \sum_{i,j,k} w_{ijk} x_{ijk}
\ef,
\ee
and $X$ is a valid matching if $C_W(X)=n$.
The numerical version is defined accordingly, where the weights
$w_{ijk}$ are integers, and we have a threshold value for the
cost~(\ref{eq.67354}).

Now we describe the reduction from One-in-Two Matching to 3DM.
An identical reduction goes from One-in-Two Assignment to
Numerical 3DM.
Call $W=\{ w_{ij} \}$ our One-in-Two Matching instance of dimension
$2n$ to be encoded, with
\be
w_{ij}=
\left\{
\begin{array}{ll}
* & i=j\leq 2n; \\
\cdot & 
\!\!\rule[-16pt]{0pt}{37pt}
  \begin{array}{l}
  i=2h, \ j=2h-1, \ h \leq n \\ 
  \textrm{or } i=2h-1, \ j=2h, \ h \leq n;
  \end{array} \\
w_{ij} \quad(\in \{0,1\}) & \textrm{otherwise.}
\end{array}
\right.
\ee
and $W^{(3)}$ our suggested output 3DM instance, also of dimension
$2n$. Our formal reduction is, calling $A(k)=\{2(k-n)-1, 2(k-n)\}$ for
$k=n+1, \ldots, 2n$,
\be
w^{(3)}_{ijk}=
\left\{
\begin{array}{ll}
1 & \!\!
  \begin{array}{l}
  i=j=2k-1, \\
  i=j=2k;
  \end{array} \\
w_{ij} & \!\!\rule{0pt}{15pt}
  \begin{array}{l}
  i \in A(k), \ j\not\in A(k);
  \end{array} \\
0 & \rule{0pt}{5mm} \textrm{otherwise.}
\end{array}
\right.
\ee
or, more pictorially, call $\vec{e}_i$ and $\vec{w}_i$ the vectors
\begin{subequations}
\begin{align}
\vec{e}_i &= (0, \ldots, 0, 
\stackrel{\textrm{$i$-th}}{1}, 0, \ldots, 0) \ef;
\\
\vec{w}_{2i-1} &= (w_{2i-1,1}, \ldots, w_{2i-1,2i-2}, 0, 0,
w_{2i-1,2i+1}, \ldots, w_{2i-1,2n} ) \ef;
\\
\vec{w}_{2i} &= (w_{2i,1\phantom{-1}}, \ldots, w_{2i,2i-2\phantom{-1}}, 0, 0,
w_{2i,2i+1\phantom{-1}}, \ldots, w_{2i,2n\phantom{-1}} ) \ef;
\end{align}
\end{subequations}
then $W^{(3)}$, written as a matrix on indices $(i,k)$, of vectors
on index $j$, looks like
\be
\label{eq.8747668}
W^{(3)}
=
\left(
\begin{array}{cccccc}
\vec{e}_1 & 0 & \hdots & \vec{w}_{1} & 0 & \hdots \\
\vec{e}_2 & 0 & \hdots & \vec{w}_{2} & 0 & \hdots \\
0 & \vec{e}_3 & \hdots & 0 & \vec{w}_{3} & \hdots \\
0 & \vec{e}_4 & \hdots & 0 & \vec{w}_{4} & \hdots \\
\vdots & \vdots & \ddots & \vdots & \vdots & \ddots 
\end{array}
\right)
\ef.
\ee
Indeed, remark that in the planes $(i,j)$ for $k=1,\ldots, n$ there
are only two allowed entries, whose $(i,j)$ coordinates correspond to
the ones of the $*$s in the $k$-th block of the original instance.
Mimicking the One-in-Two constraint, for each block $k$ we are forced
to choose a value $\sigma_k \in \{ 0, 1 \}$, with (say) $0$ and $1$
selecting respectively the entry with even and odd indices.  Then, in
all $(i,j)$ layers with $k= n+1, \ldots, n$ there are only two
non-empty $i$-rows, (which are empty in all the other layers with
$k=n+1, \ldots, 2n$).  Because of the choice of the vector
$\vec{\sigma}$, exactly one of them is now forbidden.  So, the
three-dimensional constraint of choosing one element for each index
$i$, $j$ and $k$ is at this point reduced to a two-dimensional
matching constraint, as there is a bijection between unmatched layers
$k$ and non-empty unmatched rows $i$. I.e.~the 3-dimensional 
$2n \times 2n \times 2n$ array of equation (\ref{eq.8747668}) is now
restricted to the $n \times n \times n$ array
\begin{align}
\label{eq.8747668bis}
{W'}^{(3)}(\vec{\sigma})
&=
\left(
\begin{array}{cccc}
\vec{w}_{2- \sigma(1)}^{(\vec{\sigma})} & 0 & 0 & \hdots \\
0 & \vec{w}_{4- \sigma(2)}^{(\vec{\sigma})} & 0 & \hdots \\
0 & 0 & \vec{w}_{6- \sigma(3)}^{(\vec{\sigma})} & \hdots \\
\vdots & \vdots & \vdots & \ddots 
\end{array}
\right)
\ef;
&
\big( \vec{w}_{i}^{(\vec{\sigma})} \big)_j
&=
\big( \vec{w}_{i} \big)_{2j-\sigma(j)}
\ef.
\end{align}
It is easily understood that the set of forbidden indices $j$ after
the choice of vector $\vec{\sigma}$ and the disposition of the entries
$w_{ij}$ are in agreement with the picture of section
\ref{ssec.oneintwodef}, where a vector $\vec{\sigma}$ determines a
$n$-dimensional minor of the original $2n$-dimensional instance, with
forbidden entries on the diagonal.

As a corollary of the construction we have that the reduction is linear 
not only for what concerns the size of the array (an $n \times n$ One-in-Two
instance goes into an $n \times n \times n$ 3DM array), but also on
the number of non-zero entries in the instance~\footnote{More
  precisely, the cardinality of $T$ for 3DM
equals the one of $E$ for One-in-Two Matching, 
plus the $2n$ ``deterministic''
entries of vectors $\vec{e}_i$, i.e.~inside a factor 2
if we understand that 3DM instances having planes with only one valid
entry can be trivially reduced in size.},
which is proportional to the length
of the bit-encoding of the instance.

Putting together this result with the one of section
\ref{sec.redproof}, we have in turn a \emph{linear} reduction from SAT
problems to 3DM, 

\begin{tabular}{lcl}
\rule{0pt}{14pt}
$\mathcal{N}_{\textrm{3DM}} = 10|E(G)| - 3 |V_{c}(G)|$
&& SAT problem; \\
\rule{0pt}{12pt}
$\mathcal{N}_{\textrm{3DM}} = 12|E(G)| - 8 |V_{c}(G)|$
&& NAE-SAT problem.
\end{tabular}

\noindent
which is much more economic of the cubic one first
presented in Karp seminal '72 paper \cite{Karp72} 
(see also~\cite{GJ}), and maybe
technically simpler.

\section{Encoding of more complex boolean patterns}
\label{sec.more}

We have seen how generic boolean disjunctive-form expressions
containing SAT and NAE-SAT clauses can be encoded into a One-in-Two
Matching instance of dimension four times the number of edges in the
factor-graph representation of the boolean formula. 

We can be more ambitious, and try to encode more complex boolean
structures, in order to further reduce the output size of One-in-Two
instances for real-life problems. For example, One-in-Three-SAT
Problem is another NP-complete problem, where the truth tables over
three literals only contain certain triplets of 
combinations (say, $(TFF)$, $(FTF)$ and $(FFT)$)~\cite{GJ}. 
A One-in-Three clause can clearly be
encoded with five 3-SAT clauses (or three 3-SAT and one NAE-3-SAT),
but this would require 15 (or 12) edges in the factor graph. If one
could produce an encoding for One-in-Three clauses \emph{directly}
with a 6-dimensional matrix, one would save a factor 5 (or 4) in the final
dimension of the matrix. This is the spirit in which, in
section~\ref{sec.redproof}, we designed explicitly compact matrices for
NAE-SAT clauses, instead of converting them into pairs of opposite SAT
clauses.

Of course this project can not be fulfilled for arbitrary clauses of
generic length: it is easily estimated that the number of inequivalent
truth tables for length-$k$ clauses diverges in $k$ much more severely
than the number of inequivalent $2k \times 2k$ matrices of zeroes and
ones. More precisely, the number of possible truth tables is
$2^{(2^k)}$. Then, we must consider gauge orbits (or \emph{classes},
$\mathcal{C}$) w.r.t.~the symmetries of the clause, which are
relabeling of literal indices (a group $\mathcal{S}_k$), times
independent negations of the literals (a group $(\mathbb{Z}_2)^k$). If
a truth table $T$ has a residual symmetry group $G(T)$, as the orbits
are a quotient, their size $g(\mathcal{C})$ equals the ratio between
the number of elements of $\mathcal{S}_k \times (\mathbb{Z}_2)^k$, and
of $G(T)$
\be
\# \{ \mathcal{C} \}=
\sum_{ \mathcal{C} } 1 =
\sum_{ \mathcal{C} } 
\frac{g (\mathcal{C})}{g (\mathcal{C})} =
\sum_{ \mathcal{C} } 
g (\mathcal{C}) \frac{|G(\mathcal{C})|}{2^k k!}
\geq
\sum_{ \mathcal{C} } 
g (\mathcal{C}) \frac{1}{2^k k!}
=
\frac{1}{2^k k!}
\# \{ T \}=
\frac{1}{k!}
2^{2^k-k}
\ef,
\ee
this estimate being asymptotically very reliable~\footnote{Indeed,
  from a refined analysis,
\be
\# \{ \mathcal{C} \}
-
\frac{2^{2^k}}{2^k k!}
=
\frac{1}{2^k k!}
\sum_{\substack{
f \in \mathcal{S}_k \times (\mathbb{Z}_2)^k, \\
f \neq e }}
\Big(
\sum_{ \mathcal{C} : f \in G(\mathcal{C}) } 
g(\mathcal{C})
\Big)
\ef,
\ee
and, calling $R_f$ the term in parenthesis, we have 
$R_f < 2^{\frac{3}{4} 2^{k}}$ for simple $(ij)$ transpositions in
  $\mathcal{S}_k$,
$R_f < 2^{\frac{9}{16} 2^{k}}$ for permutations $(ij)(k\ell)$ in
  $\mathcal{S}_k$,
$R_f < 2^{\frac{1}{2} 2^{k}}$ for $\mathbb{Z}_2$ symmetry, and for
  permutations $(ijk)$ in $\mathcal{S}_k$, and so on. So, the first
  correction is of order
  $2^{- \frac{1}{4} 2^{k}} \sim \exp(-0.173 \cdot 2^k)$.},
while the number of different matrices, also in the generous bound
which neglects gauge orbits and trivially equivalent patterns, 
is at most $2^{4 k^2}$. So, our idea of encoding in small structures
  more complex boolean patterns should be intended as
\begin{enumerate}
\item determine the encoding for special families of 
  generic-length clauses, as we already did for SAT, NAE-SAT and
  truth-setting structures.
  \label{puntoppp}
\item determine the encoding for all the clause classes up to a given
  reasonable length. This is done in \ref{app.1}, up to size
  4. For size 5 our estimate tells us that the number of classes is
  higher than $10^6$, and one should be very motivated to pursue this
  full classification. For size 6 there are more than 
  $4.\cdot 10^{14}$ classes, so that this classification is
  structurally unfeasible.
\item determine the encoding for specially-important clauses (as, for
  example, the 5-clause used for summing integers).
\item suggest heuristics for determining a compact encoding for new
  clause patterns, not discussed at point (\ref{puntoppp}),
  which could emerge in a given real-life application.
\end{enumerate}
In this section we develop the idea of point (\ref{puntoppp}), for four
classes of clauses of generic length $k$:
\begin{description}
\item[2-false:] clauses in which only two literal assignments for
  $(u_1, \ldots, u_k)$ evaluate to False;
\item[range-T:] clauses in which $(u_1, \ldots, u_k)$ evaluates to True
  if the number of True literals is in a range 
  $\{ h_{\textrm{min}}, \ldots, h_{\textrm{max}} \}$, and variants in
  which each literal could enter negated;
\item[binary threshold:] clauses in which $(u_1, \ldots, u_k)$,
  interpreted as the binary number 
  $n(u)=2^{k-1} u_1 + 2^{k-2} u_2 + \ldots$, 
  evaluates to True if $n(u) \leq q$, with $q$ a threshold value;
\item[binary distinct ($q$-colouring):] clauses in 
  which $(u_1, \ldots, u_h, u_{h+1}, \ldots, u_{2h})$,
  interpreted as the set of two binary numbers 
  $n_1(u)=2^{h-1} u_1 + 2^{h-2} u_2 + \ldots + u_h$, 
  $n_2(u)=2^{h-1} u_{h+1} + 2^{h-2} u_{h+2} + \ldots + u_{2h}$, 
  evaluates to True if $n_1(u) \neq n_2(u)$. Allows to encode
  $q$-colouring for $q=2^h$, and if $q$ is not a power of 2, for 
  $2^{h-1} < q < 2^h$, jointly with a ``binary threshold'' clause
  per colour-variable.
\end{description}


\subsection{2-False clauses}

We expect that clauses in which almost all choices of literals are
evaluated to True, or to False, are simpler to encode. 
Given a clause, call $T$ the set of True
assignments of literals, and $|T|$ its size. The cases $|T|=0,1,2^k$
are trivial, and the case $|T|=2^k-1$ is SAT. The case $|T|=2$
corresponds either to the truth-setting structure (if the two True
vectors are opposite on the hypercube $\{ \textrm{True, False} \}^k$),
or otherwise to a clause which is trivial in the sense of Rule~4 of
\ref{app.1}, and for which an encoding is easily deduced, combining a
truth-setting structure on $k' (<k)$ blocks, with a bunch of $k-k'$
diagonal $(\begin{smallmatrix} 0 & 0 \\ 0 & 1 \end{smallmatrix})$ 
blocks.

So, the first non-trivial class of clauses we will try to encode is
the set of clauses for which $|T|=2^k-2$, of which NAE-SAT is the
special case in which the two False vectors are opposite. We call them
``2-false'' clauses.  Without loss of generality, we can assume that
the two False vectors are $(FF\cdots F)$ and $(F\cdots F T\cdots T)$, 
where the number of True's is $h=1, \ldots, k$, with $h=k$
corresponding to NAE-SAT. Actually, $h=1$ is trivial in the sense of
Rule~3 of \ref{app.1}. For $1 \leq h \leq k-1$, an encoding is given
by the matrix below
\[
W =
\left(
\begin{array}{c|ccc|cccc}
\makebox[10pt][c]
  {\rule{0pt}{12pt}\raisebox{-6pt}{\rule{0pt}{12pt}}
   $\begin{smallmatrix} 0 & 0 \\ 0 & 1 \end{smallmatrix}$} 
& 
\makebox[10pt][c]
  {$\begin{smallmatrix} 0 & 1 \\ 0 & 0 \end{smallmatrix}$} 
& 
\makebox[10pt][c]
  {$\begin{smallmatrix} 0 & 1 \\ 0 & 0 \end{smallmatrix}$} 
&
\makebox[0pt][c]{$\hdots$}
& 
\makebox[10pt][c]
  {$\begin{smallmatrix} 0 & 1 \\ 0 & 0 \end{smallmatrix}$} 
& 
\makebox[10pt][c]
  {$\begin{smallmatrix} 0 & 1 \\ 0 & 0 \end{smallmatrix}$} 
& 
\makebox[10pt][c]
  {$\begin{smallmatrix} 0 & 1 \\ 0 & 0 \end{smallmatrix}$} 
&
\makebox[0pt][c]{$\hdots$}
\\
\hline
\makebox[10pt][c]
  {\rule{0pt}{12pt}\raisebox{-6pt}{\rule{0pt}{12pt}}
   $\begin{smallmatrix} 0 & 0 \\ 1 & 0 \end{smallmatrix}$} 
& 
\makebox[10pt][c]
  {$\begin{smallmatrix} 1 & 0 \\ 0 & 1 \end{smallmatrix}$} 
& 
0 
& 
\makebox[0pt][c]{$\hdots$}
&
0
& 
0 
&
0
& 
\makebox[0pt][c]{$\hdots$}
\\
\makebox[10pt][c]
  {\rule{0pt}{12pt}\raisebox{-6pt}{\rule{0pt}{12pt}}
   $\begin{smallmatrix} 0 & 0 \\ 1 & 0 \end{smallmatrix}$} 
& 
0
&
\makebox[10pt][c]
  {$\begin{smallmatrix} 1 & 0 \\ 0 & 1 \end{smallmatrix}$}  
&
&
0
& 
0 
&
0
& 
\\
\rule{0pt}{13pt}%
\raisebox{2pt}[0pt][0pt]{$\vdots$}
&
\raisebox{2pt}[0pt][0pt]{$\vdots$}
& &
\raisebox{2pt}[0pt][0pt]{$\ddots$}
&
\raisebox{2pt}[0pt][0pt]{$\vdots$}
& & &
\raisebox{2pt}[0pt][0pt]{$\ddots$}
\\
\hline
\makebox[10pt][c]
  {\rule{0pt}{12pt}\raisebox{-6pt}{\rule{0pt}{12pt}}
   $\begin{smallmatrix} 1 & 0 \\ 0 & 0 \end{smallmatrix}$} 
& 
0
& 
0 
& 
\makebox[0pt][c]{$\hdots$}
&
\makebox[10pt][c]
  {$\begin{smallmatrix} 1 & 0 \\ 0 & 1 \end{smallmatrix}$} 
& 
\makebox[10pt][c]
  {$\begin{smallmatrix} 0 & 0 \\ 1 & 0 \end{smallmatrix}$} 
& 
0
&
\makebox[0pt][c]{$\hdots$}
\\
\makebox[10pt][c]
  {\rule{0pt}{12pt}\raisebox{-6pt}{\rule{0pt}{12pt}}
   $\begin{smallmatrix} 1 & 0 \\ 0 & 0 \end{smallmatrix}$} 
& 
0
& 
0 
& 
& 
\makebox[10pt][c]
  {$\begin{smallmatrix} 0 & 0 \\ 1 & 0 \end{smallmatrix}$} 
&
\makebox[10pt][c]
  {$\begin{smallmatrix} 1 & 0 \\ 0 & 1 \end{smallmatrix}$} 
& 
\makebox[10pt][c]
  {$\begin{smallmatrix} 0 & 0 \\ 1 & 0 \end{smallmatrix}$} 
& 
\\
\makebox[10pt][c]
  {\rule{0pt}{12pt}\raisebox{-6pt}{\rule{0pt}{12pt}}
   $\begin{smallmatrix} 1 & 0 \\ 0 & 0 \end{smallmatrix}$} 
& 
0
& 
0 
& 
& 
0
&
\makebox[10pt][c]
  {$\begin{smallmatrix} 0 & 0 \\ 1 & 0 \end{smallmatrix}$} 
&
\makebox[10pt][c]
  {$\begin{smallmatrix} 1 & 0 \\ 0 & 1 \end{smallmatrix}$} 
&
\raisebox{-2pt}[0pt][0pt]{$\ddots$}
\\
\rule{0pt}{13pt}%
\raisebox{2pt}[0pt][0pt]{$\vdots$}
&
\raisebox{2pt}[0pt][0pt]{$\vdots$}
& &
\raisebox{2pt}[0pt][0pt]{$\ddots$}
&
\raisebox{2pt}[0pt][0pt]{$\vdots$}
& 
&
\raisebox{2pt}[0pt][0pt]{$\ddots$}
&
\raisebox{2pt}[0pt][0pt]{$\ddots$}
\end{array}
\right)
\raisebox{-37pt}
{$
\left.
\rule{0pt}{35pt}
\right\}
h$}
\]
Indeed, for each possible choice of the literals, we have either an easy
certificate of absence of any matching, or a matching which is valid
at sight, according to the table (notation: $X \cdots X$ means a
sequence of all $X$, while $\ldotsfitti$ means arbitrary literals in the
interval)
\[
\begin{array}{r|llll}
(\,F\cdots F\ | \ 
\makebox[41pt][l]{$ F\cdots F\,) $} &
\multicolumn{4}{l}{
  \textrm{1st row = $\vec{0}$}}
\\
\rule{0pt}{14pt}%
(\,F\cdots F\ | \ 
\makebox[41pt][l]{$ T\cdots 
T\,) $} &
\multicolumn{4}{l}{
  \textrm{1st column = $\vec{0}$}}
\\
\rule{0pt}{14pt}%
(\,T \hspace{3mm} \ldotsfitti \hspace{3mm} \ | \  
\makebox[41pt][l]{$
\quad \ldotsfitti \quad \,)$} &
  \pi(i)=i;
\\
(\,F \ldotsfitti 
\!\raisebox{1pt}{$\stackrel{\textrm{$i$-th}}{T}$}\!
\ldotsfitti \ | \  
\makebox[41pt][l]{$
\quad \ldotsfitti \quad \,)$} &
  \pi(1)=i, & \pi(i)=1, && \pi(j)=j \quad \textrm{o.w.}
\\
(\,F \cdots F\ | \  
\makebox[41pt][l]{$
\ldotsfitti 
\!\raisebox{1pt}{$\stackrel{\textrm{$i$-th}}{T}$}\!
F \ldotsfitti \,) $} &
  \pi(1)=i, & \pi(i)=i+1, & \pi(i+1)=1, & \pi(j)=j \quad \textrm{o.w.}
\\
(\,F \cdots F\ | \  
\makebox[41pt][l]{$
\ldotsfitti 
\!\raisebox{1pt}{$\stackrel{\textrm{$i$-th}}{F}$}\!
 T \ldotsfitti \,) $} 
&
  \pi(1)=i+1, & \pi(i+1)=i, & \pi(i)=1, & \pi(j)=j \quad \textrm{o.w.}
\end{array}
\]
Also remark that the number of 1s required in the encoding, 
$4k+2h-5 \ (< 6k-7) $, is linear in $k$.

\subsection{Range-T clauses}

Another family of clauses we can fully classify is what we call
``range-T'' clauses, i.e.~$k$-literal clause, in which the
number of true literals must be in the range of values
$\{h_{\textrm{min}}, \ldots, h_{\textrm{max}} \}$, 
for generic $0 \leq h_{\textrm{min}} \leq h_{\textrm{max}} \leq k$.
Special cases are SAT ($h_{\textrm{min}}=1$, $h_{\textrm{max}}=k$), 
NAE-SAT ($h_{\textrm{min}}=1$, $h_{\textrm{max}}=k-1$), and
One-in-Three-SAT ($k=3$, $h_{\textrm{min}}=h_{\textrm{max}}=1$).

Call $I$ a matrix of all 
$\begin{smallmatrix} 1 & 0 \\ 0 & 1 \end{smallmatrix}$ 
blocks on the diagonal, and $0$ outside, $I_{T}$ (resp.~$I_{F}$) a
matrix of all
$\begin{smallmatrix} 0 & 0 \\ 0 & 1 \end{smallmatrix}$ 
(resp.~$\begin{smallmatrix} 1 & 0 \\ 0 & 0 \end{smallmatrix}$) 
blocks on the diagonal, and $0$ outside; then call
$J$ a (possibly rectangular) matrix of all 1s, and $J_{FT}$
(resp.~$J_{TF}$) a (possibly rectangular) matrix of all
$\begin{smallmatrix} 0 & 1 \\ 0 & 0 \end{smallmatrix}$
(resp.~$\begin{smallmatrix} 0 & 0 \\ 1 & 0 \end{smallmatrix}$) blocks.

Then, for given $h_{1,2,3}$, $h_1 + h_2 + h_3 = k$, consider the
matrix
\[
\begin{array}{rl}
W=
&
\left(
\begin{array}{ccc}
I_T & J_{FT} & J_{FT} \\
J_{TF} & I & J_{FT} \\
J_{TF} & J_{TF} & I_F
\end{array}
\right)
\!\!
\begin{array}{l}
\left. \right\} h_1 \\
\left. \right\} h_2 \\
\left. \right\} h_3
\end{array}
\\
&
\ \ \,
\begin{array}{ccc}
\underbrace{}_{h_1}
&
\underbrace{}_{h_2}
&
\underbrace{}_{h_3}
\end{array}
\end{array}
\]
For a given assignment $\vec{\tau}$ 
of literals, call $h'_1, \ldots, h'_3$ the
number of true assignments in the three blocks. 
Also call $h''_i=h_i-h'_i$.
Because of permutation
invariance inside the first block, we can equivalently assume they are
the first ones of each block, and
find a $k \times k$ reduced matrix 
\[
W_{\vec{\tau}}
=
\left(
\begin{array}{cccccc}
I & 0 & 0 & 0 & 0 & 0 \\
0 & 0 & J & 0 & J & 0 \\
0 & J & I & 0 & 0 & 0 \\
0 & 0 & 0 & I & J & 0 \\
0 & J & 0 & J & 0 & 0 \\
0 & 0 & 0 & 0 & 0 & I
\end{array}
\right)
\!\!
\begin{array}{l}
\left. \right\} h'_1 \\
\left. \right\} h''_1 \\
\left. \right\} h'_2 \\
\left. \right\} h''_2 \\
\left. \right\} h'_3 \\
\left. \right\} h''_3
\end{array}
\]
Up to a permutation both on rows and columns
\[
W_{\vec{\tau}}
=
\left(
\begin{array}{cccccc}
I & 0 & 0 & 0 & 0 & 0 \\
0 & I & 0 & J & 0 & 0 \\
0 & 0 & 0 & J & J & 0 \\
0 & J & J & 0 & 0 & 0 \\
0 & 0 & J & 0 & I & 0 \\
0 & 0 & 0 & 0 & 0 & I
\end{array}
\right)
\!\!
\begin{array}{l}
\left. \right\} h'_1 \\
\left. \right\} h'_2 \\
\left. \right\} h'_3 \\
\left. \right\} h''_1 \\
\left. \right\} h''_2 \\
\left. \right\} h''_3
\end{array}
\]
Then, if $h''_1 > h'_2 + h'_3$ or 
$h'_3 > h''_2 + h''_1$ we have a certificate that no matching exists,
because the rows corresponding resp.~to entries $h''_1$ or $h'_3$ have
too many empty columns. Conversely, if none of the two facts above
happens, calling $h'=h'_1 + h'_2 + h'_3$ and $h=\min(h''_1, h'_3)$,
the permutation $\pi(h'-\ell)=h'+\ell+1$, $\pi(h'+\ell+1)=h'-\ell$ for
$\ell=0, \ldots, h-1$ and $\pi(j)=j$ otherwise is a valid matching.

Thus the conditions for an assignment of literals to be evaluated to True
are
\begin{align}
h''_1 &\leq h'_2 + h'_3
\ef;
&
h'_3 &\leq h''_2 + h''_1
\ef;
\end{align}
which can be restated as
\begin{align}
h' &\geq h_1
\ef;
&
h' &\leq k-h_3
\ef;
\end{align}
and, as $h'$ is the number of True literals, we recognize the
``range-T'' constraint, with $h_{\textrm{min}}=h_1$ and
$h_{\textrm{max}}=k-h_3$. Remark that our choice of matrix encoding
for SAT and NAE-SAT coincides with the general recipe described here.

\subsection{Binary threshold clauses}

A third class of clauses we encode is the ``binary threshold'' clause.
For a set of $k$ literals $(u_1, \ldots, u_k)$, denote $u$ the
binary number $[u_1 \cdots u_k]=2^{k-1} u_1 + 2^{k-2} u_2 + \ldots$. 
The clause evaluates to True if $u \leq q$, with 
$q=[q_1 \cdots q_k]$ a threshold value with binary entries
$q_i$. The clause is trivially reduced if $q<2^{k-1}$, but, as our
proof of the encoding is inductive, we consider the general $q<2^k$
case.

A simple computer procedure which would determine if $u \leq q$ is
the following~\footnote{``Return[x]'' means 
``quit the whole procedure, giving $x$ as result''.}

\begin{verbatim}
  for(i=1,..,k-1)
  { if(q[i]==1 AND u[i]==0) Return[Yes];
    if(q[i]==0 AND u[i]==1) Return[No];  }
  if(q[k] >= u[k])          Return[Yes];
  else                      Return[No]; 
\end{verbatim}
We will show that our matrix encoding reproduces this code.
Consider the shortcuts
\begin{subequations}
\begin{align}
0 &= \begin{smallmatrix} 0 & 0 \\ 0 & 0 \end{smallmatrix}
&
1 &= \begin{smallmatrix} 1 & 1 \\ 1 & 1 \end{smallmatrix}
&
A(0) &= \begin{smallmatrix} 1 & 0 \\ 0 & 0 \end{smallmatrix}
&
A(1) &= \begin{smallmatrix} 0 & 0 \\ 0 & 1 \end{smallmatrix}
\\
B(0) &= \begin{smallmatrix} 0 & 0 \\ 0 & 0 \end{smallmatrix}
&
B(1) &= \begin{smallmatrix} 1 & 1 \\ 0 & 0 \end{smallmatrix}
&
C(0) &= \begin{smallmatrix} 1 & 0 \\ 0 & 0 \end{smallmatrix}
&
C(1) &= \begin{smallmatrix} 1 & 0 \\ 0 & 1 \end{smallmatrix}
\end{align}
\end{subequations}
and denote $A_i=A(q_i)$, $B_i=B(q_i)$, $C=C(q_k)$. A valid
encoding is given by the matrix
\be
\label{eq.06774}
\begin{pmatrix}
A_1 & 0 & 0 & \cdots & 0 & B_1 \\
1 & A_2 & 0 &        & 0 & B_2 \\
0 & 1 & A_3 & \raisebox{0pt}[0pt][0pt]{$\ddots$}
                     & 0 & B_3 \\
\vdots & \ddots & \ddots & \ddots & \vdots & \vdots \\
0 & \cdots & 0 & 1 & A_{k-1} & B_{k-1} \\
0 & \cdots & 0 & 0 & 1       & C       
\end{pmatrix}
\ee
where the matrix minor associated to a binary number is the one in
which the odd (resp.~even) rows and columns are deleted if the binary
digit is 1 (resp.~0).

As anticipated, we prove this inductively. First notice that $C$
encodes the clause with $k=1$: if $q=1$ every literal assignment is
valid, while if $q=0$ only $u=0$ is accepted. This encodes the third
``if'' of the program.

Then, in the induction assume that the minor of~(\ref{eq.06774}) in
which the first two rows and columns are removed encodes the clause on
the $k-1$ digits $[q_2 \cdots q_k]$. 
Then, if $q_1=1$ we have
\be
\label{eq.06774-1}
\begin{pmatrix}
\begin{smallmatrix} 0 & 0 \\ 0 & 1 \end{smallmatrix}
 & 0 & 0 & \cdots & 0 & 
             \begin{smallmatrix} 1 & 1 \\ 0 & 0 \end{smallmatrix}
 \\
\begin{smallmatrix} 1 & 1 \\ 1 & 1 \end{smallmatrix}
  & A_2 & 0 &        & 0 & B_2 \\
0 & 1 & A_3 & \raisebox{0pt}[0pt][0pt]{$\ddots$}
                     & 0 & B_3 \\
\vdots & \ddots & \ddots & \ddots & \vdots & \vdots \\
0 & \cdots & 0 & 1 & A_{k-1} & B_{k-1} \\
0 & \cdots & 0 & 0 & 1       & C       
\end{pmatrix}
\ee
If $u_1=1$, the first row of the reduced matrix is $(1,0,\ldots,0)$,
so that one is forced to have $\pi(1)=1$, and the existence of a valid
matching is reduced to the constraint 
$[u_2 \cdots u_k] \leq [q_2 \cdots q_k]$.  
If $u_1=0$, the first row of the reduced matrix is $(0,\ldots,0,1)$,
and, as, regardless to $\{ u_2, \ldots, u_k \}$, we have a whole
subdiagonal filled with 1s, the permutation $\pi(1)=k$ and
$\pi(i)=i-1$ for $i=2, \ldots, k$ is a valid matching. This encodes
the first ``if'' of the program.  Instead, if $q_1=0$ we have
\be
\label{eq.06774-0}
\begin{pmatrix}
\begin{smallmatrix} 1 & 0 \\ 0 & 0 \end{smallmatrix}
 & 0 & 0 & \cdots & 0 & 
             \begin{smallmatrix} 0 & 0 \\ 0 & 0 \end{smallmatrix}
 \\
\begin{smallmatrix} 1 & 1 \\ 1 & 1 \end{smallmatrix}
  & A_2 & 0 &        & 0 & B_2 \\
0 & 1 & A_3 & \raisebox{0pt}[0pt][0pt]{$\ddots$}
                     & 0 & B_3 \\
\vdots & \ddots & \ddots & \ddots & \vdots & \vdots \\
0 & \cdots & 0 & 1 & A_{k-1} & B_{k-1} \\
0 & \cdots & 0 & 0 & 1       & C       
\end{pmatrix}
\ee
If $u_1=1$, the first row of the reduced matrix is $(0,0,\ldots,0)$,
so that no matching is possible. If $u_1=0$ the first row of the
reduced matrix is $(1,0,\ldots,0)$, and again one is forced to have
$\pi(1)=1$, and the existence of a valid matching is reduced to the
constraint $[u_2 \cdots u_k] \leq [q_2 \cdots q_k]$. This
encodes the second ``if'' of the program, and completes the proof.
The number of 1s in the matrix is $5k-4+2 \sum_i q_i \ (<7k-4)$, again
only linear in $k$.

\subsection{Binary distinct clauses}

Finally, we will encode into a matrix with $k$ blocks
the clause which compares two binary numbers with $h=k/2$ digits
(``binary-distinct'' clause). More precisely, given the literals
$(u_1, \ldots, u_{h}, u_{h+1}, \ldots, u_{k})$,
associate the two binary numbers 
\begin{subequations}
\begin{align}
n_1(u) &= [u_1 \cdots u_h]= 2^{h-1} u_1 + 2^{h-2} u_2 + \ldots + u_{h}
\ef,
\\
n_2(u) &= [u_{h+1} \cdots u_{2h}] = 
  2^{h-1} u_{h+1} + 2^{h-2} u_{h+2} + \ldots + u_{2h}
\ef,
\end{align} 
\end{subequations}
and say that the clause is satisfied whenever
$n_1(u) \neq n_2(u)$. This clause is an ``OR'' of binary ``XOR''s,
i.e.~$u$ evaluates to
\[
C(u)=\bigvee_{i=1, \ldots, h} (u_i \xor u_{h+i})
\ef,
\]
and is encoded by the matrix 
(it is intended 
that matrix entries are 0 where not specified)
\be
\left(
\begin{array}{cccc|cccc}
\rule{0pt}{11pt}%
\makebox[6pt]{$\begin{smallmatrix} 0 & 0 \\ 0 & 0 \end{smallmatrix}$}
& \makebox[6pt]{$\begin{smallmatrix} 1 & 1 \\ 1 & 1
  \end{smallmatrix}$} & && 
          \makebox[6pt]{$\begin{smallmatrix} 0 & 1 \\ 1 & 0
          \end{smallmatrix}$} &&&\\
\rule{0pt}{11pt}%
  & \makebox[6pt]{$\begin{smallmatrix} 1 & 0 \\ 0 & 1
  \end{smallmatrix}$} &
  \makebox[6pt]{\raisebox{0pt}[0pt][0pt]{$\ddots$}} &&&
          \makebox[6pt]{$\begin{smallmatrix} 0 & 1 \\ 1 & 0
          \end{smallmatrix}$} &&\\
\rule{0pt}{11pt}%
  &   & \makebox[6pt]{\raisebox{0pt}[0pt][0pt]{$\ddots$}} &
  \makebox[6pt]{$\begin{smallmatrix} 1 & 1 \\ 1 & 1
  \end{smallmatrix}$} &&
             & \makebox[6pt]{\raisebox{0pt}[0pt][0pt]{$\ddots$}} &\\
\rule{0pt}{11pt}%
  &   &   & \makebox[6pt]{$\begin{smallmatrix} 1 & 0 \\ 0 & 1
  \end{smallmatrix}$} &&&& 
          \makebox[6pt]{$\begin{smallmatrix} 0 & 1 \\ 1 & 0
          \end{smallmatrix}$} 
          \raisebox{-5pt}{\rule{0pt}{5pt}}  \\
\hline
\rule{0pt}{11pt}%
  &&&& \makebox[6pt]{$\begin{smallmatrix} 1 & 0 \\ 0 & 1
  \end{smallmatrix}$} & \makebox[6pt]{$\begin{smallmatrix} 1 & 1 \\ 1
    & 1 \end{smallmatrix}$} &   &   \\
\rule{0pt}{11pt}%
  &&&&   & \makebox[6pt]{\raisebox{0pt}[0pt][0pt]{$\ddots$}} &
             \makebox[6pt]{\raisebox{0pt}[0pt][0pt]{$\ddots$}} & \\
  &&&&   &   & \makebox[6pt]{$\begin{smallmatrix} 1 & 0 \\ 0 & 1
  \end{smallmatrix}$} & \makebox[6pt]{$\begin{smallmatrix} 1 & 1 \\ 1
  & 1 \end{smallmatrix}$} \\
\rule{0pt}{11pt}%
\makebox[6pt]{$\begin{smallmatrix} 1 & 1 \\ 1 & 1 \end{smallmatrix}$}
&&&&   &   &   & \makebox[6pt]{$\begin{smallmatrix} 0 & 0 \\ 0 & 0
  \end{smallmatrix}$}
\end{array}
\right)
\ee
Indeed, the only part of the matrix sensible to literal assignment is
the diagonal of the top-right quadrant, which has 1s in correspondence
of the unequal-digit indices, and thus is totally empty iff the two
binary integers are equal. In this latter case we have an easy
certificate, for example the first $h$ rows have complexively only
$h-1$ columns not totally empty. Conversely, if the $i$-th element of
the diagonal is 1 (that is, if $u_i \neq u_{h+i}$), a valid
permutation is
\be
\left\{
\begin{array}{lll}
\pi(i)=h+i \\
\pi(2h)=1  \\
\pi(j)=j+1 & 1 \leq j < i, & h+i \leq j < 2h; \\
\pi(j)=j   & i < j < h+i.
\end{array}
\right.
\ee
This proves the encoding. Again remark the linear number of 1s:
$7k-8$.

\section{The roots of NP-hardness: a negative remark}
\label{sec.roots}

An astonishing fact of Complexity Theory is the frequent similarity in
the pictorial description of pairs of problems, which are indeed one
proven to be polynomial, the other to be NP-complete. Examples are
Chinese Postman vs.~Travelling Salesman, or Arc Covering vs.~Vertex
Covering, or Min-Cut-Max-Flow vs.~Max-Cut-Max-Flow. In other cases, a
natural threshold value seems to exists, for example, 2-SAT,
2-Coloring, Minimum Spanning Tree and 2-Dimensional Matching are
polynomial, while $K$-SAT, $K$-Coloring, Minimum Spanning
($K$-)hypertree and $K$-Dimensional Matching are NP-hard for any 
$K \geq 3$. A generic lesson is that we lack for a reasonable intuition
of which problems are hard, and which not.

On the other side, a hypothetic problem of cost minimization, in which
an instance consists of a random extraction of $2^n$ i.i.d.~``costs''
variables for the $2^n$ feasible solutions in $\{0,1\}^n$ (called
\emph{REM}, Random Energy 
Model~\cite{REMcites, REMcites2, REMcites3}), 
would certainly be
exponentially hard: no strategy is possible, behind the trivial one of
just reading all the costs, and finding the smallest one, and this
would take $2^n$ operations.  Similarly, a decision problem in which
each of the $2^n$ configurations is a solution \emph{independently
from the others}, with probability $p=2^{-\alpha n}$, $0 < \alpha <
1$, would be exponentially hard: no strategy is possible, behind the
trivial one of just searching sequentially for a solution, the average
number of operations required being $2^{(1-\alpha)n}$.

Of course, these problems are out of the NP class, as the mere
encoding of the instance is exponential in size (if you want, this is
the content of the celebrated Shannon entropy theorem). This
``physical'' intuition corresponds to a well-known statement of
Complexity Theory, namely that P$\neq$NP ``in an oracle
setting''~\cite{Baker}.

The widespread idea behind (average-case) NP-hardness is that,
although an ensemble of random instances of a given NP-complete
decision problem only visits an infinitesimal fraction of the
$2^{(2^n)}$ possible SAT-UNSAT truth tables (a subset of the ones
which can be optimally encoded with a polynomial entropy, as the
polynomal encoding of the instance is a bound for this value), it is
possible that they are both ``sufficiently well distributed'' and
``sufficiently typical'', such that the resulting ensemble of tables
is ``well approximated'' by the one of the REM (or ``oracle'')
ensemble.

Some interesting efforts have been recentely done in the direction of
proving that certain NP-complete problems have statistical properties
strongly similar (although not identical) to the ones of
REM~\cite{NPPmertens, mertensborgs, mertensborgs2}, and this fact has
been interpreted as a hint toward NP-completeness.  Some other
observations have been raised on the fact that the same kind of
properties arises also in well-known polynomial problems, such as the
determination of the rank of a matrix~\cite{nostroXorSat}.

An observation of this last nature can be done also in this case. 
That is, we want to state that \emph{not only} One-in-Two Matching is
formulated in an apparently similar way to other Matching problems
which are indeed polynomial, \emph{but also} the statistical
properties of the spectra of costs for random instances of these
problems share many common features.

We know that pure Matching is polynomial. We have just proven that
One-in-Two-Matching is NP-complete, and we gave at the end of
section~\ref{ssec.oneintwodef} an argument for this fact: changing
also a single ``spin'' determination $\sigma_i$ translates into the
substitution of a whole row and a whole column in the induced matching
subproblem.
What ``almost always''~\footnote{In the sense of probability, that is,
  for $N$ the size of the problem, always up to a probability $p_N$, such that 
  $p_N \to 0$ for $N \to \infty$.} happens for a random instance of,
say, i.i.d.~real positive entries $w_{ij}$ with measure
$p(w)=e^{-w}$, is that at fixed $\vec{\sigma}$ the optimal
permutation $\pi^*(\vec{\sigma})$ is such that 
$C_W(\vec{\sigma},\pi^*(\vec{\sigma})) \simeq \zeta(2)$
\cite{MezPar, MezPar2, MezPar3}
and all the summands are of order $1/N$, the largest one being of
order $\ln(N)/N$. Then, taking a $\vec{\sigma}'$ which differs from
$\vec{\sigma}$ on a single entry, resamples both a whole row and a
whole column. In correspondence with the old permutation we will now
have a huge energy,
$\eval{C_W(\vec{\sigma}',\pi^*(\vec{\sigma}))} \simeq \zeta(2)+2$,
well inside the high-density part of the spectrum.
Then, correlations among optimal permutations 
$\pi^*(\vec{\sigma})$ and $\pi^*(\vec{\sigma}')$, and among energies
$C_W(\vec{\sigma},\pi^*(\vec{\sigma})) - \zeta(2)$ and
$C_W(\vec{\sigma}',\pi^*(\vec{\sigma}')) - \zeta(2)$,
can be studied, also by mean of the exact results by 
Nair, Prabhakar and Sharma~\cite{chnair, chnair2}
(this being the subject of a paper in preparation~\cite{ourChnair}).

Then, it can be seen that similar results arise from the analysis of a
further variant of the problem, that we can call ``One-in-Four''
Matching. We restrict the ensemble of valid matchings to permutations
$\pi$ such that, for each $i=1, \ldots, n$, either $\pi(2i-1)$ or
$\pi(2i)$ are in $\{ 2i-1, 2i \}$, i.e.~in the graphical language of
section~\ref{sec.redproof} we have four $*$s in the four entries of
the $2 \times 2$ diagonal blocks. For this problem, we can describe
the configurations by mean of triplets $(\vec{\sigma}, \vec{\tau},
\pi)$, with $\vec{\sigma}, \vec{\tau} \in \{0,1\}^n$ and $\pi \in
\mathcal{S}_n$ with no fixed points, and we can repeat the argument
above, such that the best over $\pi$ of $(\vec{\sigma}, \vec{\tau})$
pairs differing also for a single entry are strongly decorrelated, as
the sub-instances differ by a whole row or a whole
column. Nonetheless, One-in-Four Matching is trivially reduced to
Matching, and the same holds for Assignment. Indeed, consider now
classes of configurations $(\vec{\sigma}, \vec{\tau}, \pi)$ sharing
the same $\pi$. The set of $2^{2n}$ costs for various $(\vec{\sigma},
\vec{\tau})$ is
\[
\Big\{ \sum_i w_{2i-\sigma(i), 2\pi(i)-\tau(\pi(i))}
\Big\}_{\vec{\sigma}, \vec{\tau} \in \{0, 1\}^n }
\]
and is trivially minimized, as each variable $\sigma_i$ or $\tau_j$
enters only once in the cost function
\be
\label{eq.375487}
\min_{\vec{\sigma}, \vec{\tau} \in \{0, 1\}^n }
\Big( \sum_i w_{2i-\sigma(i), 2\pi(i)-\tau(\pi(i))}
\Big)
=
\sum_i \min_{\sigma,\tau \in \{0,1\} }
w_{2i-\sigma, 2\pi(i)-\tau}
\ee
that is, the minimization problem is reduced to a traditional
assignment problem, where $2 \times 2$ non-diagonal blocks are
replaced by the minimum weight among the four entries, and the
diagonal blocks by infinite weights. 

The same construction can \emph{not} be repeated for One-in-Two
Assignment, as the spin variables enter in the cost function in
different points.

Another derivation of this property is given by the reduction of
One-in-Four Assignment to a Min-Cut-Max-Flow
Problem~(cfr.~\cite{PapaStei}
for all definitions), which is a polynomial problem.  In a similar
fashion to the mechanism above, the reduction can be done because of
the factorization of the choices over $\sigma$ and $\tau$, the
corresponding network being
\[
\setlength{\unitlength}{60pt}
\begin{picture}(4,2)(0,0)
\put(0.0625,0.){\includegraphics[scale=1.5, bb=0 0 155 80, clip=true]
{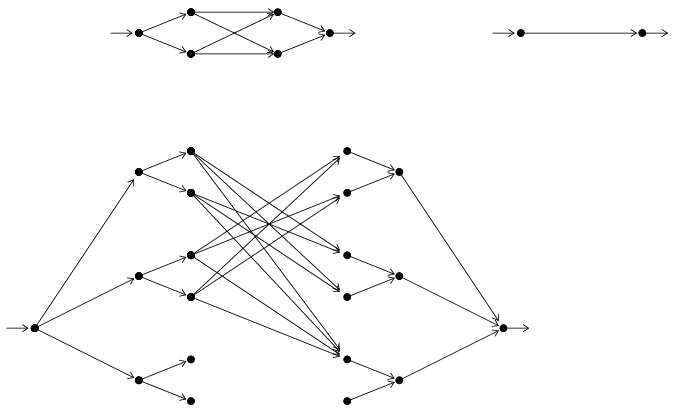}}
\put(0.65,0.1){$\vdots$}
\put(3.35,0.1){$\vdots$}
\put(2,0.3){$\vdots$}
\end{picture}
\]
all the edges have capacity 1, all the edges except the ones in the
middle layer have cost zero, and, for the middle layer, an edge
connects the $i$-th vertex on the left to the $j$-th vertex on the
right only if $\lceil i/2 \rceil \neq \lceil j/2 \rceil$, 
and in this case it has cost $w_{ij}$. Checking that this instance of
Min-Cost-Max-Flow reproduces the One-in-Four Assignment, where the
maximum flow is $n$, is a simple exercise, while the procedure of
equation~(\ref{eq.375487}) corresponds to the ``resolution'' of the
Min-Cost-Max-Flow problem on the subgraphs
\[
\setlength{\unitlength}{60pt}
\begin{picture}(4.125,0.5)(0,0)
\put(0.,0.){\includegraphics[scale=1.5, bb=30 100 195 120, clip=true]
{fig_MCMF.eps}}
\put(0.6,0.18){$\scriptstyle{w_{21}}$}
\put(0.6,0.3){$\scriptstyle{w_{12}}$}
\put(0.7,0.45){$\scriptstyle{w_{11}}$}
\put(0.7,0.02){$\scriptstyle{w_{22}}$}
\put(2.28,0.22){$=$}
\put(3.15,0.33){$\scriptstyle{\min(w_{ij})}$}
\end{picture}
\]



\section{A Quantum Adiabatic Algorithm for One-in-Two Matching}
\label{sec.qaa}

Here we describe a quantum Hamiltonian reproducing a Quantum
Adiabatic Algorithm for One-in-Two Matching or Assignment,
which is a very natural one for condensed-matter systems. 
We assume our instance is composed of a matrix of size $2n$, with
weights $w_{ij}$ out of the diagonal blocks, all much larger than zero
(this is always the case, up to a trivial gauge transformation on the
instance).

Consider two layers, each consisting of $2n \times 2n$ discrete
positions on a square lattice. We call them the row- and the
column-layers. We should imagine them geometrically ``facing'', like
two plates of a condenser.

In each row of the row-layer there is one particle, which could jump
from one column to a neighbouring one. If particle in row $i$ is in
layer position $(ij)$, we say that it is in the state
$\ket{j}_{r(i)}$. Introduce the set of quantum oscillators $a_{i,j}$,
$a^{\scriptscriptstyle \dagger}_{i,j}$.

Similarly, in the column-layer we have one particle per column, which
can hop on neighbouring rows, and the state of column $j$ is 
$\ket{i}_{c(j)}$ if the particle occupies position $i$. We call
$b_{j,i}$, $b^{\scriptscriptstyle \dagger}_{j,i}$
the quantum oscillators for this layer.

A kinetic ``diffusion'' term 
would have the form 
$a^{\scriptscriptstyle \dagger}_{i,j\pm 1} a_{i,j}$.
So the purely kinetic part of the Hamiltonian, $\widehat{H}_0$, is
\be
\widehat{H}_0
=
\sum_{i,j} 
\left(
a^{\scriptscriptstyle \dagger}_{i,j+1} a_{i,j}
+
a^{\scriptscriptstyle \dagger}_{i,j-1} a_{i,j}
+
b^{\scriptscriptstyle \dagger}_{j,i+1} b_{j,i}
+
b^{\scriptscriptstyle \dagger}_{j,i-1} b_{j,i}
\right)
\ef.
\ee
On the boundaries of the chains, you can assume either open or
periodic boundary conditions (the first one is more ``physical'',
the second one makes the analysis simpler, anyhow the difference is
negligible). Then, under this Hamiltonian, the particles are
non-interacting. At sufficiently low temperatures, 
scaling with $n^{-2}$, the 
single-particle system (say, of row $i$), up to an exponentially small
fraction of the time, will occupy the ground state 
with zero momentum, $(2n)^{-1/2} \sum_j \ket{j}_{r(i)}$, as it has a
polynomially-small but finite gap, given by the square of the
smallest non-zero momentum on the lattice.

The ground state of the system under the Hamiltonian 
$\widehat{H}_0$ is
\begin{multline}
\ket{\psi}_{\textrm{GS}}
= (2n)^{-2n}
\big( \ket{1} + \cdots + \ket{2n} \big)_{r(1)} 
\otimes \cdots \otimes 
\big( \ket{1} + \cdots + \ket{2n} \big)_{r(2n)} 
\\
\otimes 
\big( \ket{1} + \cdots + \ket{2n} \big)_{c(1)} 
\otimes \cdots \otimes 
\big( \ket{1} + \cdots + \ket{2n} \big)_{c(2n)} 
\end{multline}
and it is a reasonable initial state,
for a quantum adiabatic paradigm in which the coordinate-space
representation reproduces the set of feasible solutions, although it
is probably a poor initial state for a traditional quantum computing
approach, as it is a ``tree state'' with $\mathcal{O}(n^2 \ln n)$
symbols (cfr.~\cite{Aaronson_thesis}, ch.~13).

Up to now, we have one particle per row on the row-layer, and
similarly on the column-layer, but we have not implemented neither the
matching constraint, that row-particle $i$ is in column $j$ iff
column-particle $j$ is in row $i$, nor the one-in-two constraint.

At this aim we introduce an interaction Hamiltonian $\widehat{H}_1$,
which couples the particles on the two layers.
Furthermore, we will have special terms on the diagonal $2 \times 2$
blocks. Introduce the shortcut
\begin{align}
(a^{\scriptscriptstyle \dagger} a \,
b^{\scriptscriptstyle \dagger} b)_{i,j}
& :=
a^{\scriptscriptstyle \dagger}_{i,j} a_{i,j}
b^{\scriptscriptstyle \dagger}_{j,i} b_{j,i}
\ef;
&
(a^{\scriptscriptstyle \dagger} a \,
b^{\scriptscriptstyle \dagger} b)_{i}
:=
(a^{\scriptscriptstyle \dagger} a \,
b^{\scriptscriptstyle \dagger} b)_{i,i}
\ef.
\end{align}
For pairs $(ij)$ not in a diagonal block, we would have a term
of the form 
$-w_{ij} (a^{\scriptscriptstyle \dagger} a \,
b^{\scriptscriptstyle \dagger} b)_{i,j}$,
while for a block on entries $\{2i-1, 2i\}$ we
would have a combination of the form
$
W 
\,
\big(
(a^{\scriptscriptstyle \dagger} a \,
b^{\scriptscriptstyle \dagger} b)_{2i}
- c
\big)
\big(
(a^{\scriptscriptstyle \dagger} a \,
b^{\scriptscriptstyle \dagger} b)_{2i-1}
- c
\big)
$,
with $c$ a value between 0 and 2 (for example, 1, but notice how the
fine-tuning of $c$ \emph{exactly} equal to 1 is not required), 
and $W$ an energy
value larger than the typical $w_{ij}$. Remark that the procedure
works also if $W$ and $c$ are slightly varying from block to block, as
expected from an experimental realization.

Summarizing, the Hamiltonian
$\widehat{H}_1$ (notation: $\sum_{i,j}'$ means ``pairs $(ij)$ not in a
diagonal block'', i.e.~such that 
$\lceil \frac{i}{2} \rceil \neq \lceil \frac{j}{2} \rceil$)
\be
\widehat{H}_1
=
- {\sum_{i,j}}' w_{ij} (a^{\scriptscriptstyle \dagger} a \,
b^{\scriptscriptstyle \dagger} b)_{i,j}
- W \sum_{i=1}^{n}
\big(
(a^{\scriptscriptstyle \dagger} a \,
b^{\scriptscriptstyle \dagger} b)_{2i}
- c
\big)
\big(
(a^{\scriptscriptstyle \dagger} a \,
b^{\scriptscriptstyle \dagger} b)_{2i-1}
- c
\big)
\ee
is diagonal in coordinate space, as it only involves ``number''
operators, and has the optimal one-in-two
matchings as ground states. Indeed, violating a matching constraint
would have a finite positive cost $\sim w_{ij}$, while violating 
a one-in-two constraint would have a cost $\sim W$. Finally, for a
valid matching, the energy of the state reproduces exactly the cost
function.

For $w_{ij}$ not scaling with $n$ (which
is the proper choice in order to have an extensive energy), in the
case of matching we have a gap of order 1, while in the case of random
assignment we have a gap only algebraically small in $n$, and not
exponentially (cfr.~\cite{chnair, ourChnair}).

So, the adiabatic interpolation, for time $t \in [0,1]$,
given by $\widehat{H}_t= (1-t) \widehat{H}_0 + t \widehat{H}_1$, could
be a good prescription for a quantum adiabatic algorithm for the
one-in-two matching, as, at least at the two endpoints, there are no
exponentially-small energy gaps. A further analysis of what happens at
intermediate times would be too long and difficult, and out of the
aims of this paper.

\section{Conclusion and perspectives}
\label{sec.conc}

In this paper we introduced a problem, the \emph{One-in-Two Matching},
which lies in between the (polynomial) Matching Problem and the
(NP-complete) Hamiltonian Circuit Problem~(HC). Despite, at a first
glance, it could seem much more similar to Matching than to HC, 
we prove 
that indeed it is NP-complete (section~\ref{sec.redproof}). We thus
expect that it captures the core reasons of complexity of HC, although
largely simplifying the model.

The reduction is relatively simple and specially compact.  As a side
result, combination with a reduction proof from One-in-Two Matching to
3-Dimensional Matching~(3DM) given in section~\ref{sec.3DM} provides
a simple reduction proof for 3DM, which could be preferred for
pedagogical purposes to the original one by Karp in 1972.

Everything above applies to the ``weighted'' variants of the
problems: (One-in-Two) Matching $\to$ (One-in-Two) Assignment,
Hamiltonian Circuit $\to$ Travelling Salesman, 3DM $\to$ numerical
3DM.

In the introduction, and in section~\ref{sec.qaa}, 
we also give very preliminary arguments towards
the fact that One-in-Two Matching is a viable candidate for an
experimental realization of 
a Quantum Adiabatic Algorithm protocol,
devoted to the solution of NP-complete problems. In two
words, the main advantages are the structural simplicity of the
problem formulation, and the unusual compactness of the reduction.

At the aim of still improving this second point for real-life
applications (in which short clauses could appear, that,
when reduced to SAT, would largely increase the instance size), in
section~\ref{sec.more} we give compact encodings of a larger family of
clauses, including SAT, NAE-SAT, One-in-Three-SAT, Colouring and much
more.
Furthermore, in \ref{app.1} we find an encoding for all clauses
involving up to four literals.

Further directions of investigation will include
a better understanding of the statistical properties of Random
One-in-Two Assignment, at the light of the results of Nair, Prabhakar
and Sharma~\cite{chnair, chnair2}, and also with Cavity Method
techniques~\cite{mezrev}, in relation to the important question of
determining the geometrical characteristics of the hard instances of
TSP~\cite{ourTSPinprep, ourChnair}.

\appendix

\section{Truth-table dictionary}
\label{app.1}

In this Appendix we classify the encoding of all truth tables up to 4
literals.  Given $k$ literals $u_1$, \ldots, $u_k$, in principle there
are $2^{(2^k)}$ possible truth tables, corresponding to the set of
subsets of $\{ \textrm{True}, \textrm{False} \}^k$ (the satisfying
assignments of the literals).  More precisely, a clause $C_T$ is
identified by the set $T$ of literal configurations evaluated to True
\begin{align}
T&=\{ \vec{\tau} \} \subseteq 
\{ \textrm{True}, \textrm{False} \}^k
\ef;
&
C_T(\vec{\tau})
&=\mathrm{True} 
\quad \Leftrightarrow \quad
\vec{\tau}
\in T
\ef.
\end{align}
Still in section~\ref{sec.more} we discuss the ``gauge'' invariance
which allows to consider truth tables inside the same class as
equivalent, this fact allows to reduce the number of tables to be
analyzed, for a fixed value of the length
$k$. This number is further reduced by the fact that certain tables
are ``trivial'' inside a boolean decision problem, in a sense which is
depicted by the following rules:
\begin{description}
\item[Rule 1:] The clause is always UNSAT, i.e.~$T=\emptyset$.
  The instance is determined to be
  unsatisfiable. 
\item[Rule 2:] The clause is always SAT, 
  i.e.~$T=\{ \textrm{True}, \textrm{False} \}^k$.
  One can reduce to a smaller
  instance, where this clause is removed.
\item[Rule 3:] The clause can be encoded in logical form with no use of a
  literal (say, $u_i$). On the truth table, this means that, defining
  the symmetry operation
\[
R_i (\tau_1, \ldots, \tau_k)=
(\tau_1, \ldots, \negg{\tau}_i, \ldots, \tau_k)
\ef,
\]
  $T = R_i (T)$. 
  The clause is equivalent
  to a smaller one, and thus the whole instance can be reduced in
  size.
\item[Rule 4:] The clause forces the assignment of a given literal (say,
  $u_i=\textrm{True}$, or $u_i=\textrm{False}$). 
  One can simply regard $u_i$ as a synonimous of the fixed boolean
  variable, and the instance is reduced in size
  (the new instance has one literal less, and all the
  clauses in which $u_i$ appears are shorter).
\item[Rule 5:] The clause forces the relative assignment of a given pair of
 literals (say, $u_i \xor u_j=\textrm{True}$, or
 $u_i \xor u_j=\textrm{False}$). 
  One can simply regard $u_j$ as a synonimous of $u_i$ (or of
 $\negg{u}_i$), and the instance is reduced in size
  (the new instance has one literal less, and all the
  clauses in which both $u_i$ and $u_j$ appear are shorter).
\end{description}
Actually, Rule~5 is not so rigorous: sometimes synonima could be
useful in certain constructions, and for example we made large use of
them in our reduction proof from SAT to One-in-Two Matching (and we
also described the ``consistency check'' clause, which indeed violates
Rule~5 on \emph{all} the pairs). 
So, in our enumeration we will include all clauses violating
Rule~5, although in this case we will point out the fact.


We make a complete enumeration ``by hand'' up to length 3, where the
truth tables are easy to handle. We skip both clauses which are
trivial w.r.t.~rules 1-3, and clauses of the kind ``range-T'',
for which a general recipe has already be given in
section~\ref{sec.more}.

For length 4, the manual enumeration would have been a formidable
task, so we preferred to use a randomized search method: we just tried
to fill randomly matrices $W$, with a given density, and classified
the resulting clauses, up to when each class had at least one
representative.\footnote{Although this was not assured to happen, we
  were lucky, and found a representative per each class in a few hours
  of computer time.}

Another difficulty is that, as the same truth tables are difficult to
handle, our results would be hardly fruible if the user would not have
a recipe to identify his clause of interest in our list. For this
reason, we decided an easy algorithm for the determination of an
``algebraic signature'' of the gauge orbit of a given truth
table. This signature is an integer number, and thus, as in a
``dictionary'', decides the sorting of the entries of our enumeration.
For fruibility to the user, also the clauses violating rules 3 and 4
are listed, with a side annotation. 
On the other side, the cases $|T| \leq 2$ and $|T| \geq 14
= 2^4-2$, which fall in the general framework of
section~\ref{sec.more}, have been omitted.

More detailed explanations are given below.

\subsection{Clauses of length 2}

The cases $|T|=0,1,4$ are trivial. The case $|T|=3$ is SAT. The case
$|T|=2$, which is both NAE-SAT and XOR, i.e.~truth-setting structure, 
is already in the framework of section~\ref{sec.redproof}.
So, there are no new cases.

\subsection{Clauses of length 3}

\begin{table}
\caption{\label{tab.k3}Encoding matrices for the new clauses of length
  3. For each clause are given: a boolean description, a suggested
  One-in-Two encoding, and a graphical description of the truth table
  (bottom vertex: $(TTT)$; bottom row, left to right:
  $(TTF)\cdots(FTT)$; filled disk: $\vec{\tau}\in T$.)}
\begin{center}
\begin{tabular}{|c|ccc|}
\hline
\rule{0pt}{4mm}\raisebox{-2mm}{\rule{0pt}{5mm}}%
$|T|$ & \multicolumn{3}{|c|}{clauses} \\
\hline
\rule{0pt}{5mm}\raisebox{-3mm}{\rule{0pt}{5mm}}%
&
$u_1 \bor (u_2 \band u_3)$ &
NAE $\band (u_1 \bor \negg{u}_2 \bor u_3)$ & 
$ \negg{\textrm{XOR}} \bor (u_1 \band u_2 \band u_3)$
\\ 
5
&
{\scriptsize
$
\begin{pmatrix}
 0 & 0 & 0 & 0 & 0 & 1 \\
 0 & 1 & 0 & 0 & 0 & 0 \\
 0 & 0 & 1 & 0 & 0 & 0 \\
 1 & 0 & 0 & 1 & 0 & 0 \\
 0 & 0 & 0 & 0 & 1 & 0 \\
 0 & 0 & 0 & 1 & 0 & 1
\end{pmatrix}
$
}
&
{\scriptsize
$
\begin{pmatrix}
 0 & 0 & 0 & 0 & 0 & 1 \\
 0 & 1 & 0 & 0 & 0 & 0 \\
 0 & 0 & 1 & 0 & 0 & 1 \\
 1 & 0 & 0 & 1 & 0 & 0 \\
 0 & 0 & 0 & 0 & 1 & 0 \\
 1 & 0 & 1 & 0 & 0 & 0
\end{pmatrix}
$
}
&
{\scriptsize
$
\begin{pmatrix}
 1 & 0 & 0 & 0 & 0 & 0 \\
 0 & 0 & 0 & 1 & 0 & 1 \\
 0 & 0 & 1 & 0 & 0 & 0 \\
 0 & 1 & 0 & 0 & 0 & 1 \\
 0 & 0 & 0 & 0 & 1 & 0 \\
 0 & 1 & 0 & 1 & 0 & 0
\end{pmatrix}
$
}
\\
&
\includegraphics[bb=0 0 40 40, clip=true, scale=1.]{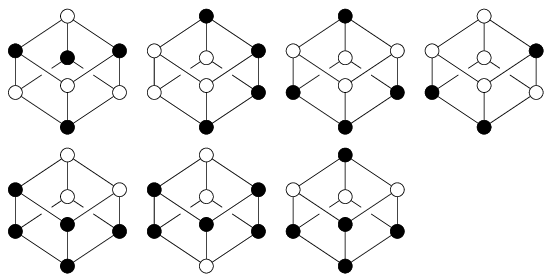}
&
\includegraphics[bb=40 0 80 40, clip=true, scale=1.]{fig_cubi.eps}
&
\includegraphics[bb=80 0 120 40, clip=true, scale=1.]{fig_cubi.eps}
\\
\hline
\rule{0pt}{5mm}\raisebox{-3mm}{\rule{0pt}{5mm}}%
&
$(u_1 \xor u_2 \xor u_3)$
&
$(\negg{u}_1 \band \negg{u}_2) \xor (u_2 \band u_3)$
& 
$u_2 \xor (u_1 \bor u_3)$
\\ 
4
&
{\scriptsize
$
\begin{pmatrix}
 0 & 0 & 0 & 0 & 1 & 1 \\
 0 & 0 & 1 & 1 & 0 & 0 \\
 1 & 0 & 0 & 0 & 1 & 0 \\
 1 & 0 & 0 & 0 & 0 & 1 \\
 0 & 1 & 0 & 1 & 0 & 0 \\
 0 & 1 & 1 & 0 & 0 & 0
\end{pmatrix}
$
}
&
{\scriptsize
$
\begin{pmatrix}
 1 & 0 & 1 & 0 & 0 & 0 \\
 0 & 1 & 0 & 0 & 0 & 0 \\
 1 & 0 & 0 & 0 & 0 & 0 \\
 1 & 0 & 0 & 0 & 0 & 1 \\
 0 & 0 & 0 & 0 & 1 & 0 \\
 0 & 0 & 0 & 1 & 0 & 1
\end{pmatrix}
$
}
&
{\scriptsize
$
\begin{pmatrix}
 1 & 0 & 0 & 0 & 0 & 0 \\
 0 & 0 & 0 & 1 & 0 & 0 \\
 0 & 0 & 1 & 0 & 0 & 0 \\
 0 & 1 & 0 & 0 & 0 & 1 \\
 0 & 0 & 0 & 0 & 1 & 0 \\
 0 & 1 & 0 & 1 & 0 & 0
\end{pmatrix}
$
}
\\
&
\includegraphics[bb=0 40 40 80, clip=true, scale=1.]{fig_cubi.eps}
&
\includegraphics[bb=40 40 80 80, clip=true, scale=1.]{fig_cubi.eps}
&
\includegraphics[bb=80 40 120 80, clip=true, scale=1.]{fig_cubi.eps}
\\
\hline
\rule{0pt}{5mm}\raisebox{-3mm}{\rule{0pt}{5mm}}%
&
\multicolumn{2}{|l}
{$(u_1 \xor \negg{u}_2) \band (u_1 \!\bor\! u_2 \!\bor\! u_3)$} & \\
3 
&
{\scriptsize
$
\begin{pmatrix}
 0 & 0 & 1 & 0 & 0 & 0 \\
 0 & 0 & 0 & 1 & 1 & 0 \\
 1 & 0 & 0 & 0 & 0 & 0 \\
 0 & 1 & 0 & 0 & 0 & 0 \\
 0 & 0 & 0 & 1 & 0 & 0 \\
 0 & 0 & 0 & 0 & 0 & 1
\end{pmatrix}
$
}
&&
\\
&
\includegraphics[bb=120 40 160 80, clip=true, scale=1.]{fig_cubi.eps}
&&
\\
\hline
\end{tabular}
\end{center}
\end{table}

The cases $|T|=0,1,8$ are trivial. The cases $|T|=2,6,7$ have been
studied in sections~\ref{sec.redproof} and~\ref{sec.more}.

Now consider the case $|T|=3$, and call 
$T=\{ \vec{\tau}_1, \vec{\tau}_2, \vec{\tau}_3 \}$. The set of
``distances'', where the distance is defined as
$d(\vec{\tau},\vec{\tau}')= \# \{ i : \tau_i \neq \tau'_i\}$, can form
the triangles $(1,1,2)$, $(2,2,2)$ and $(1,2,3)$. The clause $(1,1,2)$
violates Rule~4, because leaves a sub-hypercube (just a square in this
case) outside.  The clause $(2,2,2)$ is the One-in-Three clause, which
is a special case of range-T clause.  The clause $(1,2,3)$ violates
Rule~5. 
An example is $(u_1 \xor \negg{u}_2) \band (u_1 \bor u_2 \bor u_3)$, 
where violation of Rule~5 is made explicit. Anyhow, we give a
matrix encoding in Table~\ref{tab.k3}.

The case $|T|=5$ has a similar case study: if we consider the
complementary of $T$, it has cardinality three, and with the same
reasoning above we have the three cases with sets of distances
$(1,1,2)$, $(2,2,2)$ and $(1,2,3)$. The difference is that now all
these cases are new.

In the case $|T|=4$, either all the four vectors are at distance 2
(and in this case we deal with the important XOR clause) or there are
at least two neighbouring vectors (say, $\vec{\tau}_{1,2}$). 
Then, six vertices remain on the cube, of which
four are neighbours of $\vec{\tau}_{1,2}$, so either
$\vec{\tau}_{3,4}$ fill the pair of non-neighbouring sites, 
and we have the trivial
clause $(u_1 \xor u_2)$, which violates many rules, or there is a
vertex with two neighbours, say, $\vec{\tau}_{1}$
is at distance one from $\vec{\tau}_{2}$ and $\vec{\tau}_{3}$. If it
is at distance one also from $\vec{\tau}_{4}$, we have an
``At-least-2-in-3'' clause, i.e.~a special case of range-T clause
with $k=h_{\textrm{max}}=3$ and $h_{\textrm{min}}=2$, for
which we have a general recipe. Otherwise, if $\vec{\tau}_{4}$ is at
the last vertex of the square with vertices $\vec{\tau}_{1,2,3}$, we
have a trivial clause of length 1. There are only two other choices
left, either the pairs of neighbouring vertices form an open path of
length 3 (the ``first-F-then-T'' clause, 
$(\negg{u}_1 \band \negg{u}_2) \xor (u_2 \band u_3)$, 
with truth table $\{(TTT),(FTT),(FFT),(FFF)\}$), or the last
vertex $\vec{\tau}_{4}$ is antipodal to $\vec{\tau}_{1}$.

This completes the case study for $k=3$: we have three new cases at
$|T|=4$ and three at $|T|=5$, and a case at $|T|=3$ which violates
Rule~5. The suggested encoding matrices are listed in
Table~\ref{tab.k3}.

\subsection{Clauses of length 4}

Each entry of our dictionary is encoded in the form

\noindent
\rule{0pt}{15pt}%
\raisebox{-10pt}{\rule{0pt}{14pt}}%
{\small $|T|$ \quad Algebraic signature \quad Truth table 
\quad Encoding matrix $W$ \quad
$\# \{w_{ij}=1 \}$ \quad (Comments)}

\noindent
The \emph{Algebraic signature} allows to determine the gauge
class of a given truth table. This is the reason why the truth tables
are sorted according to that.
In a sense, it is a univoque name for a
clause: if you want to encode a given truth table with 4 literals and
$|T|$ True assignments into a One-in-Two pattern, and this truth table
does not follow into one of the general classes studied in
sections~\ref{sec.redproof} and~\ref{sec.more}, you should calculate
the algebraic signature (a simple and fast algebraic procedure), and
then search the corresponding entry in this dictionary; then, the other
fields of the entry will allow you to construct the proper matrix.

Finding a good algebraic signature recipe is in principle a
complicated task, because asking whether two truth tables are
isomorphic subgraphs of the $k$-dimensional hypercube is indeed a
graph-isomorphism problem, which is a hard problem
(cfr.~\cite{graphiso}). Nonetheless, in our case it is certainly
affordable, because the coordination is bounded by 4. Furthermore, as
fortunately the hard examples of graph-isomorphism are only a narrow
set with special geometrical properties, it is plausible that also 
the na\"ive algebraic techniques, consisting of a calculation of
a determinant, allow to discriminate the classes. Consider the
distance $d(\vec{\tau},\vec{\tau}')= \# \{ i : \tau_i \neq \tau'_i\}$,
which is an integer in $\{0, 1, \ldots, k\}$, and 0 only for
coincident vectors; then consider $k+1$ relatively-trascendent real
values $r(i)$. We have chosen
\be
\left\{
\begin{array}{rl}
r(0) &= 1
\ef;
\\
r(1) &= 
\zeta(3)^{-1}
= 0.831907372580707469
\ef;
\\
r(2) &= -\gamma_{\textrm{Euler}} = -0.577215664901532861
\ef;
\\
r(3) &= e^{-1} = 0.367879441171442322
\ef;
\\
r(4) &= -\pi^{-1} = -0.318309886183790672
\ef.
\end{array}
\right.
\ee
The choice is, of course, arbitrary. We choose decreasing real numbers
in $[0,1]$, with alternating signs, in order to have determinants of
the same order of magnitude. Then, we build the square matrix $M$ of size
$|T|$, with 
$M_{\vec{\tau}, \vec{\tau}'}=r(d(\vec{\tau},\vec{\tau}'))$. Just to reduce
computation times, if $|T| > 2^{k-1}$ we use instead of $T$ its complementary
set. The chosen ``algebraic signature'' of truth table $T$
is the integer part of $10^{16} \det M(T)$.

The truth table is simply encoded in a compact way, the $2^{16}$
possible vectors $\vec{\tau}$ being ordered lexicografically, and the
table $T$ being denoted by the integer 
$n(T) \in \{ 0, \ldots, 2^{16}-1 \}$ 
corresponding to the binary string of vectors evaluated to True
\begin{align}
n(T)&=\sum_{\vec{\tau} \in T}
2^{n'(\tau)}
\ef,
&
n'(\tau)&=\sum_{i=1}^k 2^{k-i} \delta_{\tau_i, \textrm{True}}
\ef.
\end{align}
The representative of the class is chosen in arbitrary way. Indeed, it
comes out that it is the one with lowest index $n(T)$, among the $T$s
in the same class.

Similarly, the encoding matrix is written in a more compact way as a
vector of integers, where each row of 0s and 1s, of length $2k=8$, is
replaced by the corresponding integer in $\{0, \ldots, 2^{8}-1 \}$.
Just for statistic reasons, the number of 1s in the matrix is
indicated.

Finally, if the clause is trivial in the sense of the rules~3-5 above
(failures of rules 1 and 2 are excluded trivially), we report the
fact, and the rule it fails (if it is trivial for more than one
reason, we choose a reason of smaller rule index).

Here the dictionary follows:

\noindent
\myappline
\mydictitem{3 }{ -15162648248230256 }{ 8226 }{ ( 
  1 , 64 , 32 , 17 , 8 , 0 , 
   0 , 144 ) }{ 8 }{R4: $u_3 = \textrm{True}$ }
\mydictitem{3 }{ -5138871171782917 }{ 8195 }{ ( 
  8 , 1 , 32 , 8 , 144 , 0 , 
   2 , 65 ) }{ 9 }{R4: $u_3 = \textrm{True}$ }
\mydictitem{3 }{ -3841648050655253 }{ 8210 }{ ( 
  34 , 1 , 1 , 16 , 192 , 0 , 
   8 , 8 ) }{ 9 }{R4: $u_3 = \textrm{True}$ }
\mydictitem{3 }{ -1235584894668443 }{ 8197 }{ ( 
  2 , 16 , 32 , 8 , 9 , 4 , 
   128 , 64 ) }{ 9 }{R5: $u_2 \xor u_4 = \textrm{False}$ }
\mydictitem{3 }{ 202153147298516 }{ 8198 }{ ( 
  32 , 16 , 132 , 64 , 8 , 2 , 
   128 , 1 ) }{ 9 }{R5: $u_3 \xor u_4 = \textrm{True}$ }
\mydictitem{3 }{ 2399162187229386 }{ 8201 }{ ( 
  6 , 8 , 128 , 64 , 48 , 32 , 
   8 , 1 ) }{ 10 }{R5: $u_1 \xor u_2 = \textrm{False}$ }
\myappline
\mydictitem{4 }{ -64417460349479835 }{ 8738 }{ ( 
  128 , 64 , 32 , 16 , 1 , 0 , 
   0 , 8 ) }{ 6 }{R3: $u_3 \not\in C$}
\mydictitem{4 }{ -55489630922495513 }{ 8242 }{ ( 
  3 , 1 , 128 , 192 , 24 , 0 , 
   8 , 48 ) }{ 11 }{R4: $u_3 = \textrm{True}$ }
\mydictitem{4 }{ -45874752110564406 }{ 8227 }{ ( 
  8 , 8 , 3 , 16 , 161 , 0 , 
   128 , 96 ) }{ 11 }{R4: $u_3 = \textrm{True}$ }
\mydictitem{4 }{ -37719695731007326 }{ 8211 }{ ( 
  40 , 8 , 136 , 16 , 3 , 0 , 
   128 , 96 ) }{ 11 }{R4: $u_3 = \textrm{True}$ }
\mydictitem{4 }{ -32798143896572570 }{ 8203 }{ ( 
  9 , 8 , 128 , 64 , 3 , 4 , 
   32 , 48 ) }{ 11 }{R5: $u_1 \xor u_2 = \textrm{False}$ }
\mydictitem{4 }{ -30712541418824951 }{ 8199 }{ ( 
  32 , 16 , 132 , 1 , 9 , 128 , 
   2 , 72 ) }{ 11 }{R5: $u_1 \xor u_2 = \textrm{False}$ }
\mydictitem{4 }{ -28706142641542548 }{ 8338 }{ ( 
  3 , 8 , 32 , 192 , 24 , 16 , 
   8 , 133 ) }{ 13 }{}
\mydictitem{4 }{ -28647676317807325 }{ 8481 }{ ( 
  8 , 64 , 128 , 8 , 48 , 0 , 
   2 , 1 ) }{ 8 }{R3: $u_4 \not\in C$}{}
\mydictitem{4 }{ -26253185098326629 }{ 8205 }{ ( 
  6 , 16 , 34 , 64 , 8 , 128 , 
   160 , 1 ) }{ 11 }{R5: $u_1 \xor u_2 = \textrm{False}$ }{}
\mydictitem{4 }{ -23870314469588248 }{ 8218 }{ ( 
  3 , 64 , 12 , 16 , 129 , 32 , 
   32 , 136 ) }{ 12 }{}
\mydictitem{4 }{ -21717939174696339 }{ 8214 }{ ( 
  10 , 64 , 129 , 8 , 144 , 4 , 
   32 , 33 ) }{ 12 }{}
\mydictitem{4 }{ -21459144106706130 }{ 8229 }{ ( 
  6 , 64 , 128 , 16 , 9 , 32 , 
   34 , 8 ) }{ 11 }{R5: $u_2 \xor u_4 = \textrm{False}$ }{}
\mydictitem{4 }{ -19374761209641485 }{ 8217 }{ ( 
  33 , 64 , 160 , 1 , 8 , 128 , 
   2 , 20 ) }{ 11 }{}
\mydictitem{4 }{ -17236351616013344 }{ 8220 }{ ( 
  128 , 16 , 5 , 64 , 9 , 32 , 
   2 , 40 ) }{ 11 }{R5: $u_2 \xor u_3 = \textrm{True}$ }{}
\mydictitem{4 }{ -15085506227251376 }{ 8580 }{ ( 
  32 , 64 , 128 , 16 , 1 , 4 , 
   2 , 8 ) }{ 8 }{R5: $u_2 \xor u_4 = \textrm{False}$ }{}
\mydictitem{4 }{ -14842292216336441 }{ 8772 }{ ( 
  2 , 1 , 128 , 16 , 8 , 4 , 
   32 , 64 ) }{ 8 }{R3: $u_3 \not\in C$}{}
\mydictitem{4 }{ -13941487138238279 }{ 8265 }{ ( 
  12 , 1 , 129 , 16 , 96 , 128 , 
   2 , 40 ) }{ 12 }{}
\mydictitem{4 }{ -9019935303803522 }{ 8520 }{ ( 
  4 , 24 , 160 , 129 , 64 , 17 , 
   2 , 40 ) }{ 13 }{R5: $u_1 \xor u_3 = \textrm{True}$ }{}
\myappline
\mydictitem{5 }{ -137319486169585905 }{ 8370 }{ ( 
  7 , 64 , 32 , 9 , 144 , 128 , 
   16 , 24 ) }{ 13 }{}
\mydictitem{5 }{ -131925251824472796 }{ 8243 }{ ( 
  3 , 16 , 32 , 17 , 8 , 0 , 
   128 , 192 ) }{ 10 }{R4: $u_3 = \textrm{True}$ }{}
\mydictitem{5 }{ -106990011212970741 }{ 8235 }{ ( 
  9 , 1 , 128 , 192 , 10 , 32 , 
   32 , 52 ) }{ 13 }{}
\mydictitem{5 }{ -106200686484487583 }{ 8467 }{ ( 
  9 , 64 , 32 , 129 , 24 , 0 , 
   2 , 136 ) }{ 11 }{R4: $u_3 = \textrm{True}$ }{}
\mydictitem{5 }{ -100711446287411196 }{ 8483 }{ ( 
  34 , 64 , 160 , 16 , 129 , 0 , 
   8 , 8 ) }{ 10 }{R4: $u_3 = \textrm{True}$ }{}
\mydictitem{5 }{ -97383480909282057 }{ 8207 }{ ( 
  32 , 8 , 128 , 1 , 24 , 4 , 
   2 , 65 ) }{ 10 }{R5: $u_1 \xor u_2 = \textrm{False}$ }{}
\mydictitem{5 }{ -93806950451816800 }{ 8215 }{ ( 
  11 , 1 , 128 , 144 , 40 , 32 , 
   22 , 72 ) }{ 15 }{}
\mydictitem{5 }{ -87252051588588716 }{ 8231 }{ ( 
  56 , 64 , 132 , 144 , 33 , 128 , 
   2 , 40 ) }{ 14 }{}
\mydictitem{5 }{ -84213109538527750 }{ 8219 }{ ( 
  130 , 16 , 9 , 72 , 49 , 4 , 
   128 , 40 ) }{ 14 }{}
\mydictitem{5 }{ -81265445872985527 }{ 8339 }{ ( 
  24 , 1 , 129 , 28 , 168 , 1 , 
   2 , 192 ) }{ 15 }{}
\mydictitem{5 }{ -79740217163681572 }{ 8269 }{ ( 
  34 , 1 , 128 , 80 , 8 , 161 , 
   6 , 20 ) }{ 14 }{}
\mydictitem{5 }{ -72810579632836741 }{ 8221 }{ ( 
  34 , 1 , 128 , 144 , 40 , 3 , 
   28 , 68 ) }{ 15 }{}
\mydictitem{5 }{ -69543625692646563 }{ 8222 }{ ( 
  128 , 16 , 11 , 8 , 17 , 4 , 
   34 , 96 ) }{ 13 }{}
\mydictitem{5 }{ -69010625791285801 }{ 8342 }{ ( 
  6 , 10 , 4 , 16 , 129 , 132 , 
   104 , 65 ) }{ 15 }{}
\mydictitem{5 }{ -67084531825011153 }{ 8267 }{ ( 
  160 , 1 , 9 , 16 , 96 , 4 , 
   2 , 136 ) }{ 12 }{}
\mydictitem{5 }{ -65196296191649679 }{ 8252 }{ ( 
  36 , 64 , 128 , 16 , 8 , 1 , 
   2 , 33 ) }{ 10 }{R5: $u_2 \xor u_3 = \textrm{True}$ }{}
\mydictitem{5 }{ -64617744063353716 }{ 8237 }{ ( 
  161 , 64 , 12 , 16 , 10 , 130 , 
   128 , 33 ) }{ 14 }{}
\mydictitem{5 }{ -62513764335079739 }{ 8297 }{ ( 
  37 , 1 , 6 , 18 , 8 , 144 , 
   128 , 224 ) }{ 15 }{}
\mydictitem{5 }{ -60805966899451609 }{ 8282 }{ ( 
  34 , 64 , 36 , 16 , 8 , 128 , 
   128 , 1 ) }{ 10 }{}
\mydictitem{5 }{ -58556655299675617 }{ 8581 }{ ( 
  128 , 8 , 32 , 1 , 66 , 4 , 
   10 , 16 ) }{ 10 }{R5: $u_2 \xor u_4 = \textrm{False}$ }{}
\mydictitem{5 }{ -57613278077157236 }{ 8582 }{ ( 
  34 , 1 , 132 , 16 , 8 , 32 , 
   128 , 65 ) }{ 11 }{}
\mydictitem{5 }{ -54873765588920526 }{ 8524 }{ ( 
  4 , 8 , 3 , 66 , 48 , 128 , 
   80 , 33 ) }{ 13 }{R5: $u_1 \xor u_3 = \textrm{True}$ }{}
\mydictitem{5 }{ -53148596289303402 }{ 8773 }{ ( 
  2 , 64 , 160 , 16 , 8 , 128 , 
   36 , 1 ) }{ 10 }{R5: $u_1 \xor u_4 = \textrm{False}$ }{}
\mydictitem{5 }{ -51297235131455269 }{ 8476 }{ ( 
  3 , 64 , 32 , 129 , 16 , 20 , 
   6 , 136 ) }{ 13 }{}
\mydictitem{5 }{ -49945641870770219 }{ 9289 }{ ( 
  136 , 33 , 131 , 10 , 96 , 4 , 
   82 , 48 ) }{ 17 }{}
\mydictitem{5 }{ -49144516830228831 }{ 8492 }{ ( 
  128 , 40 , 14 , 9 , 18 , 32 , 
   68 , 65 ) }{ 15 }{}
\mydictitem{5 }{ -48332850287105365 }{ 8521 }{ ( 
  48 , 40 , 136 , 66 , 3 , 4 , 
   192 , 17 ) }{ 15 }{}
\myappline
\mydictitem{6 }{ -264953748043115284 }{ 8755 }{ ( 
  24 , 16 , 32 , 8 , 192 , 0 , 
   2 , 1 ) }{ 9 }{R3: $u_3 \not\in C$}{}
\mydictitem{6 }{ -255903677087804212 }{ 8371 }{ ( 
  128 , 16 , 1 , 24 , 72 , 1 , 
   2 , 37 ) }{ 12 }{}
\mydictitem{6 }{ -246423116736164502 }{ 8499 }{ ( 
  29 , 12 , 8 , 81 , 65 , 133 , 
   2 , 144 ) }{ 18 }{R4: $u_3 = \textrm{True}$ }{}
\mydictitem{6 }{ -234055909760095663 }{ 8251 }{ ( 
  40 , 8 , 136 , 16 , 161 , 1 , 
   2 , 196 ) }{ 14 }{}
\mydictitem{6 }{ -221848639666228098 }{ 8247 }{ ( 
  40 , 64 , 10 , 24 , 3 , 128 , 
   132 , 144 ) }{ 14 }{}
\mydictitem{6 }{ -219506781223121329 }{ 8374 }{ ( 
  6 , 64 , 128 , 16 , 129 , 130 , 
   44 , 9 ) }{ 14 }{}
\mydictitem{6 }{ -211897911062832022 }{ 9009 }{ ( 
  8 , 8 , 145 , 16 , 103 , 0 , 
   228 , 96 ) }{ 17 }{R4: $u_3 = \textrm{True}$ }{}
\mydictitem{6 }{ -198829802885264160 }{ 8239 }{ ( 
  33 , 1 , 128 , 8 , 24 , 4 , 
   2 , 225 ) }{ 13 }{}
\mydictitem{6 }{ -196931604634552768 }{ 8223 }{ ( 
  11 , 16 , 136 , 8 , 129 , 4 , 
   48 , 96 ) }{ 14 }{}
\mydictitem{6 }{ -193105575805408406 }{ 8314 }{ ( 
  3 , 1 , 160 , 16 , 96 , 128 , 
   8 , 12 ) }{ 12 }{}
\mydictitem{6 }{ -186798665756457220 }{ 8271 }{ ( 
  38 , 1 , 128 , 16 , 8 , 3 , 
   164 , 68 ) }{ 14 }{}
\mydictitem{6 }{ -186227373079063234 }{ 9293 }{ ( 
  6 , 5 , 131 , 66 , 8 , 160 , 
   112 , 48 ) }{ 17 }{}
\mydictitem{6 }{ -184055504850097333 }{ 8299 }{ ( 
  14 , 8 , 32 , 192 , 18 , 17 , 
   132 , 129 ) }{ 15 }{}
\mydictitem{6 }{ -180868489214616524 }{ 8254 }{ ( 
  45 , 64 , 128 , 16 , 129 , 32 , 
   2 , 12 ) }{ 13 }{}
\mydictitem{6 }{ -176493283894497754 }{ 8343 }{ ( 
  12 , 10 , 2 , 16 , 66 , 164 , 
   192 , 1 ) }{ 14 }{}
\mydictitem{6 }{ -174574944498457624 }{ 8347 }{ ( 
  160 , 1 , 138 , 16 , 98 , 1 , 
   8 , 13 ) }{ 15 }{}
\mydictitem{6 }{ -164368307347587140 }{ 8477 }{ ( 
  130 , 1 , 9 , 16 , 96 , 4 , 
   128 , 40 ) }{ 12 }{}
\mydictitem{6 }{ -164359755773378736 }{ 8283 }{ ( 
  129 , 1 , 36 , 74 , 146 , 132 , 
   24 , 40 ) }{ 17 }{}
\mydictitem{6 }{ -159842111719858328 }{ 8253 }{ ( 
  35 , 1 , 128 , 16 , 194 , 32 , 
   12 , 12 ) }{ 14 }{}
\mydictitem{6 }{ -157236364066051777 }{ 8583 }{ ( 
  26 , 16 , 160 , 72 , 161 , 4 , 
   138 , 40 ) }{ 17 }{}
\mydictitem{6 }{ -155304537441776442 }{ 8286 }{ ( 
  142 , 8 , 128 , 9 , 66 , 36 , 
   48 , 17 ) }{ 16 }{}
\mydictitem{6 }{ -153275171299567776 }{ 8525 }{ ( 
  128 , 10 , 9 , 66 , 65 , 36 , 
   48 , 20 ) }{ 15 }{}
\mydictitem{6 }{ -153237592690580239 }{ 8494 }{ ( 
  162 , 66 , 12 , 8 , 65 , 129 , 
   32 , 144 ) }{ 15 }{}
\mydictitem{6 }{ -150450270880229517 }{ 8301 }{ ( 
  49 , 64 , 136 , 129 , 18 , 34 , 
   132 , 12 ) }{ 16 }{}
\mydictitem{6 }{ -147186103159461859 }{ 8652 }{ ( 
  4 , 8 , 3 , 3 , 80 , 48 , 
   160 , 192 ) }{ 14 }{R5: $u_1 \xor u_3 = \textrm{True}$ }{}
\mydictitem{6 }{ -144210284492927278 }{ 8598 }{ ( 
  18 , 68 , 4 , 11 , 16 , 224 , 
   144 , 5 ) }{ 16 }{}
\mydictitem{6 }{ -142969789437608309 }{ 8478 }{ ( 
  38 , 80 , 2 , 73 , 24 , 130 , 
   132 , 16 ) }{ 16 }{}
\mydictitem{6 }{ -142370542521970898 }{ 8523 }{ ( 
  3 , 64 , 136 , 129 , 48 , 4 , 
   18 , 40 ) }{ 14 }{}
\mydictitem{6 }{ -141698008727709048 }{ 8789 }{ ( 
  32 , 8 , 128 , 16 , 72 , 4 , 
   2 , 1 ) }{ 9 }{R3: $u_3 \not\in C$}{}
\mydictitem{6 }{ -138524570147284945 }{ 8749 }{ ( 
  37 , 33 , 14 , 8 , 88 , 32 , 
   128 , 145 ) }{ 17 }{}
\mydictitem{6 }{ -138108485317911594 }{ 8775 }{ ( 
  131 , 16 , 132 , 136 , 72 , 32 , 
   50 , 129 ) }{ 16 }{}
\mydictitem{6 }{ -135427674056010516 }{ 8589 }{ ( 
  48 , 64 , 131 , 17 , 9 , 2 , 
   132 , 136 ) }{ 15 }{}
\mydictitem{6 }{ -133261805405082912 }{ 8493 }{ ( 
  128 , 48 , 69 , 1 , 8 , 32 , 
   2 , 84 ) }{ 13 }{}
\mydictitem{6 }{ -132786334798600321 }{ 8526 }{ ( 
  20 , 48 , 130 , 68 , 8 , 34 , 
   192 , 1 ) }{ 14 }{}
\mydictitem{6 }{ -130731773807879766 }{ 8781 }{ ( 
  148 , 64 , 33 , 136 , 16 , 130 , 
   4 , 33 ) }{ 14 }{}
\mydictitem{6 }{ -130024582675614020 }{ 9291 }{ ( 
  136 , 40 , 34 , 66 , 48 , 68 , 
   148 , 1 ) }{ 16 }{}
\mydictitem{6 }{ -125686589014661632 }{ 10644 }{ ( 
  10 , 96 , 96 , 10 , 17 , 20 , 
   132 , 129 ) }{ 16 }{}
\mydictitem{6 }{ -125660450966743013 }{ 8857 }{ ( 
  6 , 1 , 32 , 16 , 8 , 128 , 
   128 , 65 ) }{ 10 }{R3: $u_3 \not\in C$}{}
\mydictitem{6 }{ -124616208289630152 }{ 8604 }{ ( 
  128 , 18 , 4 , 26 , 9 , 48 , 
   68 , 65 ) }{ 15 }{}
\mydictitem{6 }{ -123666407520778105 }{ 8508 }{ ( 
  20 , 64 , 3 , 144 , 1 , 34 , 
   4 , 136 ) }{ 13 }{}
\mydictitem{6 }{ -122864364540640491 }{ 8613 }{ ( 
  40 , 16 , 128 , 64 , 40 , 2 , 
   6 , 1 ) }{ 11 }{R5: $u_2 \xor u_4 = \textrm{False}$ }{}
\mydictitem{6 }{ -120544022323974311 }{ 8538 }{ ( 
  145 , 64 , 32 , 10 , 3 , 132 , 
   12 , 144 ) }{ 15 }{}
\mydictitem{6 }{ -119127428200296783 }{ 8790 }{ ( 
  128 , 16 , 12 , 8 , 67 , 33 , 
   18 , 33 ) }{ 14 }{}
\mydictitem{6 }{ -118936297852452994 }{ 9321 }{ ( 
  36 , 3 , 13 , 66 , 10 , 208 , 
   160 , 20 ) }{ 18 }{}
\mydictitem{6 }{ -113571267670065141 }{ 8796 }{ ( 
  33 , 18 , 36 , 192 , 72 , 128 , 
   10 , 17 ) }{ 15 }{}
\mydictitem{6 }{ -112145695798823883 }{ 8556 }{ ( 
  6 , 64 , 130 , 17 , 40 , 129 , 
   12 , 48 ) }{ 15 }{}
\mydictitem{6 }{ -110328833598895423 }{ 9306 }{ ( 
  17 , 18 , 160 , 192 , 65 , 4 , 
   10 , 40 ) }{ 15 }{}
\mydictitem{6 }{ -109077852951287423 }{ 9036 }{ ( 
  20 , 8 , 65 , 5 , 96 , 128 , 
   2 , 48 ) }{ 13 }{R5: $u_1 \xor u_3 = \textrm{True}$ }{}
\mydictitem{6 }{ -101641038122898375 }{ 9576 }{ ( 
  20 , 80 , 32 , 10 , 3 , 132 , 
   136 , 65 ) }{ 15 }{}
\mydictitem{6 }{ -101300461292717132 }{ 9156 }{ ( 
  6 , 48 , 131 , 17 , 64 , 160 , 
   12 , 24 ) }{ 16 }{R5: $u_1 \xor u_3 = \textrm{True}$ }{}
\myappline
\mydictitem{7 }{ -489610810524822682 }{ 9011 }{ ( 
  8 , 64 , 32 , 144 , 25 , 0 , 
   2 , 129 ) }{ 11 }{R4: $u_3 = \textrm{True}$}{}
\mydictitem{7 }{ -452337729517226882 }{ 8379 }{ ( 
  128 , 8 , 32 , 82 , 24 , 4 , 
   16 , 1 ) }{ 11 }{}
\mydictitem{7 }{ -408354319245236969 }{ 8375 }{ ( 
  20 , 16 , 32 , 13 , 136 , 192 , 
   2 , 68 ) }{ 14 }{}
\mydictitem{7 }{ -400582064369183280 }{ 8315 }{ ( 
  11 , 1 , 128 , 192 , 48 , 4 , 
   18 , 41 ) }{ 15 }{}
\mydictitem{7 }{ -393459705592617740 }{ 8255 }{ ( 
  136 , 1 , 32 , 192 , 24 , 4 , 
   2 , 9 ) }{ 12 }{}
\mydictitem{7 }{ -384942911261086050 }{ 8751 }{ ( 
  33 , 1 , 136 , 8 , 192 , 4 , 
   2 , 48 ) }{ 12 }{}
\mydictitem{7 }{ -378955175228864660 }{ 8479 }{ ( 
  144 , 1 , 32 , 9 , 193 , 4 , 
   2 , 24 ) }{ 13 }{}
\mydictitem{7 }{ -372730902319860600 }{ 8287 }{ ( 
  10 , 64 , 160 , 16 , 129 , 4 , 
   32 , 9 ) }{ 12 }{}
\mydictitem{7 }{ -365307444265761614 }{ 8351 }{ ( 
  128 , 12 , 32 , 11 , 16 , 20 , 
   66 , 80 ) }{ 14 }{}
\mydictitem{7 }{ -350325235410304638 }{ 8599 }{ ( 
  6 , 68 , 4 , 24 , 24 , 162 , 
   192 , 1 ) }{ 15 }{}
\mydictitem{7 }{ -346991515368820537 }{ 9295 }{ ( 
  36 , 3 , 192 , 16 , 10 , 34 , 
   140 , 65 ) }{ 16 }{}
\mydictitem{7 }{ -345602958711606536 }{ 8495 }{ ( 
  21 , 1 , 32 , 8 , 194 , 128 , 
   130 , 28 ) }{ 15 }{}
\mydictitem{7 }{ -342028055323505620 }{ 8303 }{ ( 
  168 , 8 , 33 , 17 , 144 , 4 , 
   2 , 192 ) }{ 14 }{}
\mydictitem{7 }{ -334343815080691876 }{ 8411 }{ ( 
  136 , 64 , 32 , 9 , 131 , 1 , 
   8 , 21 ) }{ 14 }{}
\mydictitem{7 }{ -331245387788784281 }{ 8527 }{ ( 
  13 , 72 , 36 , 192 , 1 , 52 , 
   2 , 144 ) }{ 16 }{}
\mydictitem{7 }{ -329455080315344929 }{ 8591 }{ ( 
  129 , 8 , 40 , 1 , 96 , 4 , 
   2 , 144 ) }{ 12 }{}
\mydictitem{7 }{ -326446789752148616 }{ 8318 }{ ( 
  130 , 8 , 134 , 16 , 11 , 36 , 
   40 , 65 ) }{ 16 }{}
\mydictitem{7 }{ -309905349059268324 }{ 8539 }{ ( 
  150 , 64 , 9 , 3 , 152 , 48 , 
   36 , 136 ) }{ 18 }{}
\mydictitem{7 }{ -305364326870046401 }{ 8510 }{ ( 
  19 , 64 , 136 , 10 , 176 , 36 , 
   128 , 5 ) }{ 16 }{}
\mydictitem{7 }{ -303440793052772106 }{ 8783 }{ ( 
  11 , 64 , 32 , 9 , 146 , 4 , 
   128 , 24 ) }{ 14 }{}
\mydictitem{7 }{ -303164240726434633 }{ 8615 }{ ( 
  40 , 16 , 168 , 192 , 137 , 1 , 
   2 , 21 ) }{ 16 }{}
\mydictitem{7 }{ -294782498316163996 }{ 8509 }{ ( 
  144 , 8 , 36 , 129 , 72 , 33 , 
   2 , 48 ) }{ 14 }{}
\mydictitem{7 }{ -293737612421108058 }{ 8654 }{ ( 
  3 , 24 , 96 , 68 , 40 , 144 , 
   130 , 5 ) }{ 16 }{}
\mydictitem{7 }{ -291593988308626710 }{ 8606 }{ ( 
  3 , 68 , 4 , 146 , 8 , 48 , 
   208 , 5 ) }{ 16 }{}
\mydictitem{7 }{ -286206440273200448 }{ 8791 }{ ( 
  10 , 64 , 160 , 1 , 40 , 4 , 
   128 , 17 ) }{ 12 }{}
\mydictitem{7 }{ -282360887598956176 }{ 9323 }{ ( 
  129 , 40 , 140 , 16 , 33 , 6 , 
   70 , 96 ) }{ 17 }{}
\mydictitem{7 }{ -281531220380667350 }{ 8653 }{ ( 
  48 , 72 , 6 , 20 , 40 , 130 , 
   192 , 1 ) }{ 15 }{}
\mydictitem{7 }{ -280171463154608827 }{ 9325 }{ ( 
  6 , 17 , 32 , 208 , 72 , 131 , 
   20 , 24 ) }{ 17 }{}
\mydictitem{7 }{ -280099211892526104 }{ 10646 }{ ( 
  38 , 3 , 40 , 136 , 81 , 20 , 
   136 , 193 ) }{ 19 }{}
\mydictitem{7 }{ -276166021834513026 }{ 8765 }{ ( 
  132 , 1 , 160 , 9 , 24 , 33 , 
   2 , 84 ) }{ 15 }{}
\mydictitem{7 }{ -276136614689952116 }{ 9310 }{ ( 
  132 , 72 , 32 , 3 , 18 , 5 , 
   80 , 136 ) }{ 15 }{}
\mydictitem{7 }{ -275318071137808221 }{ 8605 }{ ( 
  10 , 64 , 6 , 16 , 136 , 32 , 
   160 , 1 ) }{ 12 }{}
\mydictitem{7 }{ -270984298897396969 }{ 9307 }{ ( 
  132 , 5 , 40 , 136 , 96 , 81 , 
   2 , 20 ) }{ 16 }{}
\mydictitem{7 }{ -268713156635853130 }{ 8542 }{ ( 
  11 , 33 , 130 , 152 , 16 , 36 , 
   192 , 69 ) }{ 18 }{}
\mydictitem{7 }{ -263998281304105199 }{ 8797 }{ ( 
  50 , 64 , 4 , 17 , 8 , 130 , 
   36 , 129 ) }{ 14 }{}
\mydictitem{7 }{ -261677808517195389 }{ 10645 }{ ( 
  40 , 64 , 3 , 10 , 144 , 132 , 
   20 , 161 ) }{ 16 }{}
\mydictitem{7 }{ -259011690208036504 }{ 8558 }{ ( 
  5 , 20 , 160 , 129 , 24 , 3 , 
   80 , 104 ) }{ 17 }{}
\mydictitem{7 }{ -258182304328415778 }{ 8557 }{ ( 
  52 , 16 , 38 , 192 , 8 , 130 , 
   132 , 1 ) }{ 15 }{}
\mydictitem{7 }{ -251249744669286557 }{ 9037 }{ ( 
  5 , 72 , 36 , 17 , 18 , 160 , 
   10 , 192 ) }{ 16 }{}
\mydictitem{7 }{ -249858540374917035 }{ 8861 }{ ( 
  160 , 8 , 10 , 16 , 96 , 4 , 
   130 , 1 ) }{ 12 }{}
\mydictitem{7 }{ -247427194046103816 }{ 9622 }{ ( 
  25 , 5 , 130 , 193 , 136 , 98 , 
   36 , 20 ) }{ 19 }{}
\mydictitem{7 }{ -247143106478950578 }{ 8621 }{ ( 
  5 , 66 , 32 , 9 , 136 , 128 , 
   66 , 17 ) }{ 14 }{}
\mydictitem{7 }{ -243566351808186474 }{ 8813 }{ ( 
  6 , 24 , 32 , 192 , 16 , 131 , 
   14 , 17 ) }{ 16 }{}
\mydictitem{7 }{ -242068486226572112 }{ 8798 }{ ( 
  5 , 80 , 130 , 192 , 56 , 32 , 
   10 , 9 ) }{ 16 }{}
\mydictitem{7 }{ -237212856195456665 }{ 9628 }{ ( 
  20 , 48 , 5 , 129 , 72 , 98 , 
   18 , 136 ) }{ 17 }{}
\mydictitem{7 }{ -227914280620238644 }{ 9577 }{ ( 
  17 , 96 , 6 , 11 , 136 , 132 , 
   68 , 48 ) }{ 17 }{}
\mydictitem{7 }{ -226039824958208568 }{ 9580 }{ ( 
  129 , 10 , 68 , 16 , 72 , 34 , 
   36 , 129 ) }{ 15 }{}
\mydictitem{7 }{ -223022686093957866 }{ 9578 }{ ( 
  20 , 64 , 3 , 144 , 33 , 36 , 
   10 , 136 ) }{ 15 }{}
\mydictitem{7 }{ -221577833850858925 }{ 9116 }{ ( 
  20 , 80 , 32 , 192 , 8 , 3 , 
   4 , 129 ) }{ 13 }{}
\mydictitem{7 }{ -217580664595077813 }{ 9562 }{ ( 
  128 , 72 , 96 , 10 , 18 , 20 , 
   36 , 1 ) }{ 14 }{}
\mydictitem{7 }{ -217251409270906367 }{ 9673 }{ ( 
  160 , 40 , 130 , 18 , 192 , 84 , 
   12 , 1 ) }{ 16 }{}
\mydictitem{7 }{ -213455902944141485 }{ 9158 }{ ( 
  20 , 64 , 160 , 140 , 1 , 34 , 
   132 , 25 ) }{ 16 }{}
\mydictitem{7 }{ -211010290780943168 }{ 9157 }{ ( 
  6 , 72 , 7 , 144 , 33 , 161 , 
   196 , 24 ) }{ 19 }{}
\mydictitem{7 }{ -206712752788130826 }{ 9052 }{ ( 
  20 , 20 , 32 , 129 , 8 , 67 , 
   18 , 192 ) }{ 15 }{}
\mydictitem{7 }{ -205559919351541300 }{ 9164 }{ ( 
  10 , 64 , 130 , 16 , 136 , 4 , 
   32 , 1 ) }{ 11 }{R5: $u_1 \xor u_3 = \textrm{True}$ }{}
\myappline
\mydictitem{8 }{ -899283463066040377 }{ 13107 }{ ( 
  8 , 64 , 32 , 16 , 136 , 0 , 
   2 , 1 ) }{ 8 }{R3: $u_1 \not\in C$}{}
\mydictitem{8 }{ -772207269962157112 }{ 8443 }{ ( 
  128 , 64 , 9 , 16 , 3 , 4 , 
   32 , 40 ) }{ 11 }{}
\mydictitem{8 }{ -721019390851251273 }{ 8383 }{ ( 
  130 , 64 , 32 , 13 , 129 , 6 , 
   144 , 16 ) }{ 14 }{}
\mydictitem{8 }{ -686263475047731580 }{ 8415 }{ ( 
  130 , 80 , 128 , 11 , 17 , 4 , 
   98 , 16 ) }{ 15 }{}
\mydictitem{8 }{ -673241419621910315 }{ 8319 }{ ( 
  3 , 8 , 12 , 16 , 200 , 36 , 
   160 , 129 ) }{ 15 }{}
\mydictitem{8 }{ -666032231320527476 }{ 8767 }{ ( 
  132 , 1 , 160 , 16 , 8 , 32 , 
   2 , 65 ) }{ 11 }{}
\mydictitem{8 }{ -661718979480894926 }{ 8511 }{ ( 
  132 , 5 , 32 , 81 , 8 , 144 , 
   2 , 144 ) }{ 14 }{}
\mydictitem{8 }{ -650672248853117917 }{ 8631 }{ ( 
  128 , 17 , 32 , 13 , 16 , 84 , 
   2 , 68 ) }{ 14 }{}
\mydictitem{8 }{ -642756855300046707 }{ 8571 }{ ( 
  128 , 237 , 38 , 120 , 80 , 67 , 
   152 , 33 ) }{ 24 }{}
\mydictitem{8 }{ -636308968528110613 }{ 8543 }{ ( 
  130 , 64 , 9 , 16 , 33 , 4 , 
   128 , 40 ) }{ 12 }{}
\mydictitem{8 }{ -613284019874591394 }{ 9311 }{ ( 
  128 , 33 , 65 , 68 , 10 , 6 , 
   24 , 48 ) }{ 15 }{}
\mydictitem{8 }{ -602996490440398480 }{ 8607 }{ ( 
  9 , 64 , 32 , 131 , 24 , 17 , 
   6 , 20 ) }{ 15 }{}
\mydictitem{8 }{ -583125401929517341 }{ 8863 }{ ( 
  42 , 16 , 9 , 17 , 136 , 4 , 
   130 , 208 ) }{ 16 }{}
\mydictitem{8 }{ -578102585724439884 }{ 8623 }{ ( 
  128 , 8 , 34 , 16 , 65 , 4 , 
   32 , 9 ) }{ 11 }{}
\mydictitem{8 }{ -566727496236131365 }{ 9339 }{ ( 
  11 , 12 , 160 , 192 , 33 , 68 , 
   18 , 48 ) }{ 17 }{}
\mydictitem{8 }{ -564929920852814672 }{ 8823 }{ ( 
  48 , 66 , 176 , 132 , 8 , 128 , 
   2 , 1 ) }{ 13 }{R3: $u_3 \not\in C$}{}
\mydictitem{8 }{ -560433615140304576 }{ 9623 }{ ( 
  18 , 5 , 32 , 138 , 72 , 69 , 
   144 , 20 ) }{ 17 }{}
\mydictitem{8 }{ -558283470099177545 }{ 10647 }{ ( 
  7 , 80 , 32 , 13 , 136 , 144 , 
   18 , 192 ) }{ 17 }{}
\mydictitem{8 }{ -553542914918976610 }{ 8559 }{ ( 
  137 , 64 , 128 , 24 , 131 , 4 , 
   34 , 48 ) }{ 15 }{}
\mydictitem{8 }{ -553440880392537039 }{ 8655 }{ ( 
  136 , 40 , 68 , 17 , 192 , 5 , 
   2 , 48 ) }{ 15 }{}
\mydictitem{8 }{ -550329284259922678 }{ 8815 }{ ( 
  5 , 64 , 32 , 129 , 16 , 148 , 
   2 , 12 ) }{ 13 }{}
\mydictitem{8 }{ -540872899944542692 }{ 8670 }{ ( 
  128 , 9 , 69 , 80 , 56 , 3 , 
   34 , 12 ) }{ 17 }{}
\mydictitem{8 }{ -537968997943824871 }{ 9039 }{ ( 
  48 , 18 , 192 , 129 , 65 , 4 , 
   34 , 9 ) }{ 15 }{}
\mydictitem{8 }{ -525172959641607591 }{ 8799 }{ ( 
  128 , 8 , 34 , 16 , 72 , 2 , 
   36 , 1 ) }{ 11 }{}
\mydictitem{8 }{ -521929389664079637 }{ 9677 }{ ( 
  36 , 65 , 6 , 16 , 8 , 130 , 
   160 , 65 ) }{ 14 }{}
\mydictitem{8 }{ -521206989824199314 }{ 9327 }{ ( 
  144 , 24 , 32 , 92 , 66 , 130 , 
   20 , 1 ) }{ 16 }{}
\mydictitem{8 }{ -520079249630644275 }{ 27030 }{ ( 
  41 , 22 , 193 , 194 , 161 , 82 , 
   28 , 44 ) }{ 24 }{}
\mydictitem{8 }{ -513543929637538094 }{ 8574 }{ ( 
  144 , 20 , 65 , 5 , 136 , 164 , 
   2 , 104 ) }{ 17 }{}
\mydictitem{8 }{ -510932219079709161 }{ 9435 }{ ( 
  89 , 42 , 150 , 66 , 114 , 194 , 
   14 , 193 ) }{ 26 }{}
\mydictitem{8 }{ -508220186566127504 }{ 8893 }{ ( 
  22 , 20 , 65 , 208 , 8 , 162 , 
   36 , 129 ) }{ 18 }{}
\mydictitem{8 }{ -503779283191348472 }{ 9118 }{ ( 
  128 , 48 , 34 , 67 , 24 , 48 , 
   76 , 5 ) }{ 17 }{}
\mydictitem{8 }{ -494343963011923576 }{ 8830 }{ ( 
  35 , 64 , 44 , 144 , 40 , 129 , 
   20 , 129 ) }{ 17 }{}
\mydictitem{8 }{ -493714312721210462 }{ 8637 }{ ( 
  28 , 64 , 160 , 129 , 138 , 33 , 
   10 , 144 ) }{ 17 }{}
\mydictitem{8 }{ -489334924266067936 }{ 8925 }{ ( 
  130 , 8 , 32 , 16 , 66 , 4 , 
   136 , 1 ) }{ 11 }{R3: $u_3 \not\in C$}{}
\mydictitem{8 }{ -489101200908942974 }{ 9629 }{ ( 
  130 , 48 , 96 , 12 , 66 , 20 , 
   136 , 1 ) }{ 15 }{}
\mydictitem{8 }{ -488708077465335931 }{ 10709 }{ ( 
  42 , 64 , 12 , 20 , 138 , 34 , 
   144 , 1 ) }{ 16 }{}
\mydictitem{8 }{ -482042543996118538 }{ 8829 }{ ( 
  3 , 3 , 128 , 16 , 136 , 68 , 
   44 , 192 ) }{ 15 }{}
\mydictitem{8 }{ -480275171830561668 }{ 9581 }{ ( 
  6 , 5 , 32 , 208 , 8 , 130 , 
   20 , 65 ) }{ 15 }{}
\mydictitem{8 }{ -479323869615069659 }{ 10710 }{ ( 
  57 , 6 , 201 , 71 , 161 , 82 , 
   92 , 172 ) }{ 28 }{}
\mydictitem{8 }{ -477519058171793838 }{ 9566 }{ ( 
  48 , 65 , 5 , 192 , 9 , 36 , 
   2 , 152 ) }{ 16 }{}
\mydictitem{8 }{ -473957427851750723 }{ 8685 }{ ( 
  138 , 64 , 34 , 137 , 48 , 4 , 
   144 , 129 ) }{ 16 }{}
\mydictitem{8 }{ -471341320523760961 }{ 9579 }{ ( 
  11 , 96 , 36 , 12 , 3 , 130 , 
   208 , 48 ) }{ 18 }{}
\mydictitem{8 }{ -464534444202425179 }{ 9630 }{ ( 
  17 , 48 , 192 , 137 , 129 , 6 , 
   6 , 104 ) }{ 18 }{}
\mydictitem{8 }{ -463299397530769003 }{ 9053 }{ ( 
  132 , 20 , 32 , 82 , 1 , 130 , 
   80 , 9 ) }{ 15 }{}
\mydictitem{8 }{ -457924647789240028 }{ 10653 }{ ( 
  41 , 64 , 129 , 18 , 129 , 4 , 
   18 , 40 ) }{ 15 }{}
\mydictitem{8 }{ -452601408428460382 }{ 9342 }{ ( 
  51 , 64 , 10 , 144 , 161 , 4 , 
   160 , 136 ) }{ 17 }{}
\mydictitem{8 }{ -442700287585267532 }{ 9117 }{ ( 
  36 , 64 , 129 , 16 , 8 , 128 , 
   2 , 33 ) }{ 11 }{}
\mydictitem{8 }{ -441281571114611767 }{ 9706 }{ ( 
  5 , 64 , 36 , 129 , 8 , 50 , 
   6 , 144 ) }{ 15 }{}
\mydictitem{8 }{ -431649046291600881 }{ 9594 }{ ( 
  128 , 72 , 3 , 10 , 48 , 4 , 
   80 , 33 ) }{ 14 }{}
\mydictitem{8 }{ -428443387129934507 }{ 9159 }{ ( 
  144 , 20 , 32 , 136 , 65 , 6 , 
   66 , 17 ) }{ 15 }{}
\mydictitem{8 }{ -425985466039421809 }{ 9133 }{ ( 
  176 , 33 , 13 , 144 , 10 , 160 , 
   34 , 69 ) }{ 19 }{}
\mydictitem{8 }{ -424439857990218041 }{ 9675 }{ ( 
  25 , 80 , 34 , 129 , 136 , 162 , 
   6 , 68 ) }{ 18 }{}
\mydictitem{8 }{ -421336022871598134 }{ 9654 }{ ( 
  12 , 72 , 130 , 80 , 129 , 38 , 
   160 , 17 ) }{ 17 }{}
\mydictitem{8 }{ -417702864512524473 }{ 9054 }{ ( 
  128 , 3 , 65 , 25 , 16 , 36 , 
   34 , 76 ) }{ 16 }{}
\mydictitem{8 }{ -413621507243457915 }{ 9708 }{ ( 
  130 , 18 , 68 , 24 , 72 , 164 , 
   36 , 1 ) }{ 16 }{}
\mydictitem{8 }{ -410176819348621982 }{ 9166 }{ ( 
  36 , 10 , 136 , 144 , 96 , 20 , 
   66 , 1 ) }{ 15 }{}
\mydictitem{8 }{ -409845545228962591 }{ 9690 }{ ( 
  25 , 10 , 132 , 20 , 96 , 160 , 
   66 , 5 ) }{ 17 }{}
\mydictitem{8 }{ -405077671322438068 }{ 9582 }{ ( 
  129 , 24 , 130 , 80 , 65 , 36 , 
   34 , 12 ) }{ 16 }{}
\mydictitem{8 }{ -398409201621308822 }{ 10104 }{ ( 
  160 , 48 , 68 , 17 , 10 , 130 , 
   12 , 65 ) }{ 16 }{}
\mydictitem{8 }{ -393313530918741654 }{ 11730 }{ ( 
  162 , 12 , 36 , 3 , 200 , 81 , 
   14 , 144 ) }{ 20 }{}
\mydictitem{8 }{ -393246810197863372 }{ 9174 }{ ( 
  37 , 64 , 128 , 19 , 152 , 48 , 
   130 , 12 ) }{ 17 }{}
\mydictitem{8 }{ -387574329040405188 }{ 9084 }{ ( 
  17 , 64 , 40 , 18 , 33 , 132 , 
   130 , 12 ) }{ 15 }{}
\mydictitem{8 }{ -382828422517523851 }{ 9165 }{ ( 
  20 , 20 , 192 , 129 , 8 , 34 , 
   6 , 65 ) }{ 15 }{}
\mydictitem{8 }{ -380718107299096901 }{ 9705 }{ ( 
  12 , 64 , 34 , 129 , 136 , 49 , 
   6 , 144 ) }{ 16 }{}
\mydictitem{8 }{ -376733361301144475 }{ 15555 }{ ( 
  35 , 80 , 67 , 224 , 92 , 102 , 
   195 , 112 ) }{ 26 }{R3: $u_1 \not\in C$}{}
\mydictitem{8 }{ -370347967947654772 }{ 9180 }{ ( 
  36 , 64 , 128 , 16 , 3 , 6 , 
   168 , 9 ) }{ 14 }{}
\mydictitem{8 }{ -369336898403981540 }{ 9189 }{ ( 
  133 , 10 , 160 , 72 , 24 , 6 , 
   96 , 33 ) }{ 17 }{}
\mydictitem{8 }{ -369049161187396959 }{ 9660 }{ ( 
  19 , 40 , 68 , 17 , 9 , 132 , 
   6 , 224 ) }{ 18 }{}
\mydictitem{8 }{ -347268859906627992 }{ 11220 }{ ( 
  160 , 76 , 132 , 66 , 24 , 34 , 
   20 , 1 ) }{ 16 }{}
\mydictitem{8 }{ -344304718756045964 }{ 10200 }{ ( 
  138 , 24 , 69 , 14 , 67 , 193 , 
   100 , 60 ) }{ 24 }{}
\mydictitem{8 }{ -318880083885350459 }{ 10212 }{ ( 
  6 , 73 , 136 , 131 , 96 , 20 , 
   162 , 81 ) }{ 20 }{}
\mydictitem{8 }{ -284276962660566950 }{ 13260 }{ ( 
  32 , 64 , 130 , 16 , 8 , 4 , 
   130 , 1 ) }{ 10 }{R3: $u_1 \not\in C$}{}
\myappline
\mydictitem{9 }{ -489610810524822682 }{ 56524 }{ ( 
  3 , 1 , 160 , 16 , 40 , 4 , 
   8 , 192 ) }{ 12 }{}
\mydictitem{9 }{ -452337729517226882 }{ 57156 }{ ( 
  12 , 72 , 128 , 16 , 129 , 164 , 
   2 , 65 ) }{ 14 }{}
\mydictitem{9 }{ -408354319245236969 }{ 57160 }{ ( 
  160 , 8 , 130 , 16 , 74 , 4 , 
   40 , 1 ) }{ 13 }{}
\mydictitem{9 }{ -400582064369183280 }{ 57220 }{ ( 
  129 , 8 , 33 , 21 , 72 , 34 , 
   18 , 148 ) }{ 17 }{}
\mydictitem{9 }{ -393459705592617740 }{ 57280 }{ ( 
  50 , 64 , 128 , 144 , 136 , 4 , 
   40 , 1 ) }{ 13 }{}
\mydictitem{9 }{ -384942911261086050 }{ 56784 }{ ( 
  12 , 64 , 34 , 1 , 136 , 128 , 
   32 , 17 ) }{ 12 }{}
\mydictitem{9 }{ -378955175228864660 }{ 57056 }{ ( 
  42 , 64 , 136 , 144 , 56 , 4 , 
   128 , 1 ) }{ 14 }{}
\mydictitem{9 }{ -372730902319860600 }{ 57248 }{ ( 
  41 , 64 , 128 , 131 , 24 , 4 , 
   18 , 16 ) }{ 14 }{}
\mydictitem{9 }{ -365307444265761614 }{ 57184 }{ ( 
  21 , 66 , 2 , 193 , 8 , 144 , 
   38 , 20 ) }{ 17 }{}
\mydictitem{9 }{ -350325235410304638 }{ 56936 }{ ( 
  19 , 80 , 36 , 129 , 8 , 129 , 
   34 , 196 ) }{ 17 }{}
\mydictitem{9 }{ -346991515368820537 }{ 56240 }{ ( 
  22 , 64 , 32 , 9 , 152 , 5 , 
   136 , 129 ) }{ 16 }{}
\mydictitem{9 }{ -345602958711606536 }{ 57040 }{ ( 
  130 , 64 , 9 , 136 , 48 , 4 , 
   144 , 33 ) }{ 14 }{}
\mydictitem{9 }{ -342028055323505620 }{ 57232 }{ ( 
  144 , 96 , 65 , 136 , 9 , 6 , 
   4 , 49 ) }{ 16 }{}
\mydictitem{9 }{ -334343815080691876 }{ 57124 }{ ( 
  6 , 3 , 32 , 80 , 136 , 192 , 
   12 , 17 ) }{ 15 }{}
\mydictitem{9 }{ -331245387788784281 }{ 57008 }{ ( 
  12 , 48 , 96 , 11 , 17 , 132 , 
   18 , 193 ) }{ 18 }{}
\mydictitem{9 }{ -329455080315344929 }{ 56944 }{ ( 
  20 , 80 , 32 , 131 , 8 , 18 , 
   134 , 192 ) }{ 16 }{}
\mydictitem{9 }{ -326446789752148616 }{ 57217 }{ ( 
  131 , 64 , 34 , 24 , 9 , 160 , 
   132 , 20 ) }{ 16 }{}
\mydictitem{9 }{ -309905349059268324 }{ 56996 }{ ( 
  42 , 9 , 134 , 16 , 73 , 160 , 
   36 , 193 ) }{ 19 }{}
\mydictitem{9 }{ -305364326870046401 }{ 57025 }{ ( 
  136 , 72 , 32 , 133 , 18 , 17 , 
   66 , 68 ) }{ 16 }{}
\mydictitem{9 }{ -303440793052772106 }{ 56752 }{ ( 
  5 , 18 , 32 , 88 , 17 , 6 , 
   192 , 136 ) }{ 16 }{}
\mydictitem{9 }{ -303164240726434633 }{ 56920 }{ ( 
  21 , 80 , 32 , 7 , 8 , 82 , 
   132 , 144 ) }{ 17 }{}
\mydictitem{9 }{ -294782498316163996 }{ 57026 }{ ( 
  17 , 12 , 32 , 192 , 9 , 68 , 
   2 , 145 ) }{ 15 }{}
\mydictitem{9 }{ -293737612421108058 }{ 56881 }{ ( 
  6 , 35 , 68 , 144 , 24 , 192 , 
   14 , 65 ) }{ 18 }{}
\mydictitem{9 }{ -291593988308626710 }{ 56929 }{ ( 
  129 , 13 , 3 , 84 , 40 , 208 , 
   66 , 48 ) }{ 19 }{}
\mydictitem{9 }{ -286206440273200448 }{ 56744 }{ ( 
  48 , 64 , 130 , 12 , 144 , 132 , 
   34 , 1 ) }{ 14 }{}
\mydictitem{9 }{ -282360887598956176 }{ 56212 }{ ( 
  3 , 12 , 191 , 88 , 210 , 41 , 
   83 , 186 ) }{ 30 }{}
\mydictitem{9 }{ -281531220380667350 }{ 56882 }{ ( 
  42 , 64 , 4 , 129 , 136 , 164 , 
   18 , 17 ) }{ 16 }{}
\mydictitem{9 }{ -280171463154608827 }{ 56210 }{ ( 
  17 , 66 , 32 , 152 , 17 , 4 , 
   66 , 136 ) }{ 15 }{}
\mydictitem{9 }{ -280099211892526104 }{ 54889 }{ ( 
  41 , 22 , 201 , 194 , 161 , 82 , 
   30 , 44 ) }{ 26 }{}
\mydictitem{9 }{ -276166021834513026 }{ 56770 }{ ( 
  132 , 80 , 34 , 66 , 40 , 20 , 
   136 , 1 ) }{ 15 }{}
\mydictitem{9 }{ -276136614689952116 }{ 56225 }{ ( 
  48 , 5 , 65 , 144 , 9 , 38 , 
   6 , 200 ) }{ 18 }{}
\mydictitem{9 }{ -275318071137808221 }{ 56930 }{ ( 
  17 , 72 , 160 , 13 , 80 , 132 , 
   2 , 176 ) }{ 17 }{}
\mydictitem{9 }{ -270984298897396969 }{ 56228 }{ ( 
  176 , 68 , 3 , 21 , 8 , 162 , 
   192 , 21 ) }{ 19 }{}
\mydictitem{9 }{ -268713156635853130 }{ 56993 }{ ( 
  25 , 68 , 36 , 13 , 129 , 98 , 
   130 , 144 ) }{ 19 }{}
\mydictitem{9 }{ -263998281304105199 }{ 56738 }{ ( 
  17 , 64 , 4 , 144 , 8 , 34 , 
   6 , 145 ) }{ 14 }{}
\mydictitem{9 }{ -261677808517195389 }{ 54890 }{ ( 
  21 , 18 , 168 , 194 , 161 , 6 , 
   104 , 24 ) }{ 21 }{}
\mydictitem{9 }{ -259011690208036504 }{ 56977 }{ ( 
  5 , 68 , 6 , 26 , 24 , 112 , 
   160 , 129 ) }{ 18 }{}
\mydictitem{9 }{ -258182304328415778 }{ 56978 }{ ( 
  11 , 66 , 32 , 136 , 17 , 20 , 
   68 , 145 ) }{ 17 }{}
\mydictitem{9 }{ -251249744669286557 }{ 56498 }{ ( 
  132 , 33 , 96 , 10 , 24 , 129 , 
   18 , 68 ) }{ 16 }{}
\mydictitem{9 }{ -249858540374917035 }{ 56674 }{ ( 
  129 , 64 , 33 , 136 , 49 , 4 , 
   2 , 24 ) }{ 14 }{}
\mydictitem{9 }{ -247427194046103816 }{ 55913 }{ ( 
  160 , 42 , 12 , 82 , 17 , 70 , 
   200 , 129 ) }{ 20 }{}
\mydictitem{9 }{ -247143106478950578 }{ 56914 }{ ( 
  160 , 64 , 13 , 24 , 129 , 36 , 
   2 , 144 ) }{ 15 }{}
\mydictitem{9 }{ -243566351808186474 }{ 56722 }{ ( 
  17 , 40 , 192 , 192 , 24 , 34 , 
   6 , 5 ) }{ 16 }{}
\mydictitem{9 }{ -242068486226572112 }{ 56737 }{ ( 
  36 , 80 , 130 , 10 , 72 , 132 , 
   48 , 1 ) }{ 15 }{}
\mydictitem{9 }{ -237212856195456665 }{ 55907 }{ ( 
  9 , 74 , 96 , 194 , 161 , 5 , 
   18 , 20 ) }{ 19 }{}
\mydictitem{9 }{ -227914280620238644 }{ 55958 }{ ( 
  5 , 18 , 32 , 137 , 24 , 6 , 
   80 , 192 ) }{ 16 }{}
\mydictitem{9 }{ -226039824958208568 }{ 55955 }{ ( 
  5 , 18 , 36 , 17 , 8 , 34 , 
   192 , 192 ) }{ 15 }{}
\mydictitem{9 }{ -223022686093957866 }{ 55957 }{ ( 
  136 , 33 , 65 , 84 , 9 , 52 , 
   2 , 148 ) }{ 18 }{}
\mydictitem{9 }{ -221577833850858925 }{ 56419 }{ ( 
  13 , 65 , 37 , 136 , 144 , 48 , 
   2 , 192 ) }{ 17 }{}
\mydictitem{9 }{ -217580664595077813 }{ 55973 }{ ( 
  130 , 18 , 96 , 137 , 24 , 4 , 
   48 , 65 ) }{ 16 }{}
\mydictitem{9 }{ -217251409270906367 }{ 55862 }{ ( 
  25 , 40 , 134 , 18 , 66 , 100 , 
   164 , 129 ) }{ 20 }{}
\mydictitem{9 }{ -213455902944141485 }{ 56377 }{ ( 
  20 , 64 , 32 , 140 , 17 , 130 , 
   6 , 9 ) }{ 15 }{}
\mydictitem{9 }{ -211010290780943168 }{ 56378 }{ ( 
  5 , 64 , 36 , 136 , 9 , 146 , 
   34 , 17 ) }{ 16 }{}
\mydictitem{9 }{ -206712752788130826 }{ 56483 }{ ( 
  20 , 33 , 34 , 144 , 72 , 5 , 
   66 , 137 ) }{ 17 }{}
\mydictitem{9 }{ -205559919351541300 }{ 56371 }{ ( 
  128 , 1 , 33 , 16 , 40 , 4 , 
   2 , 72 ) }{ 11 }{}
\myappline
\mydictitem{10 }{ -264953748043115284 }{ 56780 }{ ( 
  128 , 64 , 33 , 16 , 8 , 32 , 
   2 , 5 ) }{ 10 }{R3: $u_3 \not\in C$}{}
\mydictitem{10 }{ -255903677087804212 }{ 57164 }{ ( 
  3 , 64 , 32 , 28 , 16 , 6 , 
   144 , 129 ) }{ 14 }{}
\mydictitem{10 }{ -246423116736164502 }{ 57036 }{ ( 
  6 , 10 , 32 , 17 , 192 , 132 , 
   76 , 16 ) }{ 15 }{}
\mydictitem{10 }{ -234055909760095663 }{ 57284 }{ ( 
  137 , 64 , 32 , 140 , 3 , 6 , 
   144 , 144 ) }{ 16 }{}
\mydictitem{10 }{ -221848639666228098 }{ 57288 }{ ( 
  50 , 64 , 132 , 144 , 8 , 33 , 
   160 , 5 ) }{ 15 }{}
\mydictitem{10 }{ -219506781223121329 }{ 57161 }{ ( 
  128 , 35 , 6 , 80 , 8 , 49 , 
   70 , 68 ) }{ 17 }{}
\mydictitem{10 }{ -211897911062832022 }{ 56526 }{ ( 
  3 , 65 , 34 , 20 , 40 , 132 , 
   24 , 192 ) }{ 16 }{}
\mydictitem{10 }{ -198829802885264160 }{ 57296 }{ ( 
  128 , 16 , 33 , 80 , 41 , 4 , 
   2 , 40 ) }{ 13 }{}
\mydictitem{10 }{ -196931604634552768 }{ 57312 }{ ( 
  18 , 64 , 32 , 138 , 144 , 148 , 
   6 , 1 ) }{ 15 }{}
\mydictitem{10 }{ -193105575805408406 }{ 57221 }{ ( 
  3 , 64 , 130 , 144 , 8 , 36 , 
   20 , 49 ) }{ 15 }{}
\mydictitem{10 }{ -186798665756457220 }{ 57264 }{ ( 
  12 , 8 , 96 , 144 , 19 , 130 , 
   160 , 65 ) }{ 16 }{}
\mydictitem{10 }{ -186227373079063234 }{ 56242 }{ ( 
  136 , 7 , 133 , 74 , 48 , 50 , 
   164 , 65 ) }{ 21 }{}
\mydictitem{10 }{ -184055504850097333 }{ 57236 }{ ( 
  56 , 64 , 3 , 152 , 129 , 4 , 
   34 , 145 ) }{ 17 }{}
\mydictitem{10 }{ -180868489214616524 }{ 57281 }{ ( 
  128 , 17 , 32 , 22 , 8 , 82 , 
   20 , 65 ) }{ 15 }{}
\mydictitem{10 }{ -176493283894497754 }{ 57192 }{ ( 
  3 , 88 , 33 , 140 , 81 , 68 , 
   130 , 48 ) }{ 19 }{}
\mydictitem{10 }{ -174574944498457624 }{ 57188 }{ ( 
  6 , 18 , 160 , 192 , 40 , 5 , 
   88 , 33 ) }{ 17 }{}
\mydictitem{10 }{ -164368307347587140 }{ 57058 }{ ( 
  12 , 72 , 33 , 17 , 144 , 38 , 
   130 , 208 ) }{ 18 }{}
\mydictitem{10 }{ -164359755773378736 }{ 57252 }{ ( 
  20 , 68 , 32 , 7 , 24 , 146 , 
   192 , 9 ) }{ 17 }{}
\mydictitem{10 }{ -159842111719858328 }{ 57282 }{ ( 
  137 , 64 , 34 , 3 , 144 , 4 , 
   48 , 136 ) }{ 15 }{}
\mydictitem{10 }{ -157236364066051777 }{ 56952 }{ ( 
  22 , 64 , 32 , 133 , 8 , 19 , 
   132 , 145 ) }{ 17 }{}
\mydictitem{10 }{ -155304537441776442 }{ 57249 }{ ( 
  129 , 33 , 3 , 16 , 136 , 68 , 
   6 , 104 ) }{ 16 }{}
\mydictitem{10 }{ -153275171299567776 }{ 57010 }{ ( 
  56 , 68 , 65 , 18 , 129 , 36 , 
   130 , 137 ) }{ 18 }{}
\mydictitem{10 }{ -153237592690580239 }{ 57041 }{ ( 
  48 , 64 , 3 , 144 , 40 , 129 , 
   14 , 133 ) }{ 17 }{}
\mydictitem{10 }{ -150450270880229517 }{ 57234 }{ ( 
  165 , 64 , 44 , 5 , 136 , 144 , 
   2 , 48 ) }{ 17 }{}
\mydictitem{10 }{ -147186103159461859 }{ 56883 }{ ( 
  130 , 68 , 2 , 18 , 136 , 20 , 
   104 , 1 ) }{ 15 }{}
\mydictitem{10 }{ -144210284492927278 }{ 56937 }{ ( 
  13 , 74 , 40 , 133 , 98 , 193 , 
   18 , 148 ) }{ 22 }{}
\mydictitem{10 }{ -142969789437608309 }{ 57057 }{ ( 
  11 , 65 , 33 , 136 , 146 , 20 , 
   152 , 100 ) }{ 20 }{}
\mydictitem{10 }{ -142370542521970898 }{ 57012 }{ ( 
  44 , 48 , 6 , 3 , 24 , 81 , 
   132 , 192 ) }{ 18 }{}
\mydictitem{10 }{ -141698008727709048 }{ 56746 }{ ( 
  128 , 64 , 32 , 8 , 17 , 4 , 
   2 , 9 ) }{ 10 }{R3: $u_3 \not\in C$}{}
\mydictitem{10 }{ -138524570147284945 }{ 56786 }{ ( 
  3 , 64 , 12 , 16 , 33 , 36 , 
   130 , 136 ) }{ 14 }{}
\mydictitem{10 }{ -138108485317911594 }{ 56760 }{ ( 
  128 , 10 , 68 , 16 , 72 , 36 , 
   34 , 1 ) }{ 13 }{}
\mydictitem{10 }{ -135427674056010516 }{ 56946 }{ ( 
  42 , 64 , 6 , 16 , 9 , 37 , 
   132 , 129 ) }{ 16 }{}
\mydictitem{10 }{ -133261805405082912 }{ 57042 }{ ( 
  3 , 33 , 34 , 68 , 8 , 20 , 
   192 , 144 ) }{ 15 }{}
\mydictitem{10 }{ -132786334798600321 }{ 57009 }{ ( 
  129 , 20 , 33 , 196 , 9 , 112 , 
   2 , 28 ) }{ 18 }{}
\mydictitem{10 }{ -130731773807879766 }{ 56754 }{ ( 
  3 , 65 , 164 , 16 , 192 , 34 , 
   12 , 72 ) }{ 16 }{}
\mydictitem{10 }{ -130024582675614020 }{ 56244 }{ ( 
  237 , 251 , 35 , 164 , 47 , 
   241 , 85 , 234 ) }{ 38 }{}
\mydictitem{10 }{ -125686589014661632 }{ 54891 }{ ( 
  41 , 22 , 201 , 198 , 161 , 82 , 
   30 , 45 ) }{ 28 }{}
\mydictitem{10 }{ -125660450966743013 }{ 56678 }{ ( 
  6 , 64 , 32 , 16 , 136 , 3 , 
   140 , 129 ) }{ 14 }{R3: $u_3 \not\in C$}{}
\mydictitem{10 }{ -124616208289630152 }{ 56931 }{ ( 
  13 , 80 , 35 , 136 , 98 , 132 , 
   56 , 145 ) }{ 21 }{}
\mydictitem{10 }{ -123666407520778105 }{ 57027 }{ ( 
  45 , 64 , 6 , 16 , 9 , 160 , 
   130 , 129 ) }{ 16 }{}
\mydictitem{10 }{ -122864364540640491 }{ 56922 }{ ( 
  153 , 64 , 32 , 9 , 136 , 4 , 
   2 , 24 ) }{ 14 }{}
\mydictitem{10 }{ -120544022323974311 }{ 56997 }{ ( 
  25 , 65 , 160 , 67 , 136 , 4 , 
   130 , 48 ) }{ 17 }{}
\mydictitem{10 }{ -119127428200296783 }{ 56745 }{ ( 
  48 , 64 , 11 , 24 , 129 , 4 , 
   130 , 49 ) }{ 16 }{}
\mydictitem{10 }{ -118936297852452994 }{ 56214 }{ ( 
  21 , 65 , 32 , 66 , 136 , 145 , 
   134 , 28 ) }{ 19 }{}
\mydictitem{10 }{ -113571267670065141 }{ 56739 }{ ( 
  20 , 72 , 36 , 194 , 9 , 160 , 
   6 , 17 ) }{ 17 }{}
\mydictitem{10 }{ -112145695798823883 }{ 56979 }{ ( 
  25 , 33 , 194 , 20 , 136 , 164 , 
   34 , 68 ) }{ 19 }{}
\mydictitem{10 }{ -110328833598895423 }{ 56229 }{ ( 
  21 , 80 , 160 , 133 , 24 , 131 , 
   6 , 104 ) }{ 20 }{}
\mydictitem{10 }{ -109077852951287423 }{ 56499 }{ ( 
  17 , 10 , 32 , 84 , 9 , 6 , 
   192 , 144 ) }{ 16 }{}
\mydictitem{10 }{ -101641038122898375 }{ 55959 }{ ( 
  14 , 232 , 19 , 24 , 162 , 141 , 
   48 , 71 ) }{ 25 }{}
\mydictitem{10 }{ -101300461292717132 }{ 56379 }{ ( 
  33 , 64 , 36 , 17 , 24 , 6 , 
   130 , 136 ) }{ 15 }{}
\myappline
\mydictitem{11 }{ -137319486169585905 }{ 57165 }{ ( 
  13 , 64 , 32 , 21 , 152 , 144 , 
   2 , 136 ) }{ 16 }{}
\mydictitem{11 }{ -131925251824472796 }{ 57292 }{ ( 
  33 , 64 , 160 , 16 , 10 , 4 , 
   8 , 129 ) }{ 12 }{}
\mydictitem{11 }{ -106990011212970741 }{ 57300 }{ ( 
  168 , 64 , 130 , 26 , 144 , 4 , 
   48 , 1 ) }{ 15 }{}
\mydictitem{11 }{ -106200686484487583 }{ 57068 }{ ( 
  14 , 72 , 36 , 144 , 138 , 48 , 
   192 , 1 ) }{ 17 }{}
\mydictitem{11 }{ -100711446287411196 }{ 57052 }{ ( 
  128 , 12 , 32 , 24 , 17 , 69 , 
   2 , 68 ) }{ 14 }{}
\mydictitem{11 }{ -97383480909282057 }{ 57328 }{ ( 
  130 , 16 , 32 , 196 , 8 , 5 , 
   18 , 17 ) }{ 14 }{}
\mydictitem{11 }{ -93806950451816800 }{ 57320 }{ ( 
  18 , 80 , 32 , 135 , 8 , 69 , 
   130 , 20 ) }{ 17 }{}
\mydictitem{11 }{ -87252051588588716 }{ 57304 }{ ( 
  13 , 65 , 32 , 20 , 136 , 144 , 
   2 , 80 ) }{ 15 }{}
\mydictitem{11 }{ -84213109538527750 }{ 57316 }{ ( 
  128 , 12 , 3 , 20 , 9 , 112 , 
   34 , 69 ) }{ 17 }{}
\mydictitem{11 }{ -81265445872985527 }{ 57196 }{ ( 
  28 , 64 , 160 , 136 , 50 , 3 , 
   56 , 5 ) }{ 18 }{}
\mydictitem{11 }{ -79740217163681572 }{ 57266 }{ ( 
  132 , 96 , 67 , 16 , 10 , 160 , 
   40 , 5 ) }{ 16 }{}
\mydictitem{11 }{ -72810579632836741 }{ 57314 }{ ( 
  164 , 33 , 136 , 16 , 73 , 5 , 
   2 , 224 ) }{ 17 }{}
\mydictitem{11 }{ -69543625692646563 }{ 57313 }{ ( 
  35 , 64 , 132 , 20 , 9 , 161 , 
   18 , 140 ) }{ 18 }{}
\mydictitem{11 }{ -69010625791285801 }{ 57193 }{ ( 
  133 , 42 , 137 , 16 , 35 , 6 , 
   104 , 192 ) }{ 20 }{}
\mydictitem{11 }{ -67084531825011153 }{ 57268 }{ ( 
  137 , 72 , 98 , 129 , 19 , 4 , 
   34 , 152 ) }{ 19 }{}
\mydictitem{11 }{ -65196296191649679 }{ 57283 }{ ( 
  129 , 64 , 32 , 22 , 8 , 130 , 
   20 , 17 ) }{ 14 }{}
\mydictitem{11 }{ -64617744063353716 }{ 57298 }{ ( 
  6 , 64 , 32 , 129 , 8 , 132 , 
   22 , 17 ) }{ 14 }{}
\mydictitem{11 }{ -62513764335079739 }{ 57238 }{ ( 
  132 , 5 , 33 , 19 , 72 , 112 , 
   6 , 152 ) }{ 19 }{}
\mydictitem{11 }{ -60805966899451609 }{ 57253 }{ ( 
  3 , 64 , 130 , 144 , 136 , 4 , 
   48 , 41 ) }{ 15 }{}
\mydictitem{11 }{ -58556655299675617 }{ 56954 }{ ( 
  19 , 66 , 32 , 136 , 137 , 132 , 
   84 , 25 ) }{ 19 }{}
\mydictitem{11 }{ -57613278077157236 }{ 56953 }{ ( 
  162 , 50 , 137 , 24 , 194 , 36 , 
   84 , 65 ) }{ 21 }{}
\mydictitem{11 }{ -54873765588920526 }{ 57011 }{ ( 
  129 , 5 , 160 , 84 , 9 , 98 , 
   18 , 28 ) }{ 19 }{}
\mydictitem{11 }{ -53148596289303402 }{ 56762 }{ ( 
  5 , 66 , 32 , 144 , 8 , 192 , 
   6 , 21 ) }{ 15 }{}
\mydictitem{11 }{ -51297235131455269 }{ 57059 }{ ( 
  44 , 96 , 132 , 19 , 136 , 81 , 
   130 , 21 ) }{ 20 }{}
\mydictitem{11 }{ -49945641870770219 }{ 56246 }{ ( 
  7 , 88 , 40 , 74 , 161 , 193 , 
   22 , 148 ) }{ 23 }{}
\mydictitem{11 }{ -49144516830228831 }{ 57043 }{ ( 
  12 , 66 , 192 , 16 , 40 , 133 , 
   34 , 33 ) }{ 16 }{}
\mydictitem{11 }{ -48332850287105365 }{ 57014 }{ ( 
  128 , 166 , 172 , 72 , 107 , 
   50 , 85 , 113 ) }{ 27 }{}
\myappline
\mydictitem{12 }{ -64417460349479835 }{ 56797 }{ ( 
  128 , 64 , 3 , 1 , 8 , 4 , 
   32 , 48 ) }{ 10 }{R3: $u_3 \not\in C$}{}
\mydictitem{12 }{ -55489630922495513 }{ 57293 }{ ( 
  3 , 72 , 32 , 144 , 80 , 4 , 
   138 , 17 ) }{ 15 }{}
\mydictitem{12 }{ -45874752110564406 }{ 57308 }{ ( 
  128 , 33 , 40 , 80 , 72 , 4 , 
   2 , 17 ) }{ 13 }{}
\mydictitem{12 }{ -37719695731007326 }{ 57324 }{ ( 
  132 , 3 , 36 , 20 , 8 , 208 , 
   34 , 65 ) }{ 16 }{}
\mydictitem{12 }{ -32798143896572570 }{ 57332 }{ ( 
  25 , 64 , 32 , 129 , 152 , 4 , 
   2 , 137 ) }{ 15 }{}
\mydictitem{12 }{ -30712541418824951 }{ 57336 }{ ( 
  41 , 64 , 130 , 144 , 56 , 4 , 
   10 , 49 ) }{ 17 }{}
\mydictitem{12 }{ -28706142641542548 }{ 57197 }{ ( 
  136 , 42 , 194 , 80 , 50 , 4 , 
   224 , 1 ) }{ 18 }{}
\mydictitem{12 }{ -28647676317807325 }{ 57054 }{ ( 
  128 , 64 , 9 , 10 , 48 , 4 , 
   18 , 33 ) }{ 13 }{R3: $u_4 \not\in C$}{}
\mydictitem{12 }{ -26253185098326629 }{ 57330 }{ ( 
  5 , 64 , 33 , 152 , 9 , 134 , 
   130 , 48 ) }{ 17 }{}
\mydictitem{12 }{ -23870314469588248 }{ 57317 }{ ( 
  3 , 68 , 36 , 20 , 8 , 146 , 
   192 , 161 ) }{ 17 }{}
\mydictitem{12 }{ -21717939174696339 }{ 57321 }{ ( 
  130 , 48 , 140 , 26 , 98 , 100 , 
   76 , 1 ) }{ 20 }{}
\mydictitem{12 }{ -21459144106706130 }{ 57306 }{ ( 
  128 , 40 , 65 , 16 , 72 , 4 , 
   2 , 33 ) }{ 12 }{}
\mydictitem{12 }{ -19374761209641485 }{ 57318 }{ ( 
  29 , 64 , 40 , 137 , 35 , 132 , 
   18 , 144 ) }{ 19 }{}
\mydictitem{12 }{ -17236351616013344 }{ 57315 }{ ( 
  28 , 64 , 32 , 136 , 26 , 5 , 
   130 , 129 ) }{ 16 }{}
\mydictitem{12 }{ -15085506227251376 }{ 56955 }{ ( 
  14 , 80 , 160 , 133 , 40 , 67 , 
   146 , 21 ) }{ 21 }{}
\mydictitem{12 }{ -14842292216336441 }{ 56763 }{ ( 
  3 , 65 , 160 , 80 , 8 , 4 , 
   34 , 144 ) }{ 14 }{R3: $u_3 \not\in C$}{}
\mydictitem{12 }{ -13941487138238279 }{ 57270 }{ ( 
  131 , 100 , 134 , 24 , 24 , 98 , 
   196 , 33 ) }{ 21 }{}
\mydictitem{12 }{ -9019935303803522 }{ 57015 }{ ( 
  42 , 25 , 134 , 69 , 200 , 52 , 
   162 , 81 ) }{ 24 }{}
\myappline
\mydictitem{13 }{ -15162648248230256 }{ 57309 }{ ( 
  128 , 72 , 32 , 24 , 17 , 4 , 
   2 , 65 ) }{ 12 }{}
\mydictitem{13 }{ -5138871171782917 }{ 57340 }{ ( 
  22 , 68 , 32 , 144 , 8 , 134 , 
   192 , 1 ) }{ 15 }{}
\mydictitem{13 }{ -3841648050655253 }{ 57325 }{ ( 
  14 , 80 , 32 , 73 , 137 , 20 , 
   146 , 129 ) }{ 19 }{}
\mydictitem{13 }{ -1235584894668443 }{ 57338 }{ ( 
  128 , 5 , 33 , 17 , 24 , 68 , 
   2 , 104 ) }{ 15 }{}
\mydictitem{13 }{ 202153147298516 }{ 57337 }{ ( 
  12 , 70 , 32 , 193 , 25 , 133 , 
   18 , 208 ) }{ 20 }{}
\mydictitem{13 }{ 2399162187229386 }{ 57334 }{ ( 
  12 , 80 , 133 , 18 , 168 , 97 , 
   6 , 161 ) }{ 20 }{}
\myappline

                  \ack 

              \references  

\end{document}